%% file: TOP-16-014_temp.tex
\pdfoutput=1

\documentclass[11pt,twoside,a4paper,cmspaper,final,collab]{cms-tdr}
\def\svnVersion{463215P}\def\cmsCernNoTag{CERN-EP-2018-013}\def\cmsCernDate{\today}\def\cmsMessage{Published in the Journal of High Energy Physics as \href{http://dx.doi.org/10.1007/JHEP06(2018)002}{\doi{10.1007/JHEP06(2018)002.}}}

\begin{document}\cmsNoteHeader{TOP-16-014}

\hyphenation{had-ron-i-za-tion}
\hyphenation{cal-or-i-me-ter}
\hyphenation{de-vices}
\newcommand{\ejets}{\Pe+jets\xspace}
\newcommand{\mujets}{\Pgm+jets\xspace}
\newcommand{\Vjets}{V+jets\xspace}
\newcommand{\bquark}{\cPqb\xspace}
\newcommand{\ST}{\ensuremath{S_{\text{T}}}\xspace}
\newcommand{\WPT}{\ensuremath{p_{\text{T}}^{\PW}}\xspace}
\newcommand{\LPT}{\ensuremath{p_{\text{T}}^{\ell}}\xspace}
\newcommand{\LETA}{\ensuremath{\abs{\eta^{\ell}}}\xspace}
\newcommand{\NJET}{\ensuremath{N_{\text{jets}}}\xspace}
\newcommand{\Lum}{\ensuremath{35.9\fbinv}}
\newcommand{\com}{\ensuremath{\sqrt{s}=13\TeV}\xspace}
\newcommand{\pb}{\unit{pb}}
\newcommand{\powhegpythia}{\POWHEG{}+\PYTHIA{}\xspace}
\newcommand{\powhegherwig}{\POWHEG{}+\HERWIGpp{}\xspace}
\newcommand{\mgamc}{\textsc{mg5}\_a\textsc{mc@nlo}\xspace}
\newcommand{\mgamcMLM}{\textsc{mg5}\_a\textsc{mc@nlo-lo}\xspace}
\newcommand{\mgamcMLMpythia}{\textsc{mg5}\_a\textsc{mc@nlo}-\textsc{lo}+\PYTHIA{}\xspace}
\newcommand{\mgamcFxFx}{\textsc{mg5}\_a\textsc{mc@nlo}-\textsc{nlo}\xspace}
\newcommand{\mgamcFxFxpythia}{\textsc{mg5}\_a\textsc{mc@nlo}-\textsc{nlo}+\PYTHIA{}\xspace}
\newcommand{\LO}{LO\xspace}
\newcommand{\NLO}{NLO\xspace}
\newcommand{\NNLO}{NNLO\xspace}
\newcommand{\nnpdf}{NNPDF30\_nlo\_as\_0118\xspace}
\newcommand{\nnpdflo}{NNPDF30\_lo\_as\_0130\xspace}
\newcommand{\cuettune}{CUETP8M2T4\xspace}
\newcommand{\hdamp}{\ensuremath{h_{\text{damp}}}\xspace}
\newcommand{\bTransferFunction}{\ensuremath{x_{\mathrm{b}} = \pt(\PB)/\pt(\text{\bquark jet})}\xspace}
\newcommand{\deltaRDefn}{\ensuremath{\Delta R=\sqrt{\smash[b]{(\Delta\eta)^2+(\Delta\phi)^2}}}\xspace}
\newcommand{\Irel}{\ensuremath{I_{\text{rel}}}\xspace}
\newcommand{\chis}{\ensuremath{\chi^2}\xspace}
\newcommand{\pvalue}{\textit{p}-value\xspace}
\newcommand{\particleLifetime}{\ensuremath{30\unit{ps}}\xspace}
\ifthenelse{\boolean{cms@external}}{\providecommand{\NA}{\ensuremath{\cdots}\xspace}}{\providecommand{\NA}{\ensuremath{\text{---}}\xspace}}

\RCS$Revision: 463130 $
\RCS$HeadURL: svn+ssh://svn.cern.ch/reps/tdr2/papers/TOP-16-014/trunk/TOP-16-014.tex $
\RCS$Id: TOP-16-014.tex 463130 2018-06-04 09:19:50Z doburns $

\cmsNoteHeader{TOP-16-014}
\title{Measurements of differential cross sections of top quark pair production as a function of kinematic event variables in proton-proton collisions at \com}

\date{\today}

\abstract{
	Measurements of differential \ttbar production cross sections are presented in the single-lepton decay channel, as a function of a number of kinematic event variables.
	The measurements are performed with proton-proton collision data at \com, collected by the CMS experiment at the LHC during 2016, with an integrated luminosity of \Lum.
	The data are compared to a variety of state-of-the-art leading-order and next-to-leading-order \ttbar simulations.
}

\hypersetup{
pdfauthor={CMS Collaboration},
pdftitle={Measurement of differential cross sections of top quark
pair production as a function of kinematic event variables in pp
collisions at sqrt(s) = 13 TeV},
pdfsubject={CMS},
pdfkeywords={CMS, physics, top quarks}}

\maketitle

\section{Introduction}
\label{sec:Intro}

In 2016 the CERN LHC collided protons at \com, resulting in a data set recorded by the CMS experiment~\cite{Chatrchyan:2008zzk}, with an integrated luminosity of \Lum.
Approximately 30 million top quark-antiquark pairs (\ttbar) are present in this data set, which allows detailed studies of the production properties of \ttbar events to be performed.

Measurements of kinematic distributions in \ttbar events are important for verifying current theoretical models of \ttbar production and decay.
As \ttbar production and top quark decay can be a significant source of background events in many searches for physics beyond the standard model, for example in searches for supersymmetric models with top-quark-like signatures, it is important that \ttbar production be well understood and modeled.
In addition to physics beyond the standard model, a good understanding of \ttbar production is necessary for measurements of rare standard model processes, such as \ttbar production in association with a \PW, \PZ, or Higgs boson.

In this paper, we present measurements of differential \ttbar production cross sections, as a function of kinematic event variables that do not require the reconstruction of the \ttbar system.
Events are considered when the final state includes exactly one isolated lepton ($\ell = \Pe$ or \Pgm) with large transverse momentum \pt and at least four jets, of which at least two are tagged as originating from a bottom (\bquark) quark.
The kinematic event variables are the jet multiplicity (\NJET), the scalar sum of the jet \pt (\HT), the scalar sum of the \pt of all particles (\ST), the transverse momentum imbalance (\ptmiss), the magnitude of the \pt of the leptonically decaying \PW\ boson (\WPT), and the magnitudes of the \pt and pseudorapidity of the lepton (\LPT and \LETA).

The measurements of the differential \ttbar production cross sections are presented at particle level, \ie with respect to generated ``stable'' particles (with a mean lifetime longer than \particleLifetime), in a phase space that closely resembles that accessible by the CMS detector (the visible phase space).  This avoids the influence of large theoretical uncertainties that would be introduced by extrapolating the measurements to a larger phase space, or by presenting the measurements at parton level.

Several measurements of the differential \ttbar production cross sections as a function of the properties of the \ttbar system and of the jet activity in \ttbar events have been performed at the LHC, at 7 and 8\TeV~\cite{Diff7TeV,Diff8TeV,DileptonDoubleDiff8TeV, LjetHighPtDiff8TeV, DiffAllJet8TeV}, and 13\TeV~\cite{LJetsDiff13TeV, DileptonDiff13TeV, ATLASLJetsDiff13TeV, ATLASDileptonDiff13TeV}.  Measurements with respect to kinematic event variables in \ttbar events have been performed with the CMS detector at 7 and 8\TeV~\cite{TOP12042}.

\section{The CMS detector}
\label{sec:cms}

The central feature of the CMS apparatus is a superconducting solenoid of 6\unit{m} internal diameter, providing a magnetic field strength of 3.8\unit{T}. Within the solenoid volume are a silicon pixel and strip tracker, a lead tungstate crystal electromagnetic calorimeter (ECAL), and a brass and scintillator hadron calorimeter, each composed of a barrel and two endcap sections.
Forward calorimeters extend the $\eta$ coverage provided by the barrel and endcap detectors.
Muons are measured in gas-ionization detectors embedded in the steel flux-return yoke outside the solenoid.
A more detailed description of the CMS detector, together with a definition of the coordinate system used and the relevant kinematic variables, can be found in Ref.~\cite{Chatrchyan:2008zzk}.

Events of interest are selected using a two-tiered trigger system~\cite{CMSTrigger}.
The first level, composed of custom hardware processors, uses information from the calorimeters and muon detectors to select events at a rate of around 100\unit{kHz} within a time interval of less than 4\mus.
The second level, known as the high-level trigger (HLT), consists of a farm of processors running a version of the full event reconstruction software optimized for fast processing, and reduces the event rate to around 1\unit{kHz} before data storage.

\section{Signal sample and background simulation}
\label{sec:DataAndSim}

Two independent \ttbar samples are simulated with the \POWHEG (v2) generator~\cite{Powheg_ref2, Powheg_ref1, Powheg_ref3, Powheg_hvq},
which utilizes next-to-leading-order (\NLO) matrix-element calculations.
One sample uses \PYTHIA (v8.212)~\cite{Pythia6, Pythia8} with the CUETP8M2T4 tune~\cite{CUETP8M2T4_Tune} for the simulation of the parton shower and hadronization.
The second has parton showering and hadronization performed by \HERWIGpp (2.7.1)~\cite{Herwigpp} using the tune EE5C~\cite{EE5C}.

Two additional independent simulated \ttbar samples are produced with the \mgamc (v2.2.2) generator~\cite{MGamc}.
In the first, \mgamc\ is used to generate events at leading-order (\LO) accuracy with up to three additional partons, and \PYTHIA is employed with the CUETP8M1 tune~\cite{CUETP8M1} for parton showering and hadronization.
The MLM jet-parton matching algorithm~\cite{MLM} is used in this sample, referred to as \mgamcMLM.
In the second, \mgamc\ simulates events to \NLO accuracy with up to two additional partons, where parton showering and hadronization are performed using \PYTHIA with the CUETP8M2T4 tune.
The FxFx jet-parton matching algorithm~\cite{FxFx} is used, and this sample is referred to as \mgamcFxFx.
It is important to compare multiple \ttbar generators in order to find the current most suitable description of top quark production and decay, and to identify any discrepancies in the models.

\sloppypar{
In all simulated \ttbar samples, the top quark mass is set to 172.5\GeV.
The \nnpdf parton distribution function (PDF) set is used for the \NLO samples while the \nnpdflo set is used for the \LO samples~\cite{nnpdf}.
When comparing with reconstructed data, a cross section of $832^{+20}_{-29}\,(\text{scale})\pm{35}\,(\mathrm{PDF} + \alpS)\pb$ is used to normalize
the \ttbar samples, where \alpS is the strong coupling constant.
This \ttbar cross section is calculated to next-to-next-to-leading-order (\NNLO) accuracy in quantum chromodynamics (QCD) including resummation of next-to-next-to-leading logarithmic soft-gluon terms with \textsc{Top++} (v2.0)~\cite{ttxsec_1,ttxsec_2,ttxsec_3,ttxsec_4,ttxsec_5,ttxsec_6,ttxsec_7}.
The scale uncertainty in this \ttbar cross section comes from the independent variation of the factorization and renormalization scales.
}

The dominant background processes to \ttbar production, \ie the production of single top quarks and the production of vector bosons in association with jets, are also simulated.
Single top quark processes are generated with \POWHEG interfaced with \PYTHIA, and are normalized to cross sections that are calculated to NLO precision~\cite{stxsec_1,stxsec_2}.
Separate samples are generated for $t$- and $s$-channel production~\cite{Powheg_ST_tch_4f,Powheg_ST_sch}.
The sample of single top quarks in association with a \PW~boson is produced with \POWHEG (v1)~\cite{Powheg_ST_tW}.
In this sample, the diagram removal scheme~\cite{DiagramRemoval} is used to avoid double counting of Feynman diagrams in the production of single top quarks in association with a \PW~boson at \NLO and top quark pair production.
Samples of \PW~and \PZ~boson production with leptonic final states, in association with jets (\Vjets), are generated with \mgamcMLM.
Separate samples are generated with exactly one, two, three, and four additional jets to ensure a large sample of events that are likely to mimic the signature of \ttbar production.
These samples are normalized to their NNLO cross sections~\cite{FEWZ}.

In addition, QCD multijet events are generated with \PYTHIA for matrix-element calculations, parton shower simulation, and hadronization.
To obtain a large sample of QCD multijet events that are likely to mimic the signature of \ttbar production in the single-lepton decay channel, only events with large electromagnetic activity or containing a muon are generated.
These samples are normalized to their LO cross sections and are used to create transfer factors from a control region to the signal region for a QCD background estimate based on data in the control region.
The CMS detector response for all simulated samples is modeled using \GEANTfour~\cite{GEANT}.

\section{Event reconstruction and selection}
\label{sec:EventReco}
Parallel selection paths are defined to target \ttbar events that decay to final states containing an electron (\ejets) or a muon (\mujets).
The HLT in the \ejets channel requires at least one isolated electron candidate with $\pt>32\GeV$ and $\abs{\eta}<2.1$.
The corresponding requirements in the \mujets channel are at least one isolated muon candidate with $\pt>24\ \GeV$ and $\abs{\eta}<2.4$.

Offline reconstruction and selection uses the particle-flow (PF) algorithm~\cite{CMS-PRF-14-001} to reconstruct and identify each individual particle with an optimized combination of information from the subdetectors of CMS.
In the \ejets channel, electron candidates are required to satisfy $\pt>34\GeV$ and $\abs{\eta}<2.1$.
Electron candidates whose energy deposition in the ECAL is in the transition region between the barrel and endcap regions of the ECAL are not considered due to less efficient electron reconstruction.
Electron candidates must also satisfy several identification criteria~\cite{eIDDescription} to suppress the rate of jets and converted photons that are identified incorrectly as electron candidates.
In addition, electron candidates must be isolated.
To calculate the isolation, a cone of size \deltaRDefn$=0.3$ is constructed around the electron direction, where $\phi$ is the azimuthal angle.
The sum of the \pt of all PF candidates within this cone is calculated, excluding the lepton candidate and is corrected for the effects of additional proton-proton collisions within the same or nearby bunch crossings.
The relative isolation variable \Irel is defined as the ratio of this sum to the electron \pt, and is required to be less than 6\%.

In the \mujets channel, muon candidates are required to satisfy $\pt>26\GeV$ and $\abs{\eta}<2.4$.
Similarly to the electron candidates, muon candidates must satisfy additional identification criteria~\cite{muIDDescription}.
Muon candidates must be isolated, satisfying $\Irel<15\%$ where \Irel is defined as for electrons, but with a cone of size $\DR=0.4$.

For both electron and muon candidates, the lepton must be associated with the primary interaction vertex of the event.
The primary interaction vertex is defined as the reconstructed vertex associated with the largest sum of $\pt^2$ from physics objects that have been defined using information from the tracking detector, including jets, the associated missing transverse momentum, which was taken as the negative vector sum of the \pt of those jets, and charged leptons.

The trigger, reconstruction and identification efficiencies for both electrons and muons are measured in data, and corrected in simulation to match those seen in data. The efficiencies are calculated using the tag-and-probe method~\cite{TnP} from events containing a \PZ boson.
The total lepton correction factors are between 0.95 and 1.

Jets are clustered from PF candidates with the anti-\kt algorithm~\cite{Cacciari:2008gp} implemented in the \textsc{FastJet} package~\cite{FASTJET}, with a distance parameter of 0.4.
The jet momentum is determined as the vector sum of the \pt of all PF candidates in the jet.
A correction is applied to jet energies to take into account the contribution from additional proton-proton interactions using the charged hadron subtraction method~\cite{JECJERUnc}.
The measured energy of each jet is corrected for known variations in the jet energy response as a function of the measured jet $\eta$ and $\pt$.
The jet energy resolution (JER) is corrected in simulation to match that seen in data.
Jets are required to satisfy $\pt>30\GeV$ and $\abs{\eta}<2.4$.
Jets closer than $\Delta R = 0.4$ to identified isolated leptons are removed, as they are likely to have originated from the lepton itself.

The combined secondary vertex algorithm~\cite{BTag,BTagRun2} is used to identify jets originating from a \bquark quark.
The threshold of the algorithm is chosen such that the identification efficiency (in simulation) of genuine \bquark quark jets is $\approx$70\%,
and the probability to mistag a light quark or gluon jet is $\approx$1\%. The identification efficiency of \bquark quark jets in simulation is corrected to match that seen in data.

The distribution of the number of additional proton-proton interactions in simulation is corrected to match data.
Events must contain exactly one high-\pt, isolated electron or muon.
Events are vetoed if they contain an additional isolated lepton candidate with $\pt>15\GeV$ and $\abs{\eta}<2.4$.
Events must also contain at least four jets, at least two of which are required to be identified as originating from a \bquark quark.

\section{Cross section measurement}
\label{sec:measurement}

As stated in Section \ref{sec:Intro}, the differential \ttbar production cross sections are measured as a function of the kinematic event variables: \NJET, \HT, \ST, \ptmiss, \WPT, \LPT and \LETA.
The \NJET variable is the total number of jets in the event with $\pt>30\GeV$ and $\abs{\eta}< 2.4$.
The variable \HT is the scalar sum of the \pt of these jets.
The quantity \ptmiss is defined as the magnitude of \ptvecmiss, the transverse projection of the negative vector sum of the momenta of all reconstructed PF candidates in an event.
The \LPT and \LETA variables are magnitudes of the transverse momentum and the pseudorapidity of the lepton in the event, respectively.
The variable \ST is the sum of \HT, \ptmiss, and \LPT.
The variable \WPT is the magnitude of the transverse momentum of the leptonically decaying \PW\ boson, which is constructed from ${\vec p}_{\mathrm{T}}^{\ell}$ and \ptvecmiss.

\begin{figure}
	\centering
	\includegraphics[width=0.49\linewidth]{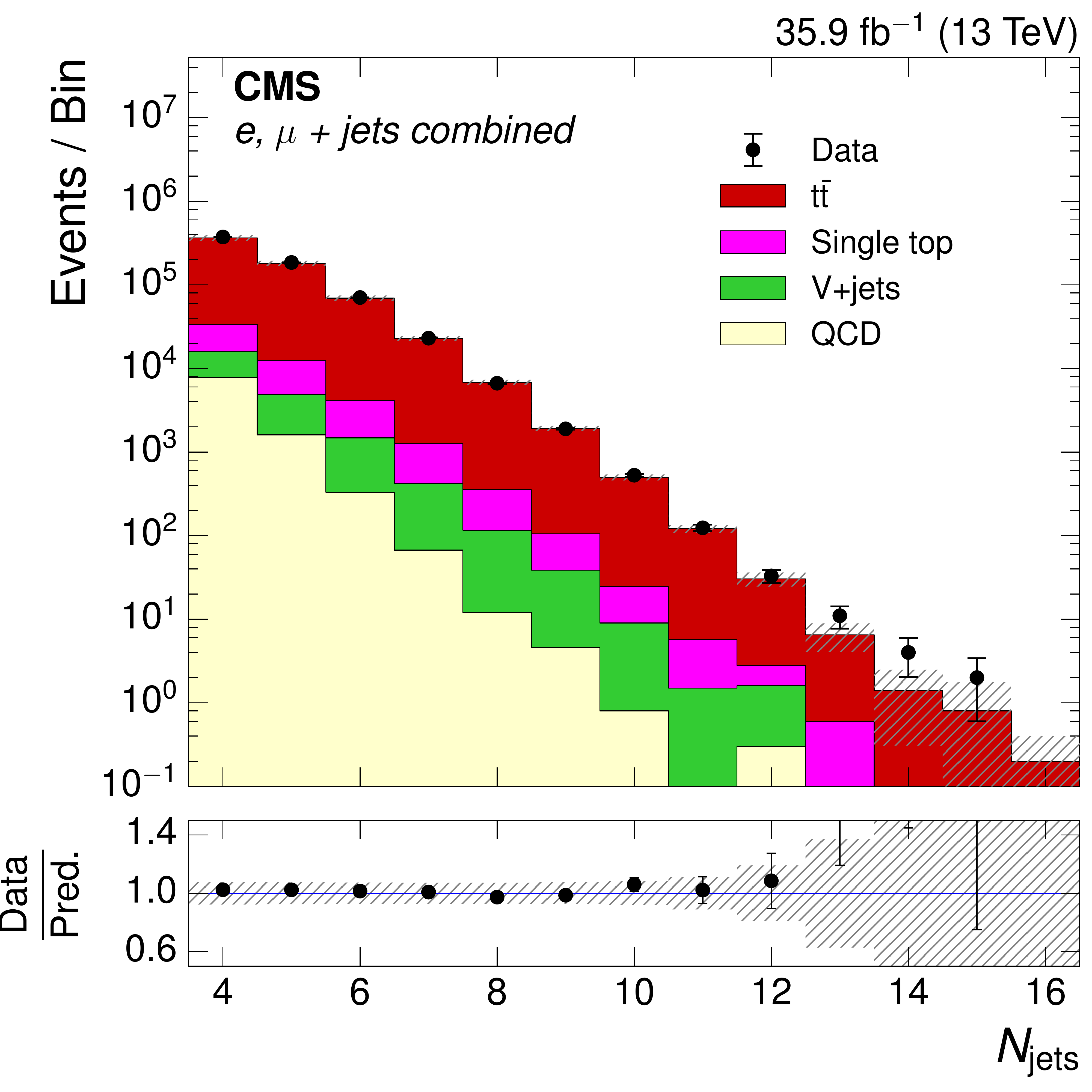} \\
	\includegraphics[width=0.49\linewidth]{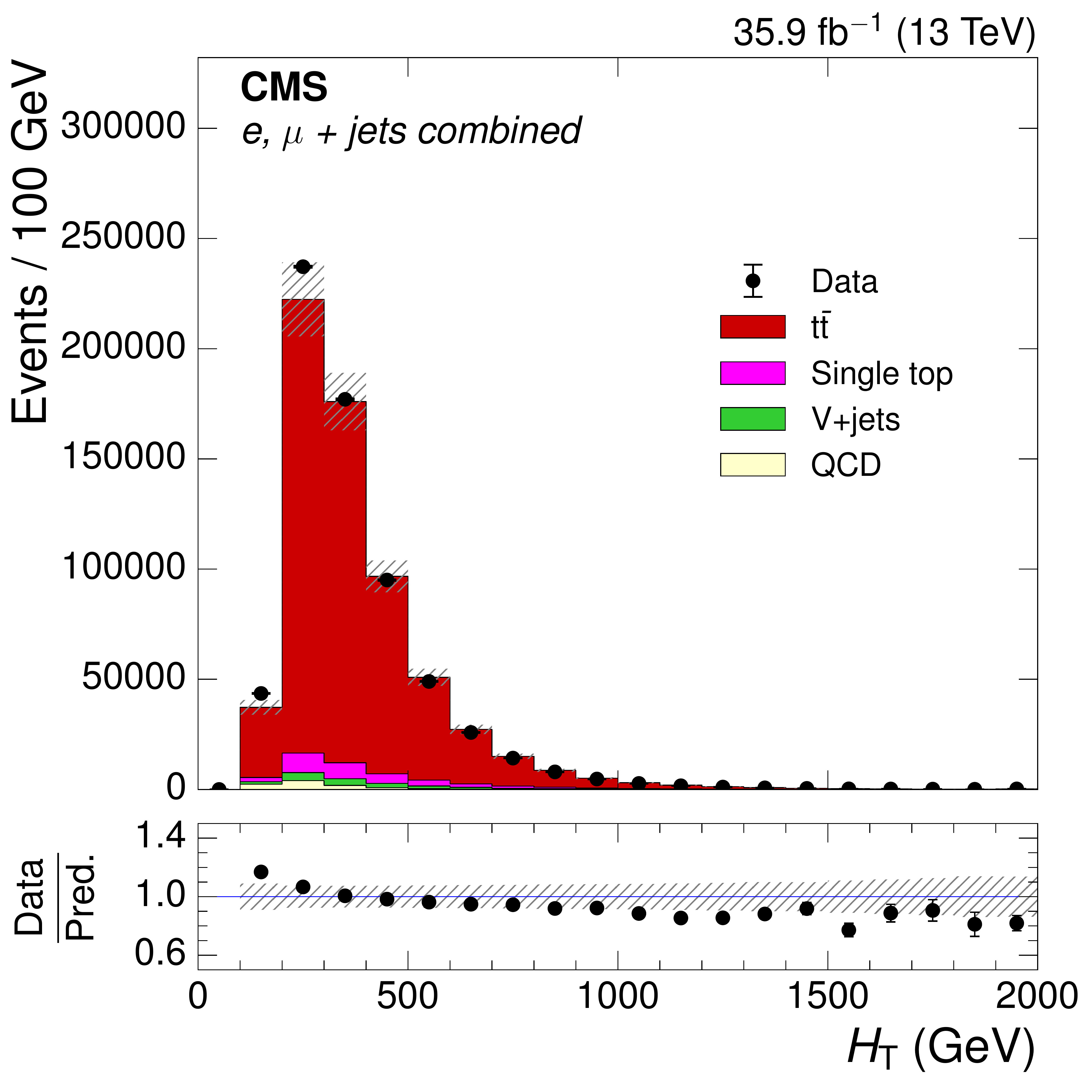}
	\includegraphics[width=0.49\linewidth]{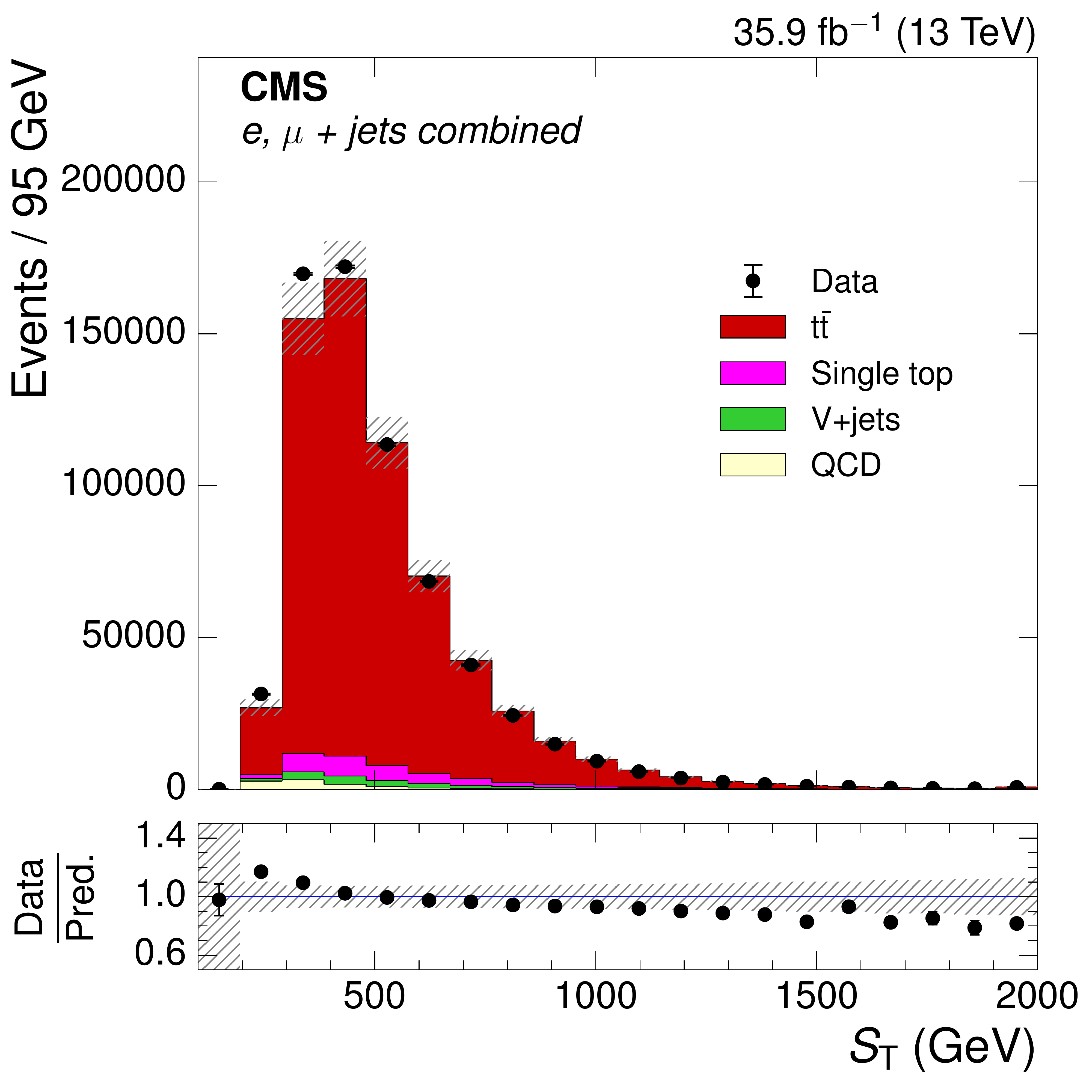}
	\caption{The distributions of \NJET, \HT and \ST after full event selection. The \ttbar simulation is normalized to the NNLO prediction. The ratio of the number of events in data to that in simulation is shown below each of the distributions, with the statistical uncertainty in the data shown by the vertical uncertainty bars.  The statistical uncertainty in the number of simulation events and the uncertainties in the modeling in simulation are shown by the hatched band.}
	\label{plt:SR1}
\end{figure}

\begin{figure}
	\centering
	\includegraphics[width=0.49\linewidth]{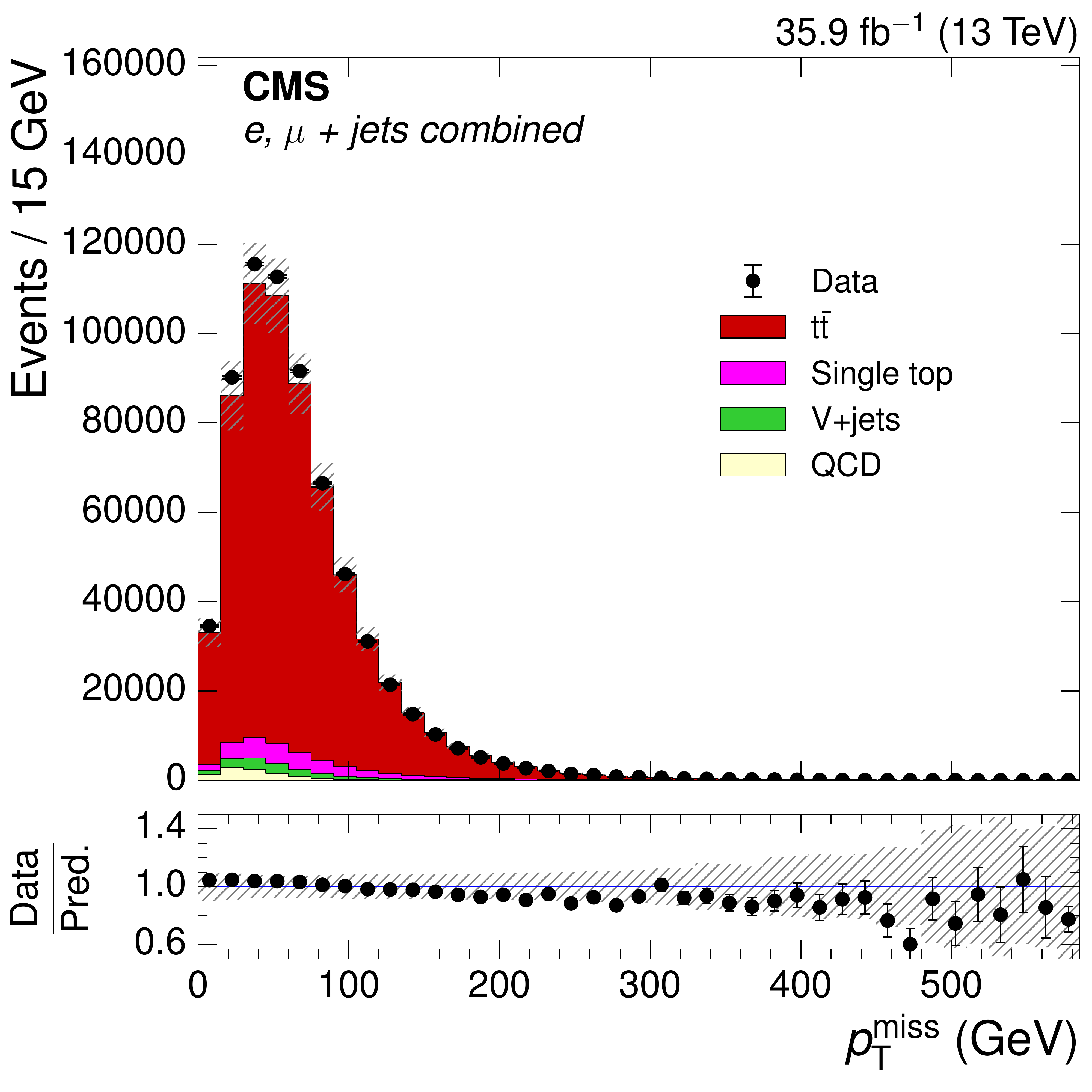}
	\includegraphics[width=0.49\linewidth]{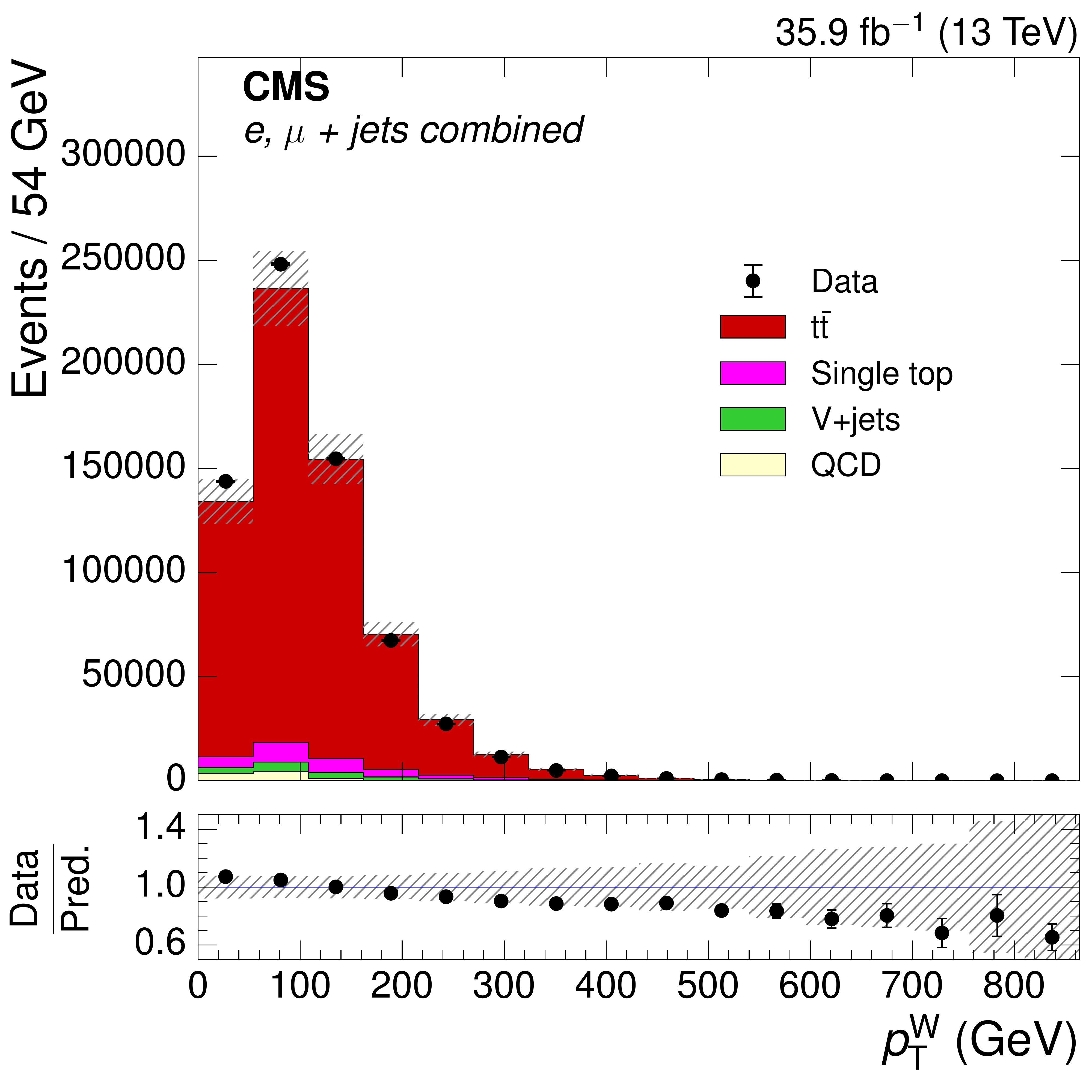} \\
	\includegraphics[width=0.49\linewidth]{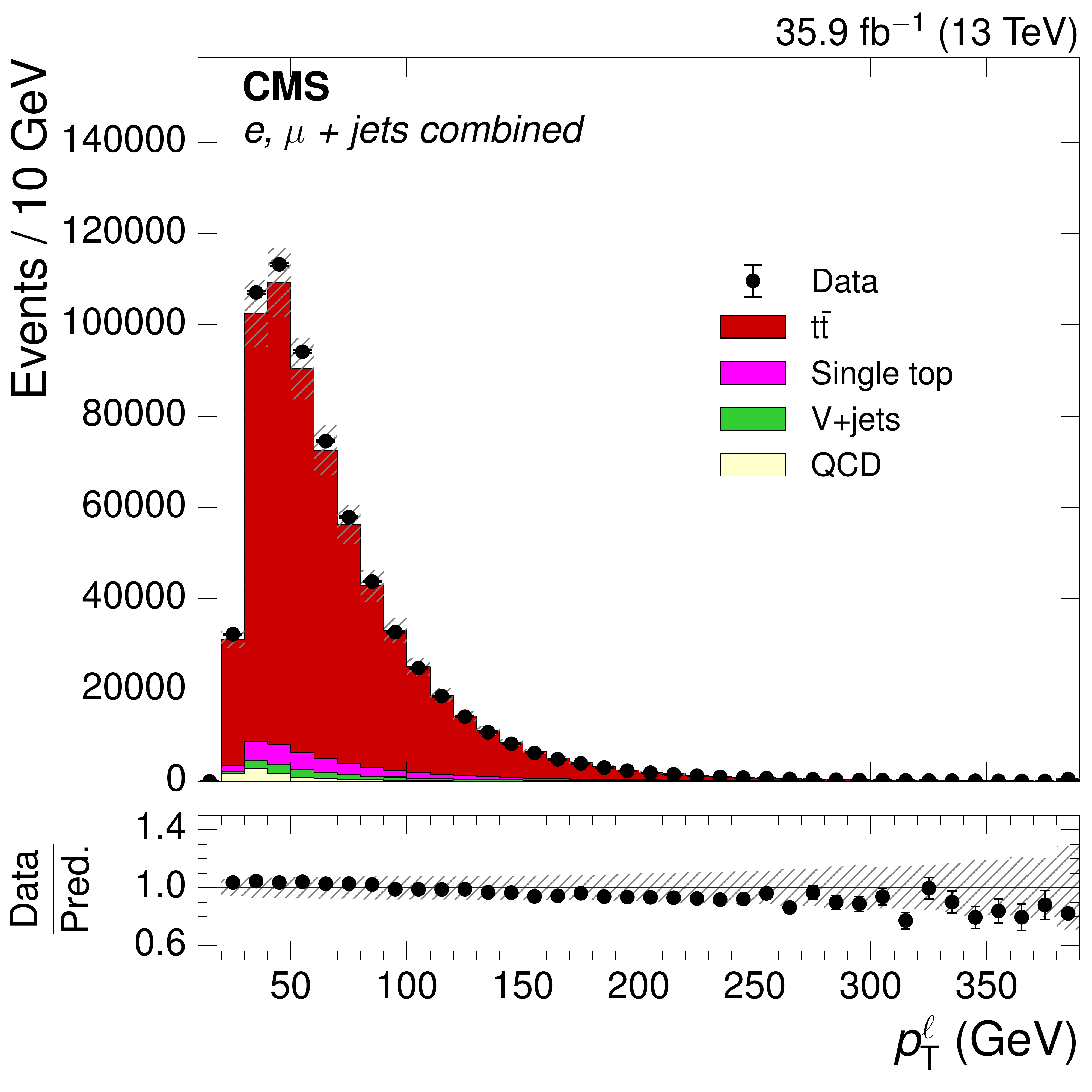}
	\includegraphics[width=0.49\linewidth]{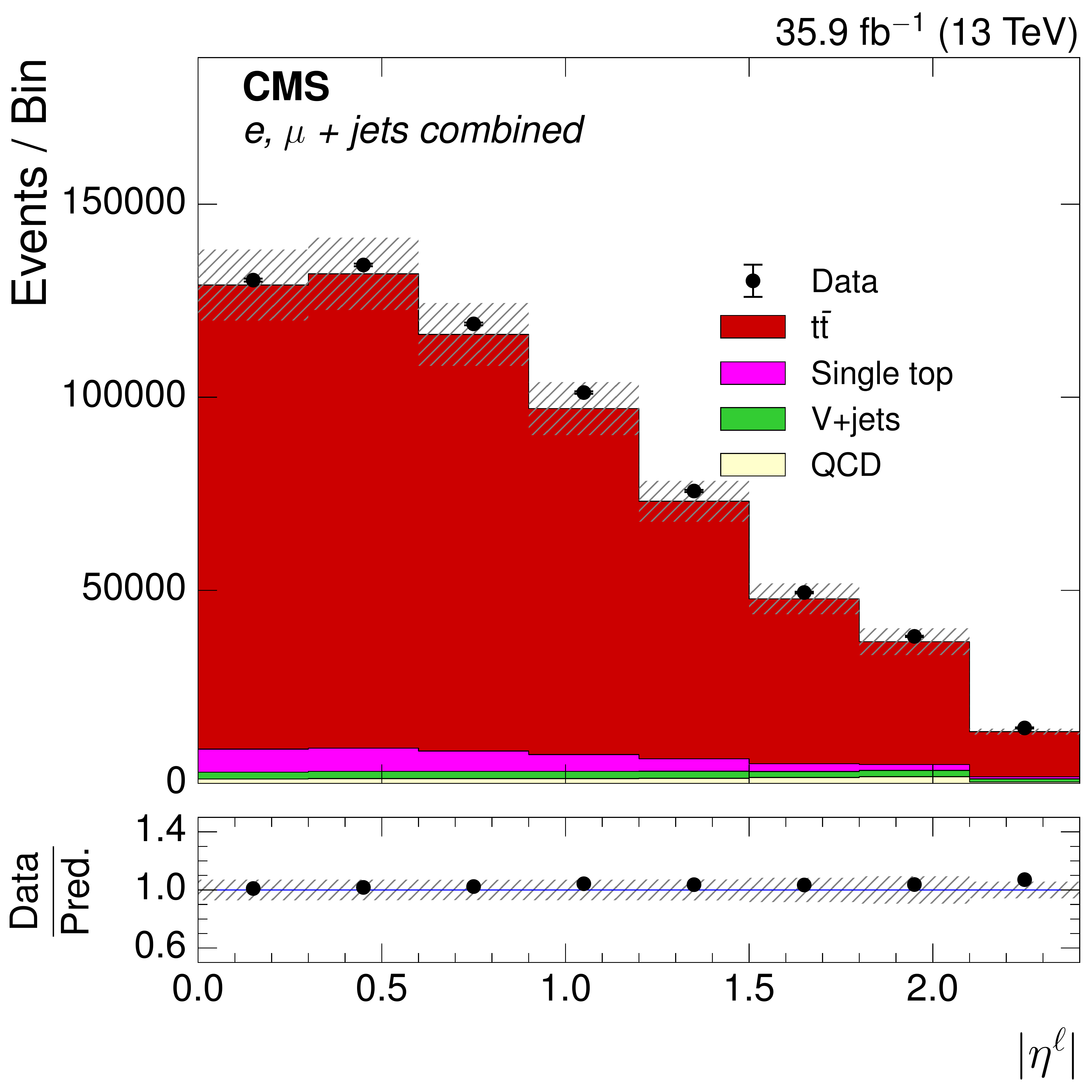}
	\caption{The distributions of \ptmiss, \WPT, \LPT and \LETA after full event selection. The \ttbar simulation is normalized to the NNLO prediction. The ratio of the number of events in data to that in simulation is shown below each of the distributions, with the statistical uncertainty in the data shown by the vertical uncertainty bars.  The statistical uncertainty in the number of simulation events and the uncertainties in modeling in simulation are shown by the hatched band.}
	\label{plt:SR2}
\end{figure}

The distributions of these variables measured in data are shown in Figs.~\ref{plt:SR1} and \ref{plt:SR2}, and are compared to the sum of signal and background events from simulation.
A total of 662\,381 events are measured in data, of which 92.1\% are predicted from the \powhegpythia\ simulation to be \ttbar events.
Single top quark production and \Vjets production contribute 4.4\% and 2.1\% to the total number of events, respectively, as estimated from simulation.
The component of multijet QCD events is estimated from control regions in the data, and comprises approximately 1.4\% of the total number of events.
The control regions are designed to obtain data samples that are enriched in
QCD multijet events that are kinematically similar to the signal region, but
with little contamination from \ttbar, single top quark, and \Vjets events.  In the \ejets channel,
the control region is obtained by inverting the isolation criterion on electron candidates.
In the \mujets channel, the control region is obtained by requiring muon candidates to satisfy $0.15<\Irel<0.30$.  In the control regions for both channels, the number of \bquark-tagged jets is also required to be exactly zero.
The contribution of \ttbar, single top quark and \Vjets events to the control regions ($\approx$15--20\%) is estimated from simulation with all corrections and subtracted from the data.  The ratio of the number of multijet QCD events in the control region to that in the signal region (the transfer factor), both predicted from simulation, is then used to scale the normalization of the data control region to obtain the multijet QCD estimate in the signal region.
Other sources of background are negligible, and are not considered in this measurement.
The level of agreement between the total event count of data and simulation, within 0.2\% , indicates that the total cross section is compatible to that stated in Section~\ref{sec:DataAndSim}.

Previous measurements~\cite{Diff7TeV,Diff8TeV,DileptonDoubleDiff8TeV,LjetHighPtDiff8TeV,DiffAllJet8TeV,LJetsDiff13TeV,DileptonDiff13TeV} report that the top quark \pt spectrum in data is softer than that predicted by NLO simulation.
This effect can be seen in some of the distributions in Figs.~\ref{plt:SR1} and \ref{plt:SR2}, where distributions correlated with the top quark \pt are also softer in data than those predicted by the simulation.

\subsection{Particle level and visible phase space definitions}

The results are presented at particle level, \ie with respect to the stable particles produced in simulation by the event generator, before detector interactions are modeled.
The generator-level definitions for the particles and visible phase space are based on the RIVET framework~\cite{Buckley:2010ar}, following the prescriptions adopted in Ref.~\cite{Collaboration:2267573}.
Generated electrons and muons not originating from a hadron or a quark are used to define electrons and muons at particle level.
Photons that are near the lepton are assumed to have radiated from it, and are clustered together with the anti-\kt algorithm with a distance parameter of 0.1.

Particle-level jets are constructed by clustering all stable particles, excluding the lepton, with the anti-\kt algorithm using a distance parameter of 0.4.
To determine if a particle-level jet originated from a \bquark quark, \bquark hadrons are included in the clustering of jets, but with the magnitude of the four-momentum of the \bquark hadron scaled to a negligible value.
The b hadrons can then be clustered into jets without affecting the kinematic properties of the jet.
A jet with a \bquark hadron among its constituents is considered to have originated from a \bquark quark.
The particle-level \ptmiss is calculated from all stable visible particles.

The differential \ttbar production cross sections are measured in a visible phase space, which is chosen to be the same for both \ejets and \mujets channels, and to closely resemble the criteria used to select events in data.
Particle-level objects are used to define the common visible phase space of \ttbar events for both \ejets and \mujets channels,
all within $\abs{\eta}<2.4$, which requires exactly one electron or muon with $\pt>26\GeV$, and no additional electrons or muons with $\pt>15\GeV$.
The event must also contain at least three particle-level jets with $\pt>30\GeV$, and one jet with $\pt>20\GeV$.
Two of these particle-level jets must also be tagged as originating from a \bquark quark.
The \HT, \ST, and \NJET variables are calculated at the particle level with respect to all particle-level jets with $\pt>20\GeV$ and $\abs{\eta}< 2.4$.
This choice of particle-level phase space is made to obtain the largest possible data sample, and the uncertainty in
the resulting extrapolation makes only a small contribution to the uncertainty in the final results.

The yield of \ttbar events for each bin in data is obtained by subtracting the contribution of each background process.
The contribution of \ttbar events that satisfy the selection criteria, but do not enter the visible phase space at particle level, is estimated from simulation and also subtracted from the data.
This amounts to approximately 7\% of all \ttbar events and are predominately those in which one of the jets fails the particle-level jet selection, but passes the reconstructed jet selection because of the resolution of the detector.
No selection is applied on the decay channel of the top quarks, so the phase space does not exclusively contain semileptonic (electron or muon) \ttbar events.
In particular, there are contributions from events where one top quark decays to a tau lepton and subsequently to an electron or muon, or where both top quarks decay leptonically but one lepton is not within the acceptance.

\subsection{Unfolding and cross section calculation}
\label{sec:Unfold}

For each kinematic event variable the yield of \ttbar events in each bin is unfolded to correct for the detector acceptance, efficiency, and bin-to-bin migrations stemming from the detector resolution to obtain the yield of \ttbar events in the visible phase space at the particle level.
The bin widths are chosen to give a low level of bin-to-bin migration, and are always greater than the detector resolution.

A response matrix, constructed using the \powhegpythia sample, relates the kinematic event distributions at reconstruction level to those at particle level.
The response matrix also includes efficiency and acceptance corrections.
Unfolding is performed by inverting the response matrix, based on a least-squares fit with Tikhonov regularization, implemented in the \textsc{TUnfold} software framework~\cite{TUnfold}.
Regularization dampens nonphysical fluctuations in the unfolded \ttbar yields, and the regularization parameter is chosen by minimizing the average global statistical correlation between the bins of each variable.
The typical regularization parameters are found to be of order $10^{-4}-10^{-3}$, and significantly lower for the \LETA variable.

The yields of \ttbar events are unfolded separately in the \ejets and \mujets channels and then combined after unfolding, giving the total number of \ttbar events at particle level in the visible phase space, $N_{\ttbar}$.
The normalized differential cross section with respect to each variable, $X$, can then be calculated using
\begin{equation} \label{eq:normalised}
\frac{1}{\sigma_{\ttbar}^\text{vis}} \frac{\rd\sigma_{\ttbar}^{i}}{\rd X} = \frac{1}{\sum_{j}N_{\ttbar}^{j}}\frac{N_{\ttbar}^{i}}{\Delta X^{i}},
\end{equation}
where $\sigma_{\ttbar}^\text{vis}$ is the total \ttbar production cross section in the visible phase space, $\sigma_{\ttbar}^{i}$ is the \ttbar production cross section in bin $i$, $N_{\ttbar}^{i (j)}$ is the number of \ttbar events in bin $i (j)$ after unfolding, and $\Delta X^{i}$ is the width of bin $i$. The absolute differential cross section can be calculated as
\begin{equation} \label{eq:absolute}
\frac{\rd \sigma_{\ttbar}^{i}}{\rd X} = \frac{N_{\ttbar}^{i}}{\Lumi\,\Delta X^{i}  },
\end{equation}
where $\Lumi$ is the integrated luminosity of the data.

\section{Systematic uncertainties}
\label{sec:syst}
Sources of systematic uncertainties are evaluated and propagated to the final result by recalculating the response matrix with a modified \ttbar simulation and/or by modifying the background predictions.

The uncertainty in the integrated luminosity of the data is estimated to be $\pm$2.5\%~\cite{CMS-PAS-LUM-17-001}.
The uncertainty in the number of additional inelastic interactions in the same or nearby bunch crossings is estimated by varying the total proton-proton inelastic cross section by $\pm$4.6\%~\cite{PileupUnc}.  This cross section is used in determining the distribution of additional inelastic interactions in data, which is used to correct the simulation.

The uncertainty in the efficiency of the \bquark quark jet identification and mistagging rate in the simulation is taken as the uncertainty in the \pt, $\abs{\eta}$, and flavor-dependent correction factors~\cite{BTagRun2}.
The uncertainties in the lepton trigger, reconstruction, and identification correction factors are similarly propagated to the final results.

The uncertainties in the jet energy scale (JES) and JER are estimated as functions of jet \pt and $\abs{\eta}$~\cite{JECJERUnc}.
The uncertainty in the JES is also propagated into the calculation of \ptmiss.
Additional uncertainties in the \pt of electrons, muons, tau leptons and other unclustered PF candidates, that are used in the calculation of \ptmiss, are considered and found to be negligible.

The uncertainties in the normalization of the single top quark and \Vjets background sources are based on measurements performed in~\cite{tchannelxsec,Wbb8TeV,Zbb8TeV} and take into account an extrapolation to the current analysis phase space.
They are estimated to be $\pm$30\% and $\pm$50\% respectively and typically result in a normalization uncertainty that is negligible.
The uncertainty in the normalization and shape of the multijet QCD background is estimated by using alternative control regions containing conversion electrons in the \ejets channel and muons with $I_{\text{rel}}>0.3$ in the \mujets channel.
This effectively varies the total normalization of the multijet QCD background by up to 60\%, and also the shape of the contribution by up to $\pm$30\% in any one bin, but is found to result in a negligible uncertainty after unfolding, except at large \LETA.

Uncertainties in the top quark mass are estimated by using simulated \ttbar samples where the top quark mass has been varied up and down by 1\GeV, which is comparable to the uncertainty in the measured top quark mass~\cite{pdg2016}.

{\tolerance=1200
The uncertainty from the PDF used in the \ttbar simulation is estimated by considering 100 independent replicas of \nnpdf. The RMS of the uncertainties originating from the variation of each replica is taken as the PDF uncertainty.  The uncertainty resulting from using the \nnpdf set derived with varied values of \alpS is combined in quadrature with the PDF uncertainty.
\par}

The uncertainty arising from the mismodeling of the top quark \pt spectrum is estimated by reweighting the \pt distribution in simulation to match that measured by the previous measurements~\cite{LJetsDiff13TeV,DileptonDiff13TeV}.
The reweighting varies the yield of simulated \ttbar events in the bins of the measurement by up to 20\%, and results in a negligible uncertainty in the measured cross section.

{\tolerance=1200
Several sources of uncertainty for the modeling of the parton shower in the simulated \powhegpythia sample are considered.
\par}

The uncertainty originating from the parton shower scale used when simulating the initial-state radiation is estimated by varying the scale up and down by a factor of two.
Similarly the uncertainty originating from the scale for final-state radiation, which is constrained by measurements made at the LEP collider~\cite{fsrTuning}, is estimated by varying the scale up and down by a factor of $\sqrt{2}$.
The renormalization and factorization scales used in the matrix-element calculations are also varied independently by factors of 0.5 and 2.
An additional variation is performed where both scales are varied simultaneously by the same factors.
The shower scale uncertainty is defined as the envelope of the parton shower scale uncertainties and the matrix-element scale uncertainties.

The systematic uncertainty in matching the matrix-element to the parton shower is determined by varying the parameter \hdamp, which regulates the high-\pt radiation by damping real emission generated in \POWHEG, within its uncertainties. The parameter is set to $\hdamp=1.58^{+0.66}_{-0.59}$ multiplied by the mass of the top quark in the CUETP8M2T4 tune~\cite{CUETP8M2T4_Tune}.
The parameters controlling the underlying event in the \cuettune tune are also varied to estimate the uncertainty in this source~\cite{CUETP8M2T4_Tune}.

The uncertainty in the modeling of the momentum transfer from \bquark quarks to \bquark hadrons is estimated by reweighting the tuned quantity \bTransferFunction for each particle-level \bquark-tagged jet within its uncertainties, where \pt(\PB) is the transverse momentum of the \bquark hadron, and \pt(\bquark jet) is the transverse momentum of the particle-level \bquark-tagged jet.
The difference when using an alternative model (the Peterson model~\cite{PetersonFragmentation}) for the fragmentation of \bquark quarks is also included as an additional uncertainty.
The energy response of \bquark jets is sensitive to the single-lepton branching fractions of \bquark hadrons, and the uncertainty originating from the choice of branching fractions in the \powhegpythia simulation is estimated by reweighting the branching fractions to those reported in Ref.~\citen{pdg2016}.

The effects of any mismodeling of the color reconnection in the simulation are estimated by comparing the cross sections obtained with samples including and excluding the effects of color reconnection on the decay products of the top quarks (Early resonance decays).
A comparison to two samples obtained with alternative models of color reconnection are also included, one where QCD color rules are considered in the simulation of the color reconnection (QCD-based)~\cite{QCDCR}, and another where gluons can be moved to different color strings during the simulation of the color reconnection (Gluon move)~\cite{GluonMoveCR}.

The statistical uncertainty arising from the finite size of the \powhegpythia sample, which is used to construct the nominal response matrix, is propagated to the final measurement.
This uncertainty is negligible.

Each source of systematic uncertainty is summarised for each variable in Table~\ref{tb:syst_condensed_combined_normalised}, where
the minimum and maximum relative uncertainty in the normalized differential cross section (over all bins) are shown.
The minimum and maximum of the total relative uncertainty over all bins are also shown.
Sources of uncertainties in the calculation of \ptmiss do not affect some distributions, and are indicated in the table by \NA.
The dominant uncertainty in the measurement of the normalized cross sections comes from the uncertainty in the JES. Other significant uncertainties come from the theoretical modeling of \ttbar production in simulation, in particular from the uncertainty in the shower scale for final-state radiation.
A similar table for the absolute differential cross section uncertainties is shown in Appendix~\ref{ap:absCondensed}.
The uncertainty in the JES is also significant in the measurements of the absolute cross sections, however the uncertainty in the final-state radiation scale becomes dominant.  The total uncertainty from all sources in the normalized cross section is typically below 5\% in each bin, and can be as large as 21\%.  For the measurements of the absolute cross section, the total uncertainty is typically 10\%, and can be as large as 22\%.

\begin{landscape}
\begin{table}
	\topcaption{ The upper and lower bounds, in \%, from each source of systematic uncertainty in the normalised differential cross section, over all bins of the measurement for each variable.  The bounds of the total relative uncertainty are also shown.}
	\label{tb:syst_condensed_combined_normalised}
	\centering
	\begin{tabular}{lccccccc}
		Relative uncertainty source $(\%)$	&	\NJET	&	\HT	&	\ST	&	\ptmiss	&	\WPT	&	\LPT	&	\LETA \vspace*{0.1cm}  \\
		\hline
	    \bquark quark tagging efficiency	&	0.1 -- 0.8	&	0.2 -- 1.1	&	0.2 -- 1.5	&	0.1 -- 1.2	&	0.1 -- 1.7	&	0.1 -- 1.9	&	0.1 -- 0.5\\
		Electron efficiency	&	0.1 -- 0.2	&	0.1 -- 0.7	&	0.1 -- 0.9	&	0.1 -- 0.7	&	0.1 -- 1.4	&	0.3 -- 2.1	&	0.1 -- 0.8\\
		Muon efficiency	&	0.1 -- 0.3	&	0.1 -- 0.2	&	0.1 -- 0.3	&	0.1 -- 0.2	&	0.1 -- 0.6	&	0.1 -- 1.0	&	0.1 -- 0.1\\
		JER	&	0.1 -- 0.6	&	0.1 -- 0.7	&	0.2 -- 1.8	&	0.6 -- 5.8	&	0.2 -- 2.1	&	0.1 -- 0.2	&	$<$0.1\\
		JES	&	0.1 -- 5.5	&	2.1 -- 13.6	&	2.1 -- 15.9	&	2.1 -- 7.1	&	0.5 -- 4.9	&	0.1 -- 2.0	&	0.1 -- 0.2\\
		Electron transverse momentum in \ptmiss	&	\NA	&	\NA	&	0.1 -- 0.3	&	0.1 -- 0.9	&	0.1 -- 0.6	&	\NA	&	\NA\\
		Muon transverse momentum in \ptmiss	&	\NA	&	\NA	&	0.1 -- 0.9	&	0.1 -- 3.5	&	0.1 -- 0.9	&	\NA	&	\NA\\
		Tau transverse momentum in \ptmiss	&	\NA	&	\NA	&	0.1 -- 1.4	&	0.1 -- 1.2	&	0.1 -- 1.4	&	\NA	&	\NA\\
		Unclustered transverse momentum in \ptmiss	&	\NA	&	\NA	&	0.1 -- 1.7	&	0.2 -- 1.9	&	0.1 -- 1.0	&	\NA	&	\NA\\
		QCD bkg cross section	&	0.1 -- 0.5	&	0.1 -- 1.0	&	0.1 -- 1.7	&	0.2 -- 0.6	&	0.1 -- 0.8	&	0.1 -- 4.5	&	0.2 -- 2.9\\
		QCD bkg shape	&	0.1 -- 0.1	&	0.1 -- 0.7	&	0.1 -- 1.0	&	0.1 -- 0.1	&	0.1 -- 1.5	&	0.1 -- 4.7	&	0.1 -- 1.5\\
		Single top quark cross section	&	0.1 -- 0.4	&	0.1 -- 2.1	&	0.1 -- 4.4	&	0.1 -- 4.9	&	0.1 -- 7.1	&	0.1 -- 6.0	&	$<$0.1\\
		V+jets cross section	&	0.1 -- 0.3	&	0.1 -- 2.5	&	0.1 -- 3.7	&	0.1 -- 2.0	&	0.1 -- 3.6	&	0.1 -- 5.4	&	0.1 -- 1.5\\
		PDF 	&	0.1 -- 0.3	&	0.1 -- 0.3	&	0.1 -- 0.6	&	0.1 -- 0.3	&	0.1 -- 0.4	&	0.1 -- 0.4	&	$<$0.1\\
		Color reconnection (Gluon move)	&	0.1 -- 2.8	&	0.1 -- 4.0	&	0.1 -- 11.7	&	0.2 -- 0.9	&	0.1 -- 1.0	&	0.2 -- 4.8	&	0.1 -- 0.4\\
		Color reconnection (QCD-based)	&	0.1 -- 2.0	&	0.1 -- 4.2	&	0.1 -- 6.6	&	0.4 -- 4.2	&	0.1 -- 3.5	&	0.1 -- 7.6	&	0.1 -- 1.2\\
		Color reconnection (Early resonance decays)	&	0.2 -- 3.9	&	0.1 -- 7.1	&	0.1 -- 4.1	&	0.1 -- 1.6	&	0.1 -- 3.8	&	0.1 -- 5.0	&	0.1 -- 1.0\\
		Fragmentation	&	0.1 -- 0.4	&	0.1 -- 0.5	&	0.1 -- 0.5	&	0.1 -- 0.6	&	0.1 -- 0.6	&	0.1 -- 0.4	&	$<$0.1\\
		\hdamp	&	0.3 -- 3.8	&	0.1 -- 3.1	&	0.2 -- 2.9	&	0.1 -- 2.3	&	0.1 -- 2.7	&	0.1 -- 2.8	&	0.2 -- 1.2\\
		Top quark mass	&	0.2 -- 1.0	&	0.1 -- 3.1	&	0.2 -- 3.5	&	0.1 -- 4.0	&	0.2 -- 1.1	&	0.2 -- 4.5	&	0.1 -- 0.6\\
		Peterson fragmentation model	&	0.1 -- 1.3	&	0.1 -- 0.6	&	0.1 -- 0.9	&	0.1 -- 1.1	&	0.1 -- 1.0	&	0.1 -- 1.3	&	$<$0.1\\
		Shower scales	&	0.4 -- 4.3	&	0.5 -- 4.5	&	0.5 -- 4.9	&	0.2 -- 2.4	&	0.3 -- 3.5	&	0.1 -- 4.5	&	0.1 -- 0.7\\
		\PB\ hadron decay semileptonic branching fraction	&	0.1 -- 0.1	&	0.1 -- 0.1	&	0.1 -- 0.1	&	$<$0.1	&	$<$0.1	&	$<$0.1	&	$<$0.1\\
		Top quark \ensuremath{\pt}	&	0.1 -- 0.7	&	0.1 -- 0.9	&	0.1 -- 1.0	&	0.1 -- 0.8	&	0.1 -- 0.9	&	0.1 -- 1.3	&	$<$0.1\\
		Underlying event tune	&	0.1 -- 2.7	&	0.1 -- 5.5	&	0.2 -- 4.4	&	0.1 -- 5.4	&	0.2 -- 2.6	&	0.2 -- 6.1	&	0.1 -- 0.9\\
		Simulated sample size	&	0.1 -- 1.6	&	0.1 -- 1.6	&	0.1 -- 1.9	&	0.1 -- 2.2	&	0.1 -- 1.4	&	0.1 -- 1.7	&	0.1 -- 0.4\\
		Additional interactions	&	0.1 -- 0.4	&	0.1 -- 1.0	&	0.1 -- 1.7	&	0.1 -- 1.5	&	0.1 -- 0.9	&	0.1 -- 1.0	&	$<$0.1\\
		Integrated luminosity	&	$<$0.1	&	$<$0.1	&	$<$0.1	&	$<$0.1	&	$<$0.1	&	$<$0.1	&	$<$0.1\\
		\hline
		Total	&	0.6 -- 9.6	&	2.7 -- 14.1	&	2.8 -- 17.4	&	2.9 -- 11.7	&	0.8 -- 12.6	&	0.7 -- 13.4	&	0.7 -- 4.4\\
	\end{tabular}
\end{table}
\end{landscape}
\clearpage

\section{Cross section results}
\label{sec:xsec}
\begin{figure}
	\centering
	\includegraphics[width=0.49\linewidth]{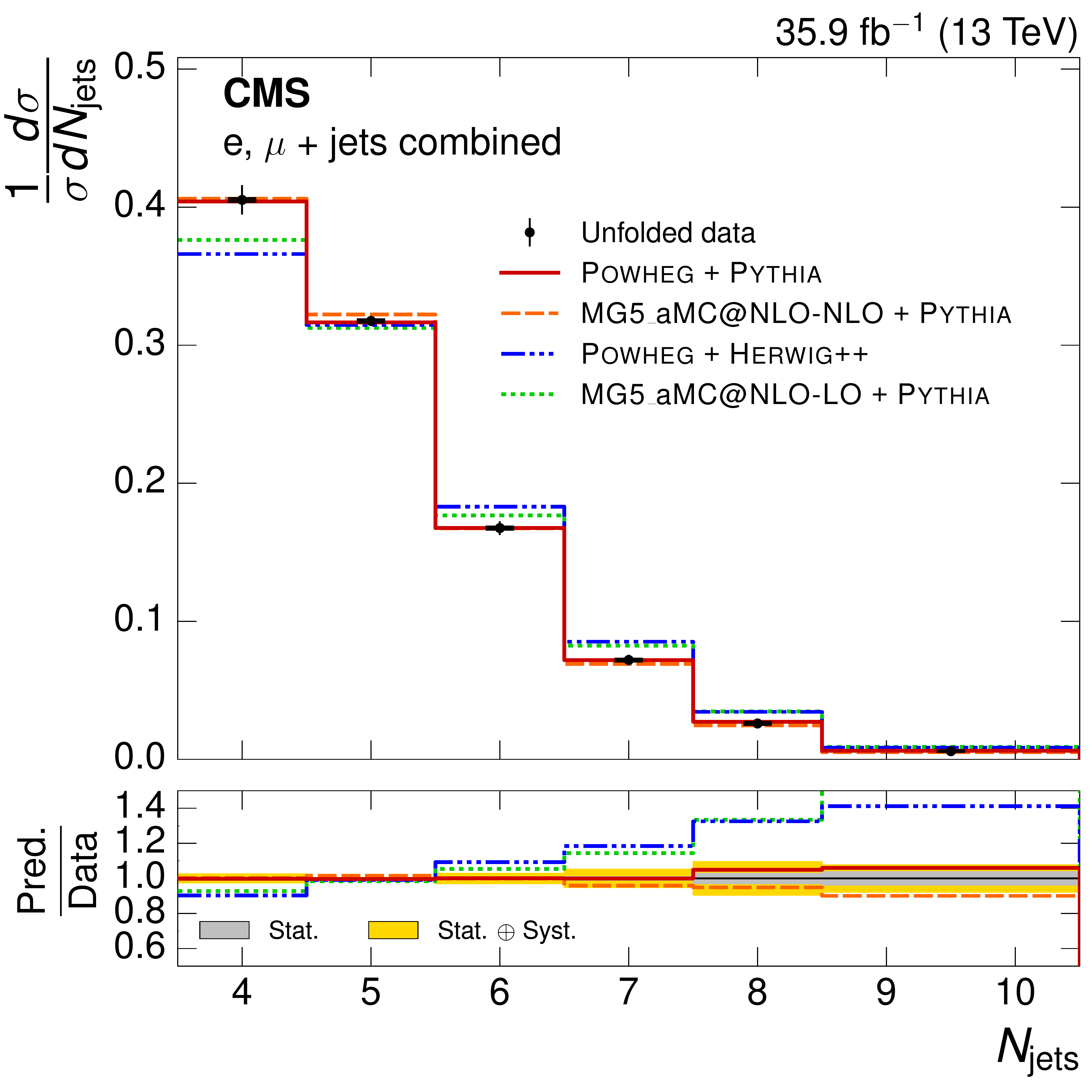}
	\includegraphics[width=0.49\linewidth]{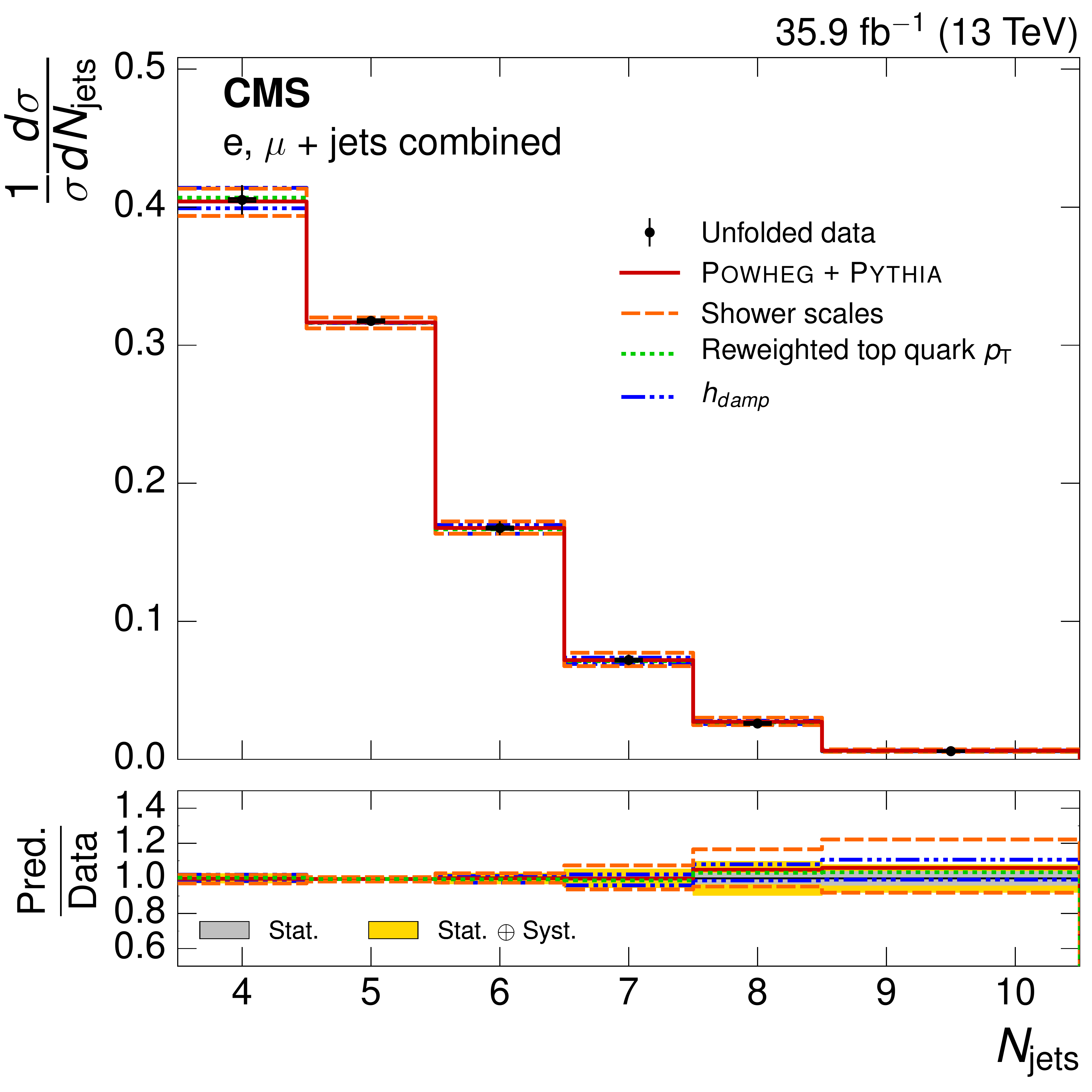}
	\caption{Normalized \NJET differential \ttbar cross section, compared to different \ttbar simulations in the left plot, and compared to the \powhegpythia simulation after varying the shower scales, and \hdamp parameter, within their uncertainties, in the right plot. The vertical bars on the data represent the statistical and systematic uncertainties added in quadrature.  The bottom panels show the ratio of the predictions to the data. }
	\label{plt:XSEC4}
\end{figure}

\begin{figure}
	\centering
	\includegraphics[width=0.49\linewidth]{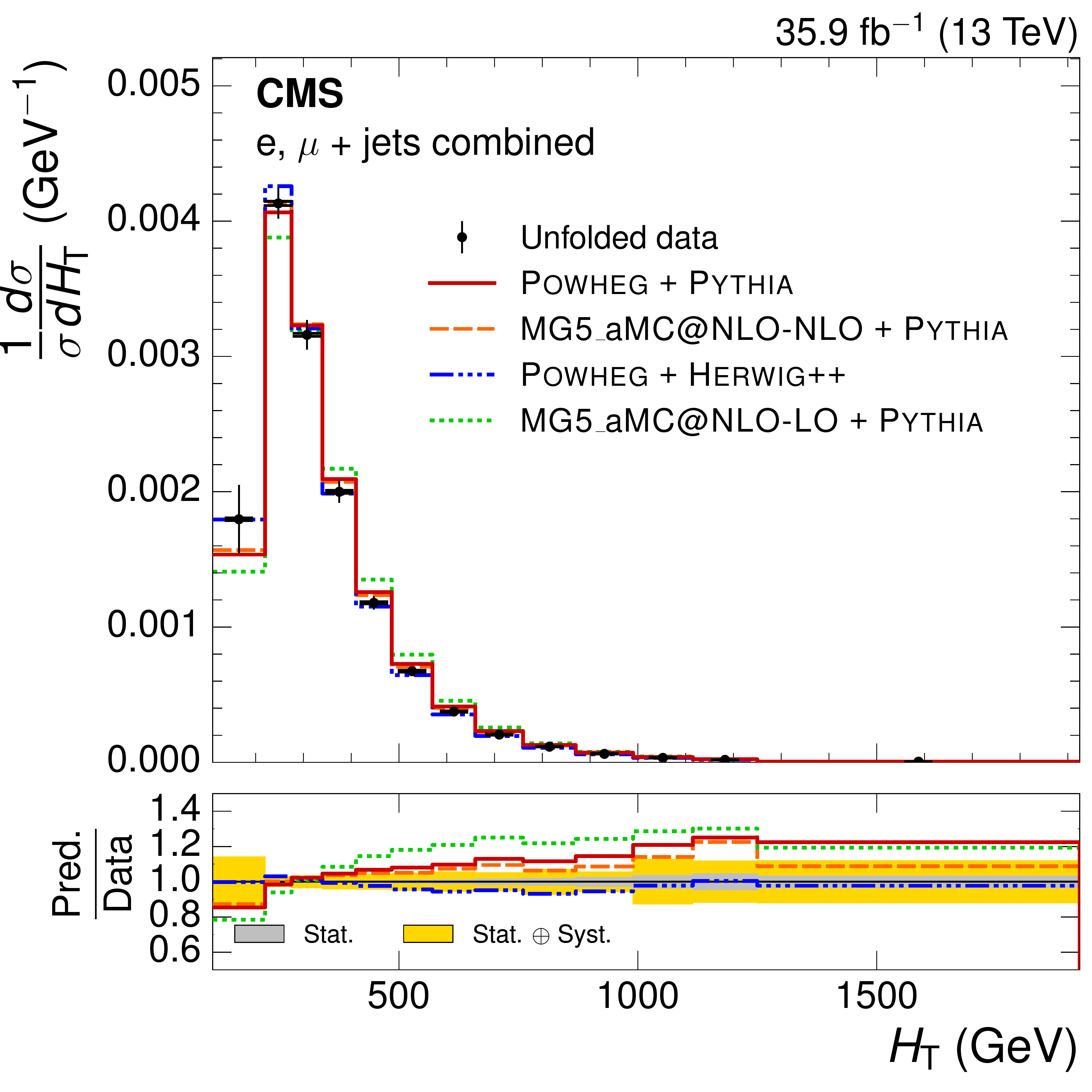}
	\includegraphics[width=0.49\linewidth]{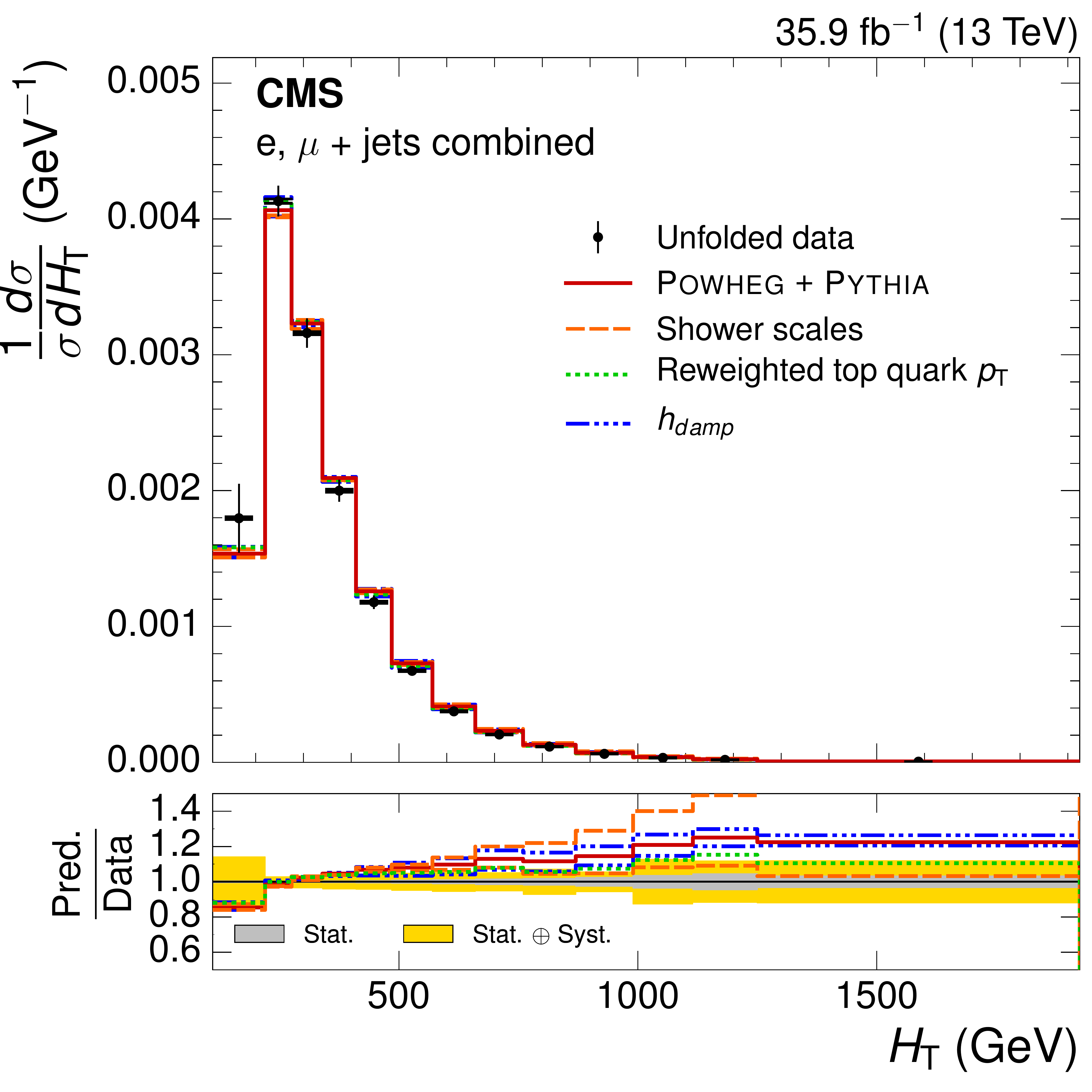} \\
	\includegraphics[width=0.49\linewidth]{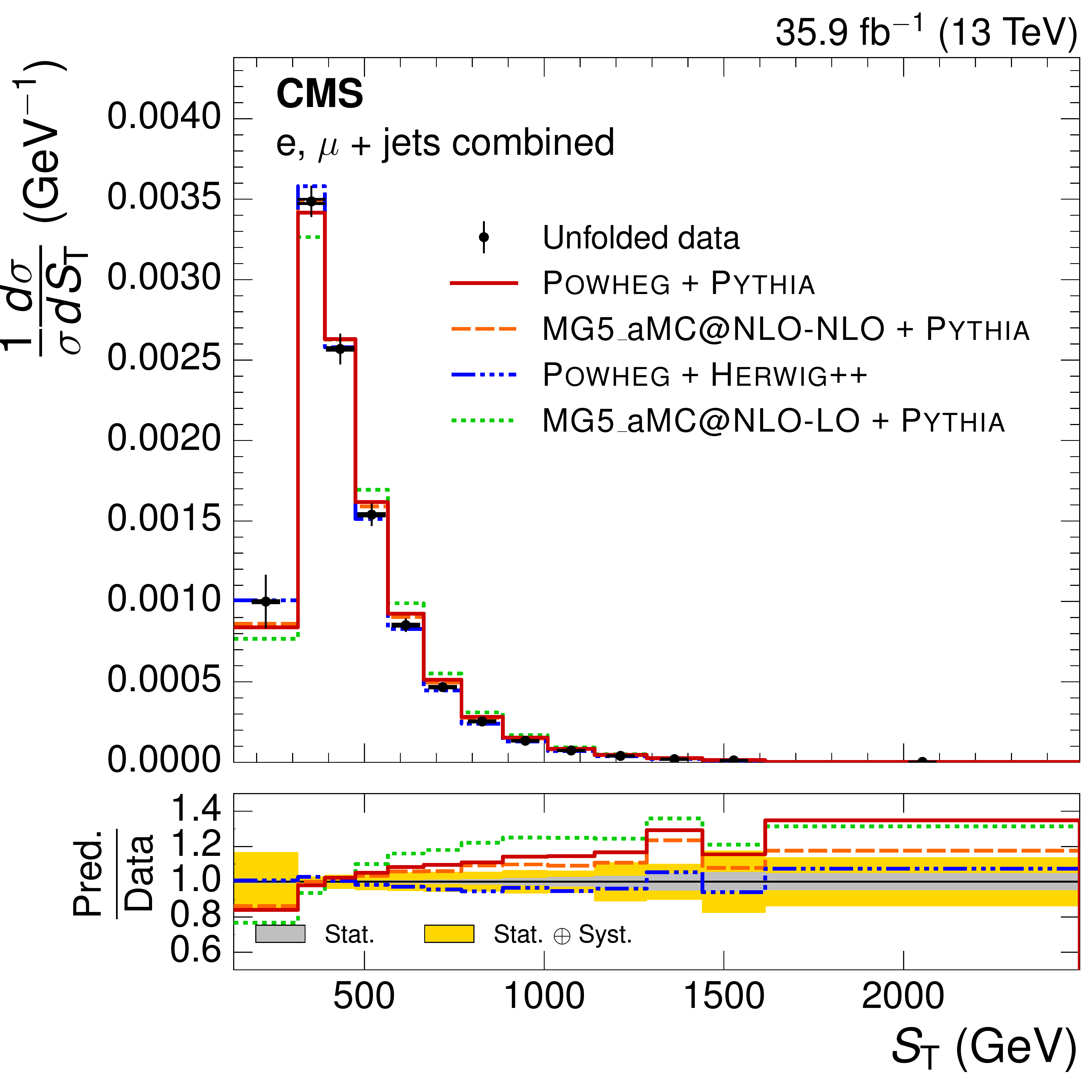}
	\includegraphics[width=0.49\linewidth]{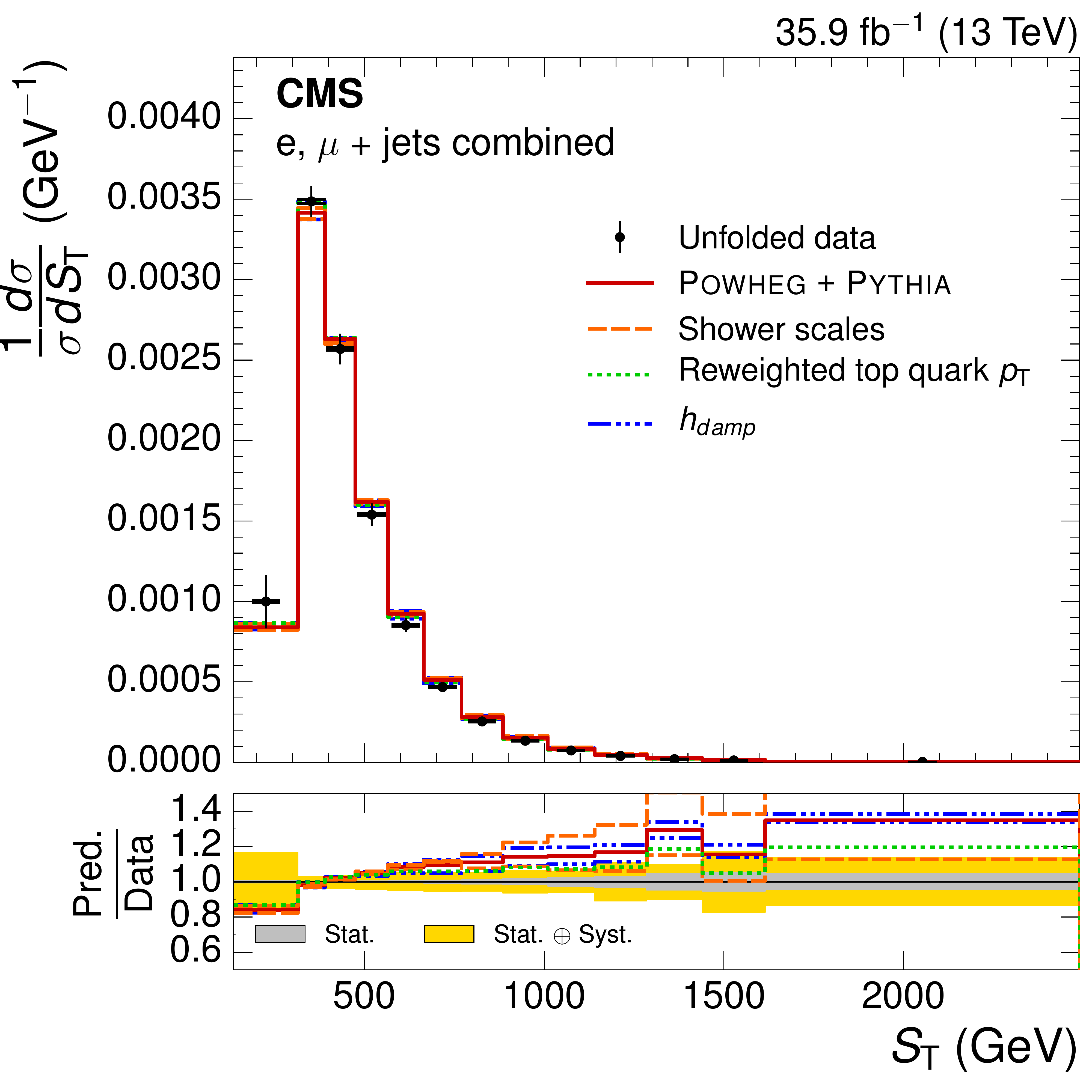}
	\caption{Normalized \HT (upper) and \ST (lower) differential \ttbar cross sections, compared to different \ttbar simulations in the left plots, and compared to the \powhegpythia simulation after varying the shower scales, and \hdamp parameter, within their uncertainties, in the right plots. The vertical bars on the data represent the statistical and systematic uncertainties added in quadrature.  The bottom panels show the ratio of the predictions to the data.}
	\label{plt:XSEC1}
\end{figure}

The normalized differential \ttbar production cross section with respect to \NJET is shown in Fig.~\ref{plt:XSEC4}, with respect to \HT and \ST in Fig.~\ref{plt:XSEC1}, with respect to \ptmiss and \WPT in Fig.~\ref{plt:XSEC2} and with respect to \LPT and \LETA in Fig.~\ref{plt:XSEC3}.
Tabulated results are listed in Appendix~\ref{ap:tablesOfResults}.
Measurements of the absolute differential \ttbar production cross sections are shown in Figs.~\ref{plt:ABSXSEC4}, \ref{plt:ABSXSEC1}, \ref{plt:ABSXSEC2} and \ref{plt:ABSXSEC3}, and tabulated in Appendix~\ref{ap:tablesOfAbsResults}.
In each figure, the measured cross section is compared with the predictions from several combinations of matrix-element and parton shower generators, namely \powhegpythia, \powhegherwig, \mgamcFxFx, and \mgamcMLM.
Each measured cross section is also compared to the \powhegpythia generator after varying the shower scales and the \hdamp parameter used in generating the sample within their uncertainties, and also after reweighting the top quark \pt as described in Section~\ref{sec:syst}.

The level of agreement between the measured and predicted differential cross sections are determined through a \chis test, where the full covariance matrix, including the correlations between the statistical and systematic uncertainties in each bin of the measurements, is taken into account.
The results, including the \pvalue of each test, are shown in Tables~\ref{tb:Chi2_normalised} and \ref{tb:Chi2_absolute}.

The predictions of the \powhegpythia model are consistent with data for the \NJET, \ptmiss, \ST, and \LPT distributions. In particular, the prediction of the \NJET distribution has a \chis per degree of freedom of
2/5 for the normalized and 2.2/6
for the absolute cross section measurement.
The jet multiplicity from previous 8\TeV measurements was used in deriving the \cuettune tune~\cite{CUETP8M2T4_Tune}, and this confirms that the tune continues to accurately describe the jet multiplicity on a larger data set with a higher $\sqrt{s}$.
On the other hand, tensions are observed for the \HT, \WPT and \LETA variables.
An additional \chis calculation between the \powhegpythia model and unfolded data is performed, where the theoretical uncertainties within the generator, described in Section~\ref{sec:syst}, are included, as well as in the unfolded data.
The correlations between the uncertainties in the prediction of the generator and the unfolded data are taken into account.
The result of this test demonstrates that the theoretical uncertainties in the \powhegpythia model cover the differences between the \powhegpythia model and the unfolded data in the phase space analyzed.

The \powhegherwig and \mgamcFxFx models are broadly consistent with the unfolded data, even without including the theoretical uncertainties in
the \chis test, with the exception of \NJET in \powhegherwig and \LETA in \mgamcFxFx.
Without these uncertainties, the \mgamcMLM model is not compatible with any kinematic event distribution in the unfolded data presented here.

The effect of the regularization in the unfolding procedure is investigated by unfolding without regularization, which typically results in a small change in the \chis.
When unfolding without regularization, the largest changes in \chis for the normalized cross sections are for the \HT distribution with the \mgamcFxFx model, where the \chis per degree of freedom increases from
11/12 to 12/12,
and for the \ptmiss distribution in the \powhegpythia model (including the model theoretical uncertainties), where the \chis per degree of freedom decreases from
2.9/5 to 2.1/5.
The effects on the \chis for all other variables and models are small.
The \chis does not change for the \LPT and \LETA distributions for any model when unfolding without regularization.

\begin{table}
	\topcaption{Results of a goodness-of-fit test between the normalized cross sections in data and several models, with values given as \chis/number of degrees of freedom (ndf)}
	\label{tb:Chi2_normalised}
	\centering
	\begin{tabular}{ccccc}
		&	 \multicolumn{2}{c}{\powhegpythia} & 	 \multicolumn{2}{c}{With MC theoretical uncertainties} \\
		\vspace*{0.02cm} &	\chis/ndf & \pvalue &	\chis/ndf & \pvalue \\			\hline
		\vspace*{0.02cm} \NJET &	2 / 5 &	 0.85 &	1.5 / 5 &	 0.91 \\
		\vspace*{0.02cm} \HT &	26 / 12 &	 $<$ 0.01 &	4.8 / 12 &	 0.97 \\
		\vspace*{0.02cm} \ST &	22 / 12 &	 0.04 &	4.2 / 12 &	 0.98 \\
		\vspace*{0.02cm} \ptmiss &	11 / 5 &	 0.06 &	2.9 / 5 &	 0.72 \\
		\vspace*{0.02cm} \WPT &	16 / 6 &	 0.01 &	2.5 / 6 &	 0.87 \\
		\vspace*{0.02cm} \LPT &	24 / 16 &	 0.09 &	14 / 16 &	 0.63 \\
		\vspace*{0.02cm} \LETA &	19 / 7 &	 $<$ 0.01 &	15 / 7 &	 0.04 \\
		\vspace*{0.2cm}
		\newline
	\end{tabular}
	\begin{tabular}{ccccccc}
		&	 \multicolumn{2}{c}{\powhegherwig} & 	 \multicolumn{2}{c}{\mgamcFxFxpythia} & 	 \multicolumn{2}{c}{\mgamcMLMpythia} \\
		\vspace*{0.02cm} &	\chis/ndf & \pvalue &	\chis/ndf & \pvalue &	\chis/ndf & \pvalue \\			\hline
		\vspace*{0.02cm} \NJET &	38 / 5 &	 $<$ 0.01 &	9.5 / 5 &	 0.09 &	78 / 5 &	 $<$ 0.01 \\
		\vspace*{0.02cm} \HT &	23 / 12 &	 0.03 &	11 / 12 &	 0.52 &	160 / 12 &	 $<$ 0.01 \\
		\vspace*{0.02cm} \ST &	21 / 12 &	 0.04 &	11 / 12 &	 0.57 &	110 / 12 &	 $<$ 0.01 \\
		\vspace*{0.02cm} \ptmiss &	1.3 / 5 &	 0.93 &	5.9 / 5 &	 0.31 &	23 / 5 &	 $<$ 0.01 \\
		\vspace*{0.02cm} \WPT &	0.81 / 6 &	 0.99 &	8.9 / 6 &	 0.18 &	30 / 6 &	 $<$ 0.01 \\
		\vspace*{0.02cm} \LPT &	11 / 16 &	 0.82 &	16 / 16 &	 0.44 &	37 / 16 &	 $<$ 0.01 \\
		\vspace*{0.02cm} \LETA &	19 / 7 &	 $<$ 0.01 &	24 / 7 &	 $<$ 0.01 &	30 / 7 &	 $<$ 0.01 \\
	\end{tabular}
\end{table}

\begin{table}
	\topcaption{Results of a goodness-of-fit test between the absolute cross sections in data and several models, with values given as \chis/number of degrees of freedom (ndf)}
	\label{tb:Chi2_absolute}
	\centering
	\begin{tabular}{ccccc}
		&	 \multicolumn{2}{c}{\powhegpythia} & 	 \multicolumn{2}{c}{With MC theoretical uncertainties} \\
		\vspace*{0.02cm} &	\chis/ndf & \pvalue &	\chis/ndf & \pvalue \\			\hline
		\vspace*{0.02cm} \NJET &	2.2 / 6 &	 0.90 &	1.7 / 6 &	 0.95 \\
		\vspace*{0.02cm} \HT &	23 / 13 &	 0.05 &	4.3 / 13 &	 0.99 \\
		\vspace*{0.02cm} \ST &	19 / 13 &	 0.11 &	4.7 / 13 &	 0.98 \\
		\vspace*{0.02cm} \ptmiss &	13 / 6 &	 0.05 &	3.1 / 6 &	 0.80 \\
		\vspace*{0.02cm} \WPT &	17 / 7 &	 0.02 &	2.7 / 7 &	 0.91 \\
		\vspace*{0.02cm} \LPT &	20 / 17 &	 0.28 &	14 / 17 &	 0.68 \\
		\vspace*{0.02cm} \LETA &	16 / 8 &	 0.04 &	15 / 8 &	 0.06 \\
		\vspace*{0.2cm}
		\newline
	\end{tabular}
	\begin{tabular}{ccccccc}
		&	 \multicolumn{2}{c}{\powhegherwig} & 	 \multicolumn{2}{c}{\mgamcFxFxpythia} & 	 \multicolumn{2}{c}{\mgamcMLMpythia} \\
		\vspace*{0.02cm} &	\chis/ndf & \pvalue &	\chis/ndf & \pvalue &	\chis/ndf & \pvalue \\			\hline
		\vspace*{0.02cm} \NJET &	39 / 6 &	 $<$ 0.01 &	12 / 6 &	 0.07 &	93 / 6 &	 $<$ 0.01 \\
		\vspace*{0.02cm} \HT &	21 / 13 &	 0.07 &	10 / 13 &	 0.66 &	150 / 13 &	 $<$ 0.01 \\
		\vspace*{0.02cm} \ST &	18 / 13 &	 0.17 &	9.3 / 13 &	 0.75 &	110 / 13 &	 $<$ 0.01 \\
		\vspace*{0.02cm} \ptmiss &	1.5 / 6 &	 0.96 &	6.6 / 6 &	 0.36 &	26 / 6 &	 $<$ 0.01 \\
		\vspace*{0.02cm} \WPT &	0.90 / 7 &	 1.00 &	9.2 / 7 &	 0.24 &	33 / 7 &	 $<$ 0.01 \\
		\vspace*{0.02cm} \LPT &	11 / 17 &	 0.87 &	15 / 17 &	 0.58 &	36 / 17 &	 $<$ 0.01 \\
		\vspace*{0.02cm} \LETA &	17 / 8 &	 0.04 &	23 / 8 &	 $<$ 0.01 &	31 / 8 &	 $<$ 0.01 \\
	\end{tabular}
\end{table}

\begin{figure}
	\centering
	\includegraphics[width=0.49\linewidth]{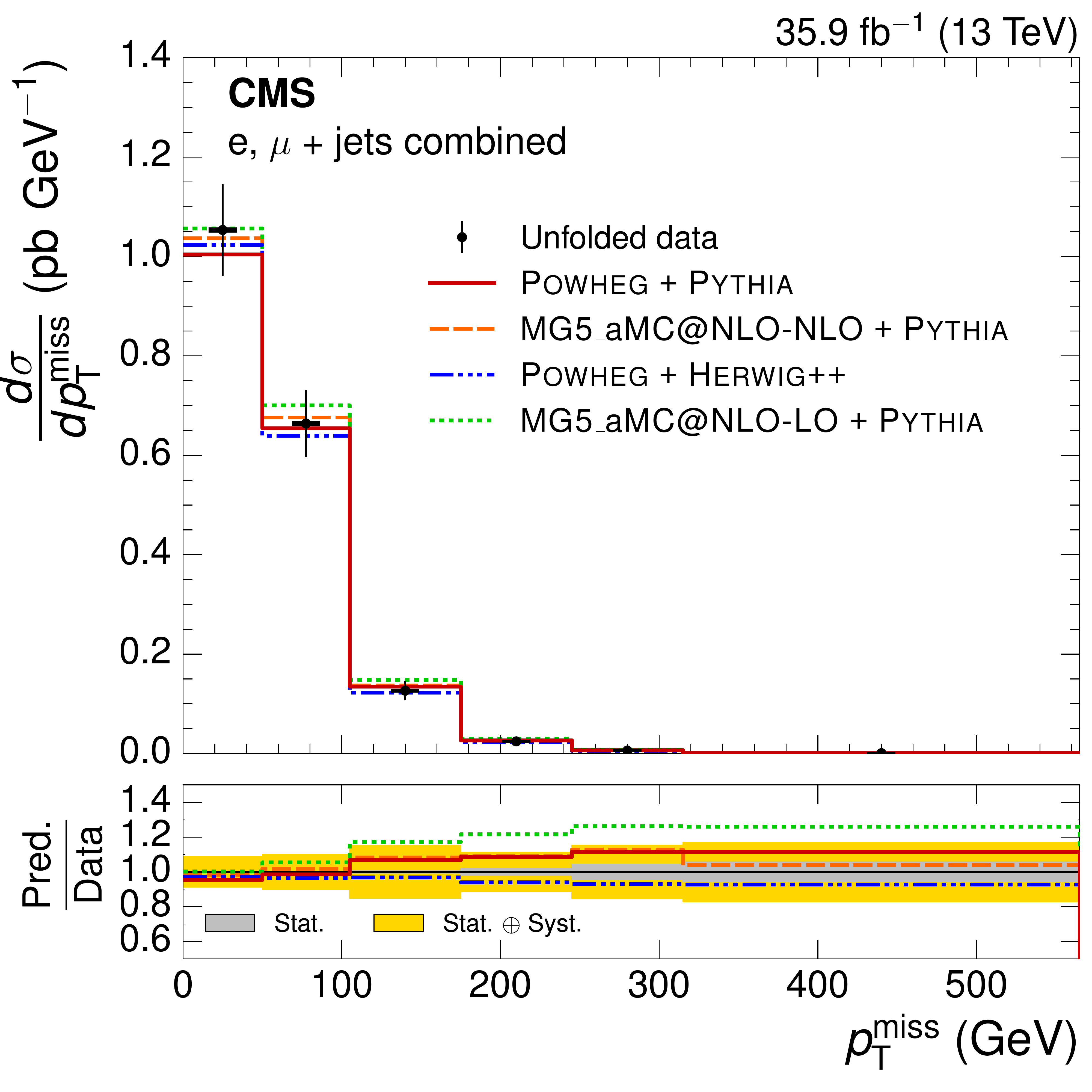}
	\includegraphics[width=0.49\linewidth]{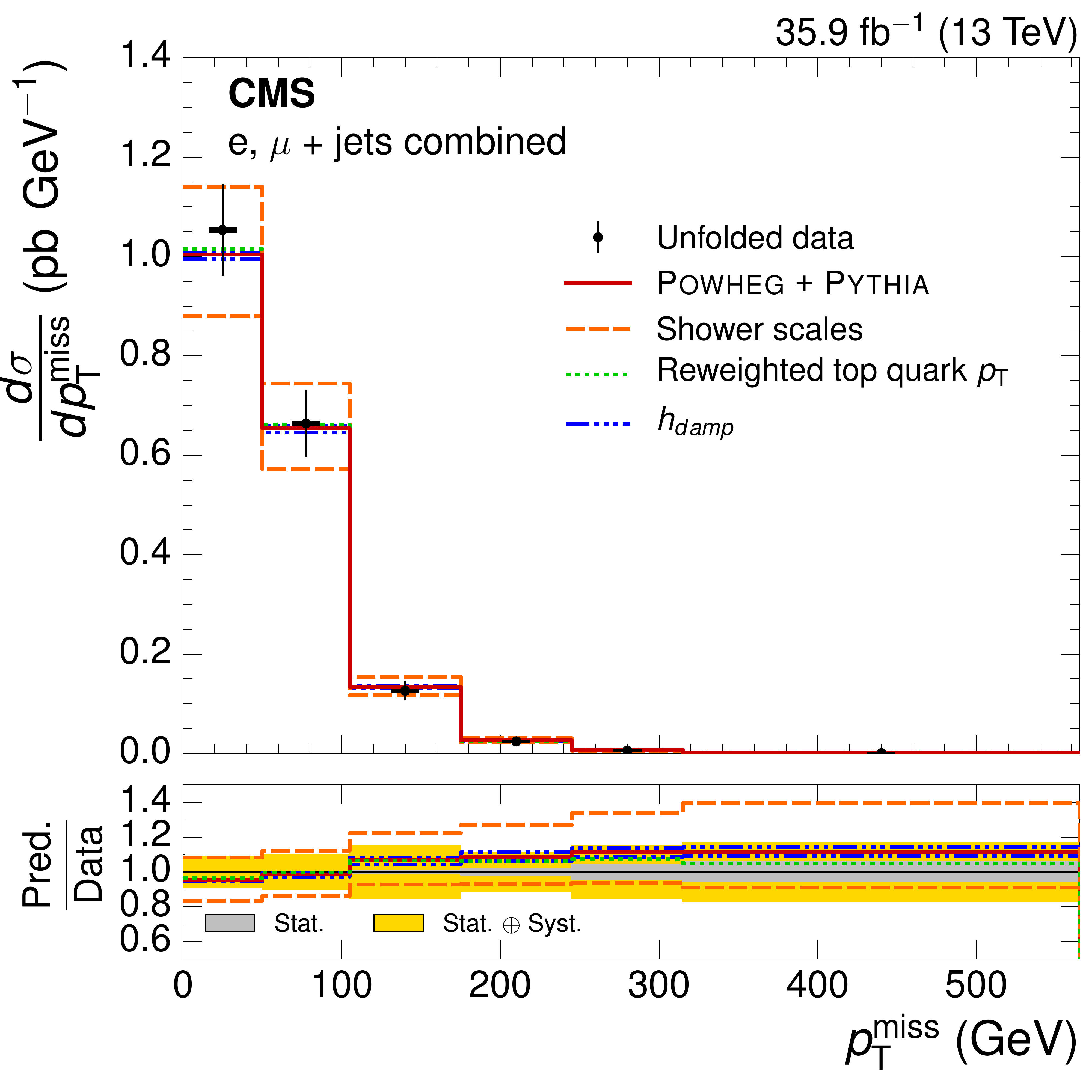} \\
	\includegraphics[width=0.49\linewidth]{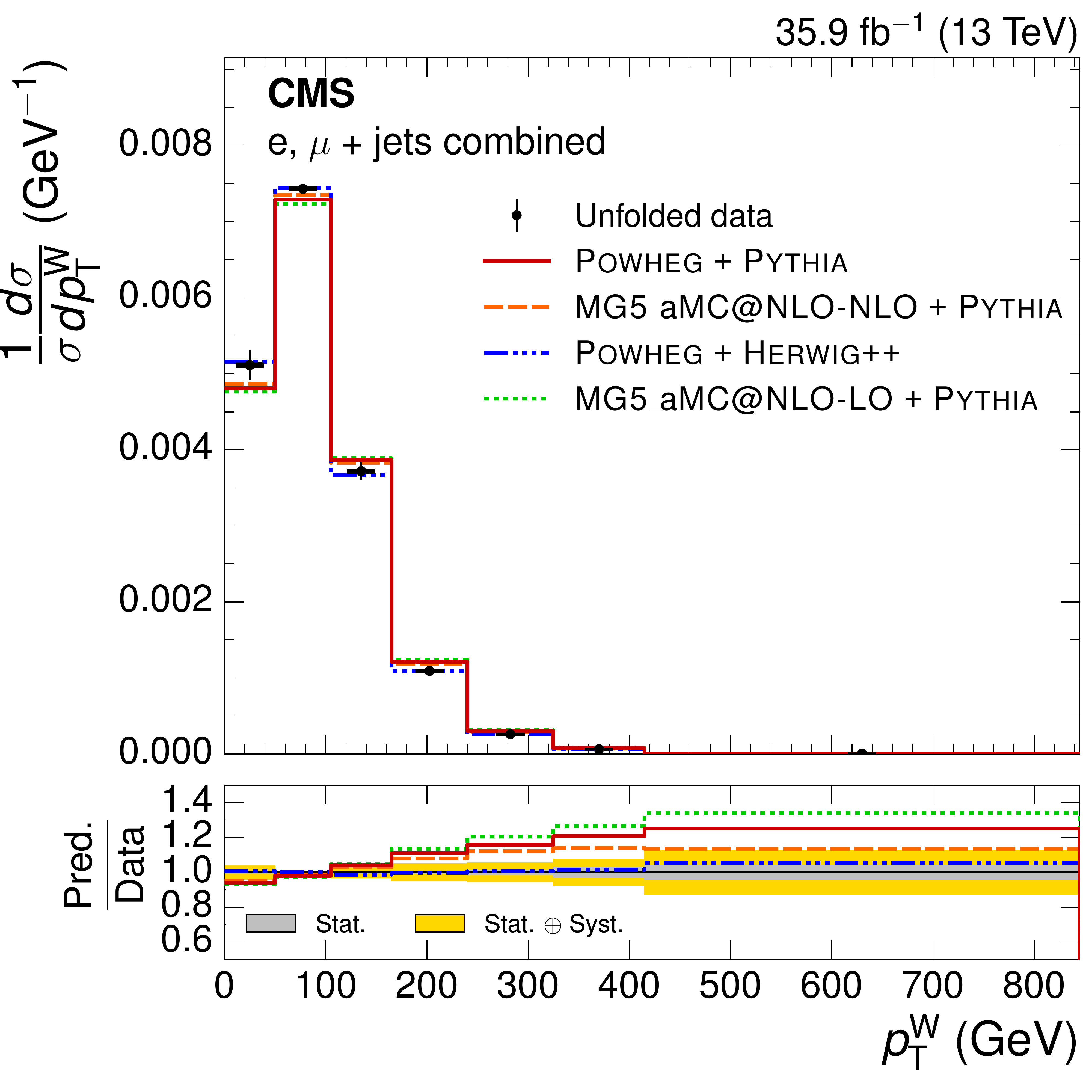}
	\includegraphics[width=0.49\linewidth]{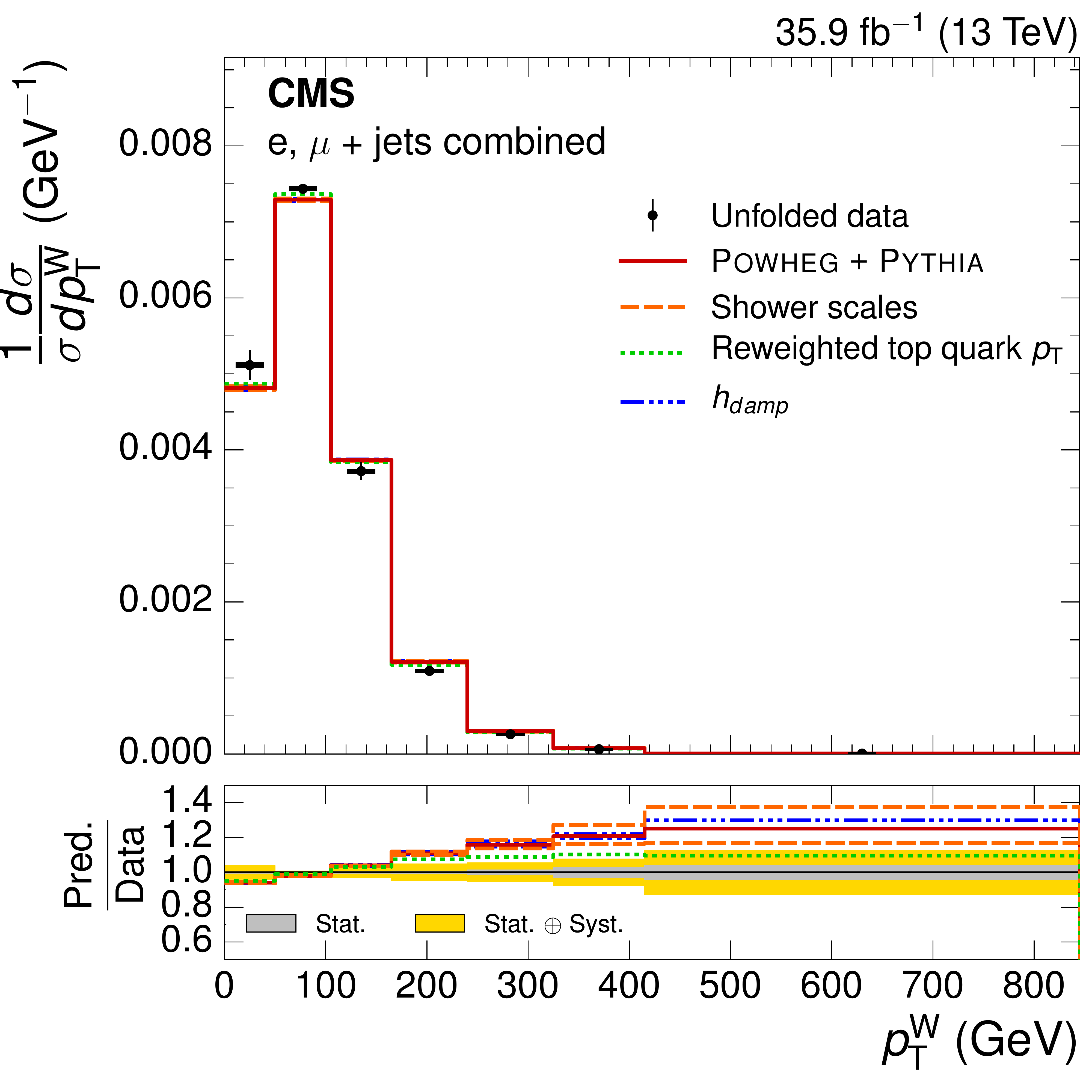}
	\caption{Normalized \ptmiss (upper) and \WPT (lower) differential \ttbar cross sections, compared to different \ttbar simulations in the left plots, and compared to the \powhegpythia simulation after varying the shower scales, and \hdamp parameter, within their uncertainties, in the right plots. The vertical bars on the data represent the statistical and systematic uncertainties added in quadrature.  The bottom panels show the ratio of the predictions to the data.}
	\label{plt:XSEC2}
\end{figure}

\begin{figure}
	\centering
	\includegraphics[width=0.49\linewidth]{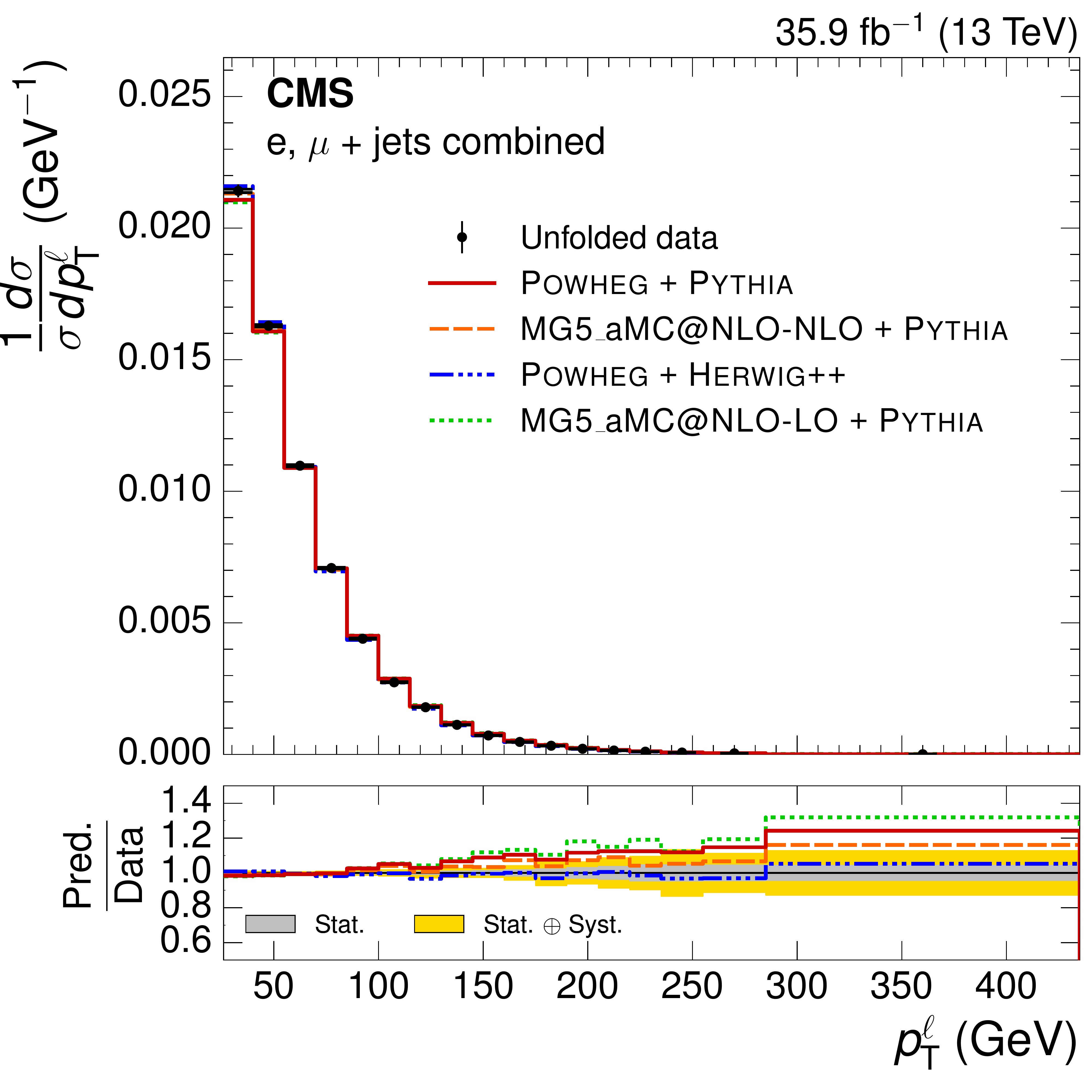}
	\includegraphics[width=0.49\linewidth]{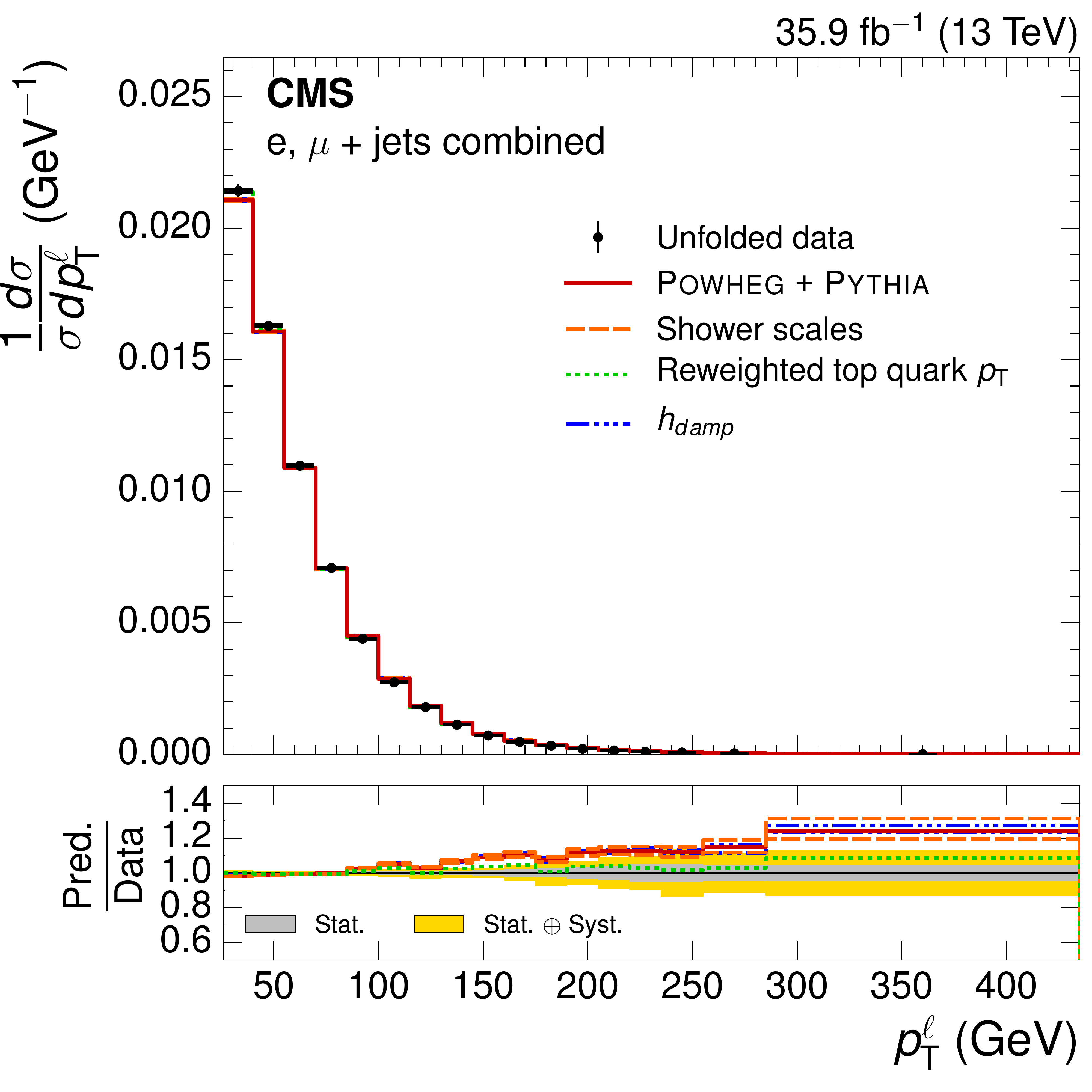} \\
	\includegraphics[width=0.49\linewidth]{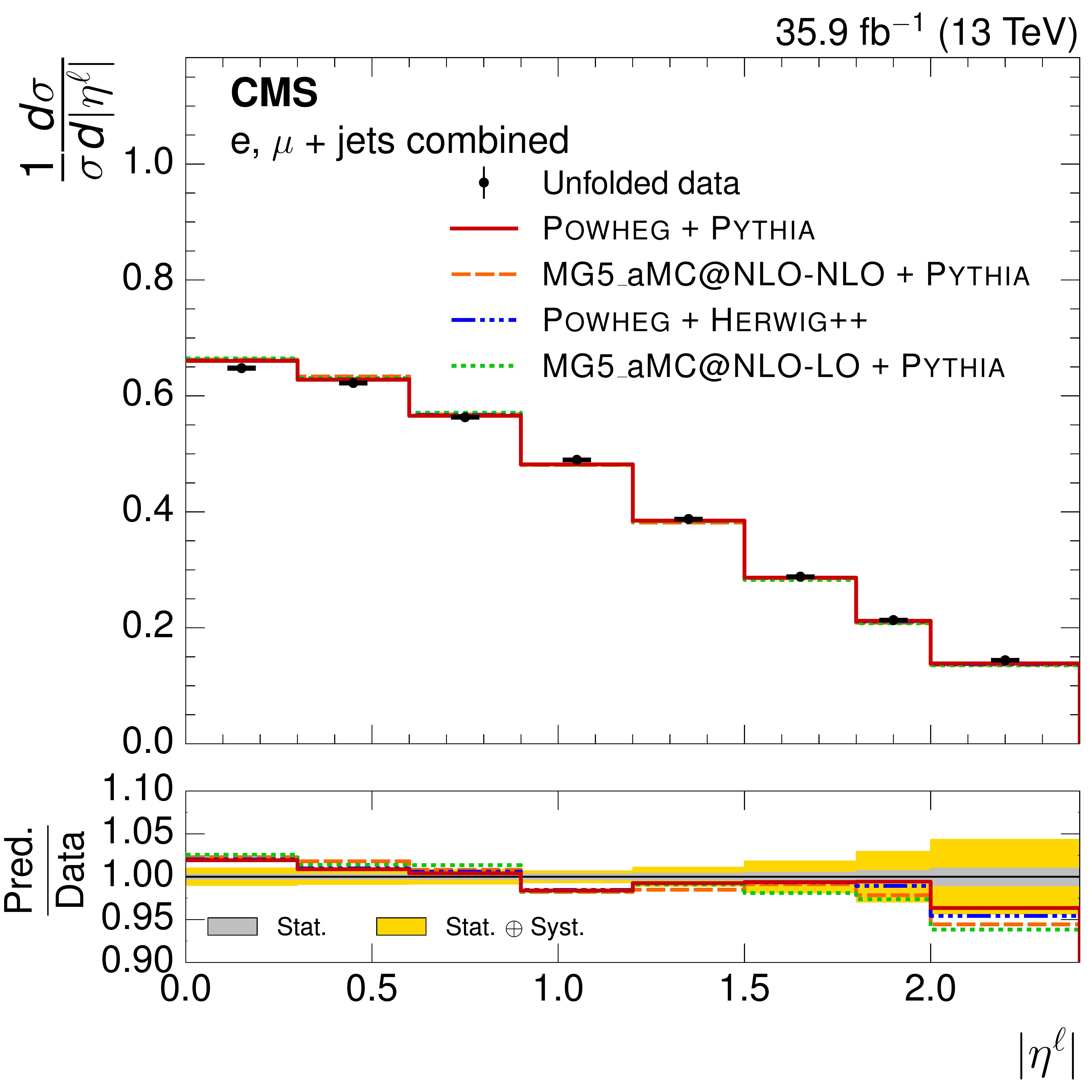}
	\includegraphics[width=0.49\linewidth]{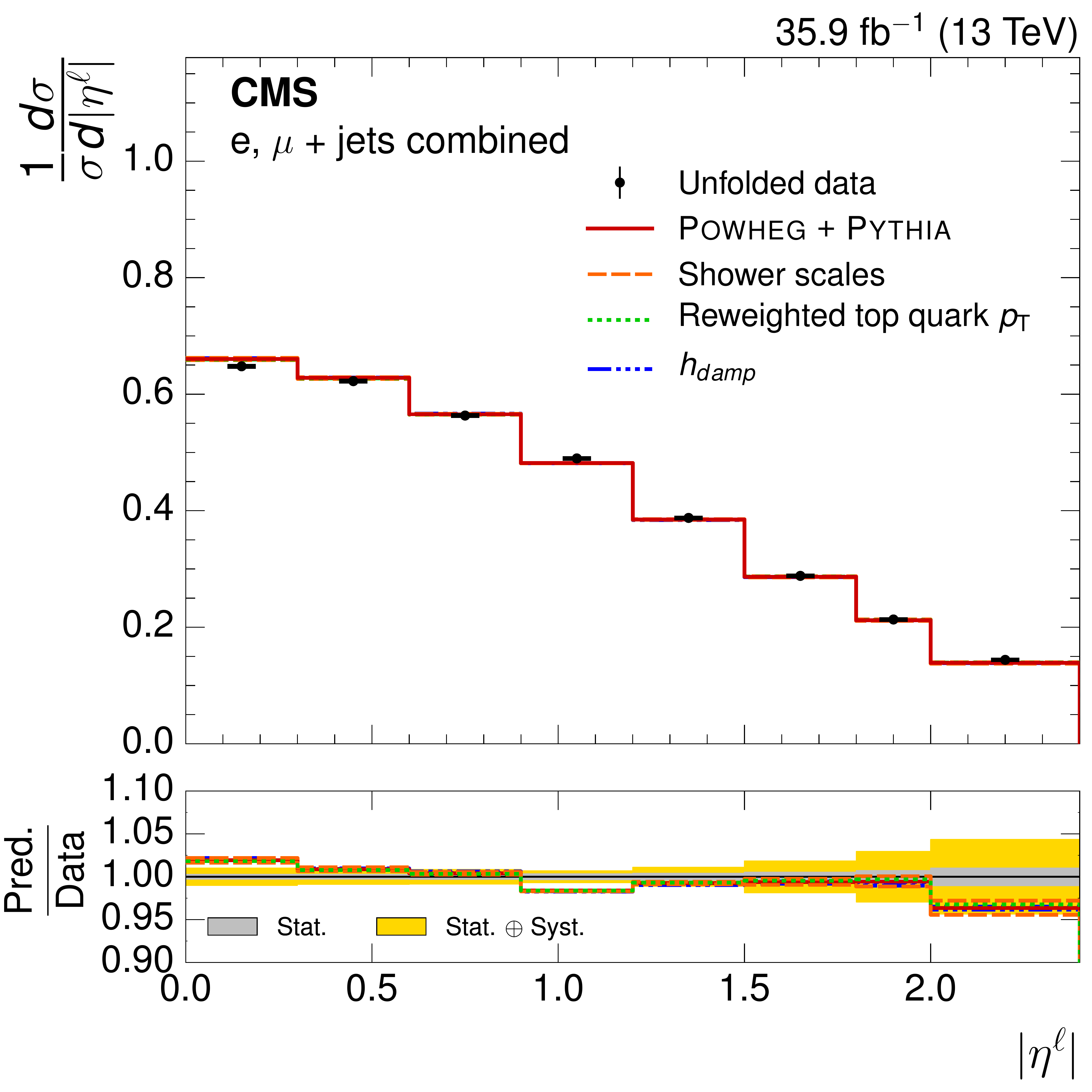}
	\caption{Normalized \LPT (upper) and \LETA (lower) differential \ttbar cross sections, compared to different \ttbar simulations in the left plots, and compared to the \powhegpythia simulation after varying the shower scales, and \hdamp parameter, within their uncertainties, in the right plots. The vertical bars on the data represent the statistical and systematic uncertainties added in quadrature.  The bottom panels show the ratio of the predictions to the data. }
	\label{plt:XSEC3}
\end{figure}

\begin{figure}
	\centering
	\includegraphics[width=0.49\linewidth]{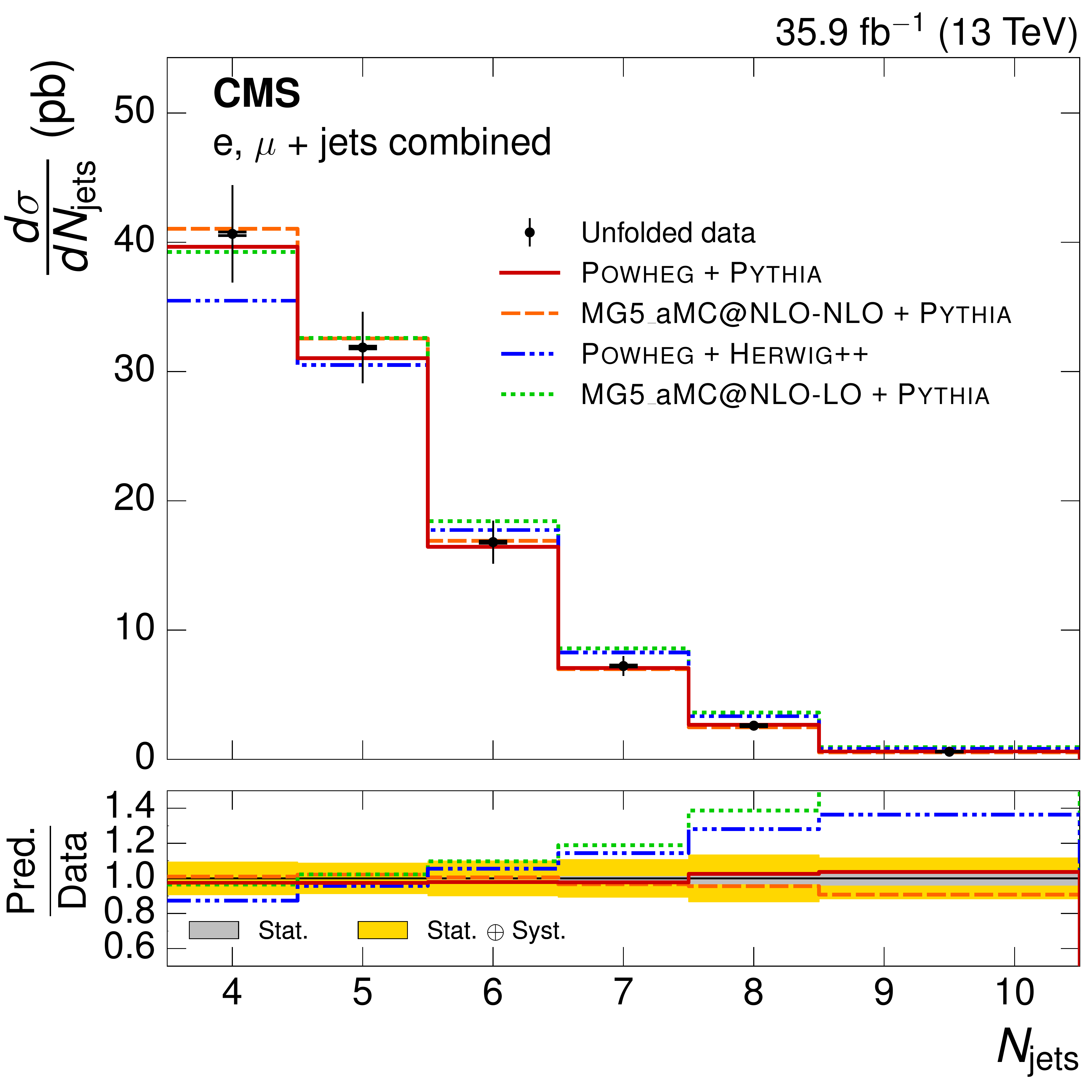}
	\includegraphics[width=0.49\linewidth]{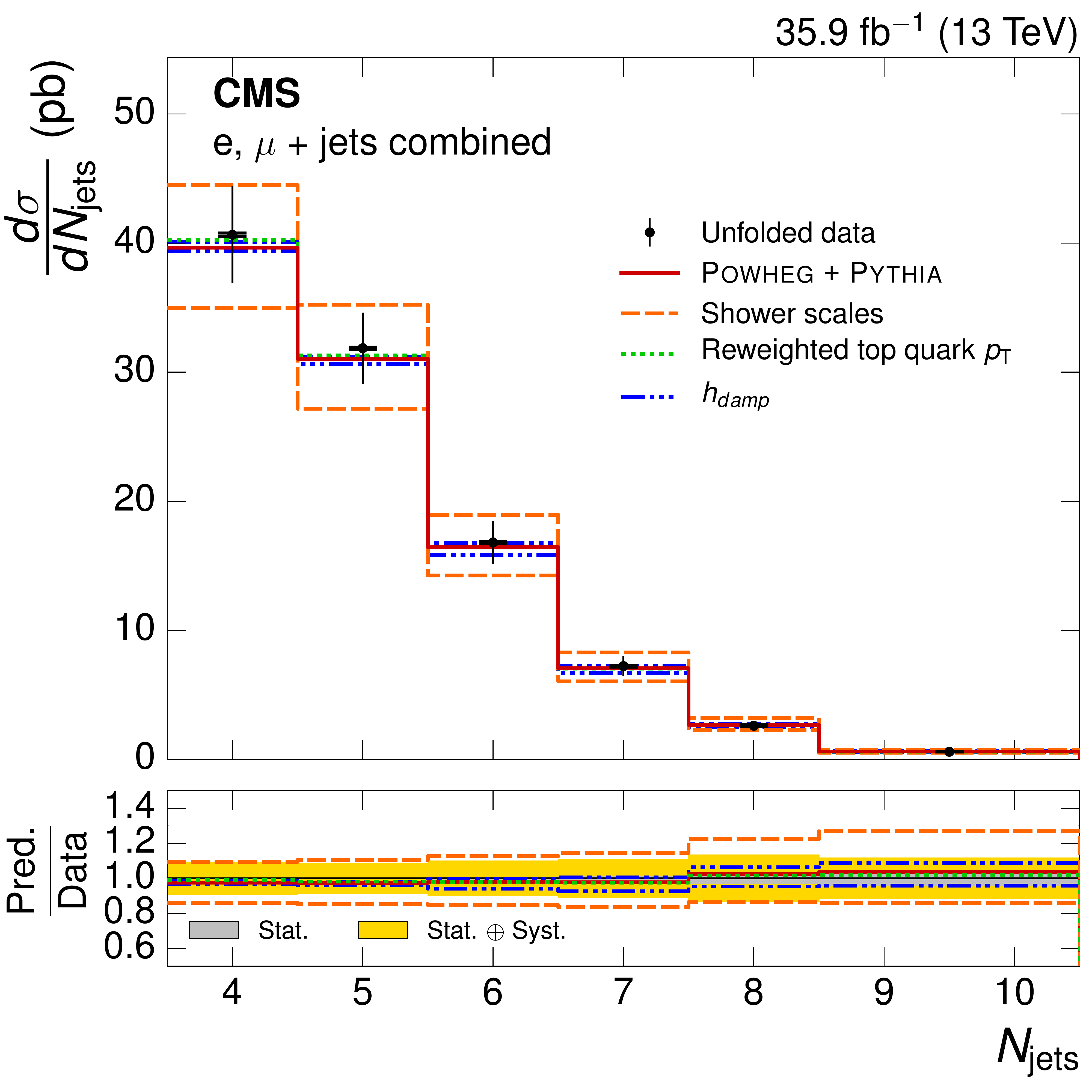}
	\caption{Absolute \NJET differential \ttbar cross section, compared to different \ttbar simulations in the left plot, and compared to the \powhegpythia simulation after varying the shower scales, and \hdamp parameter, within their uncertainties, in the right plot.  The vertical bars on the data represent the statistical and systematic uncertainties added in quadrature.  The bottom panels show the ratio of the predictions to the data. }
	\label{plt:ABSXSEC4}
\end{figure}

\begin{figure}
	\centering
	\includegraphics[width=0.49\linewidth]{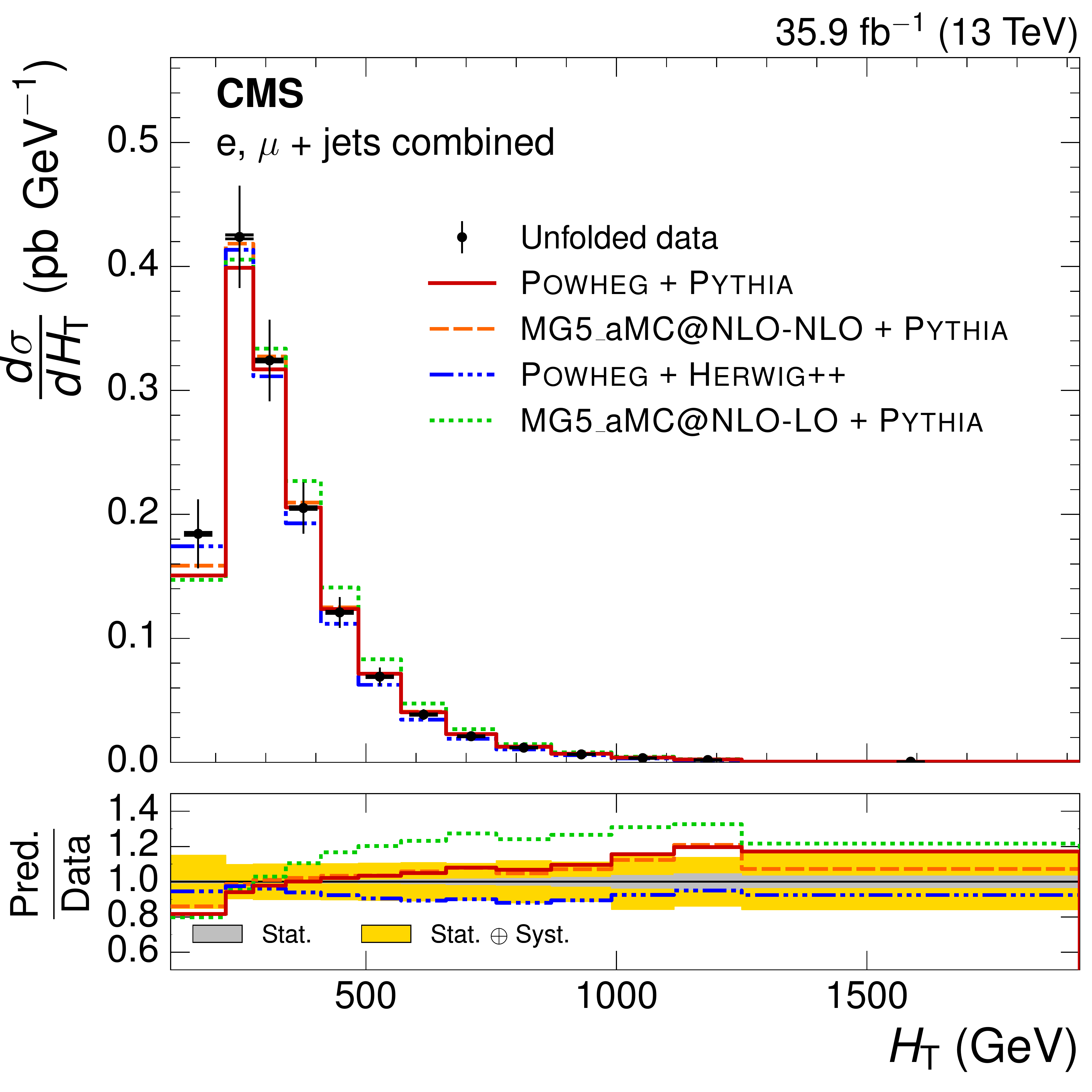}
	\includegraphics[width=0.49\linewidth]{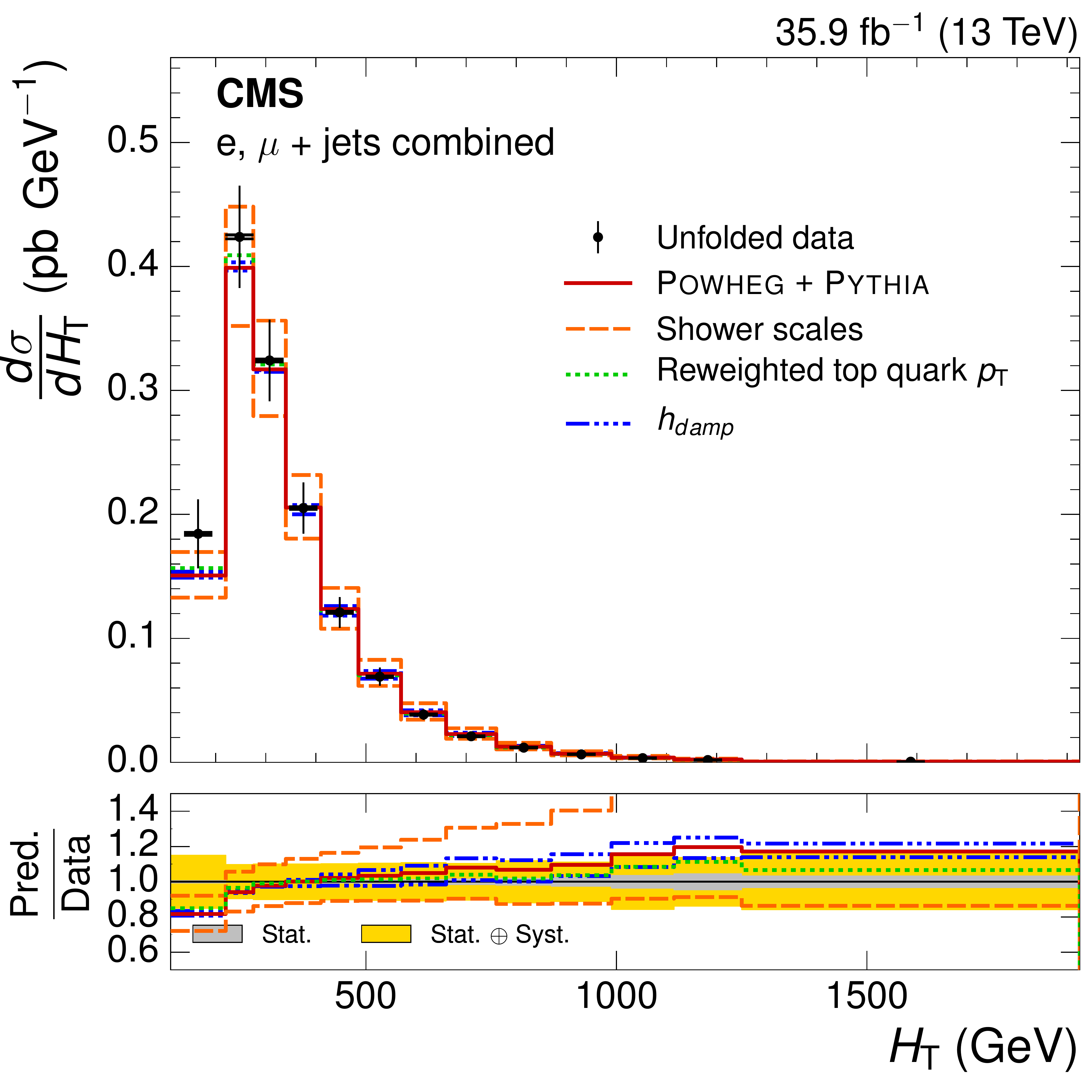} \\
	\includegraphics[width=0.49\linewidth]{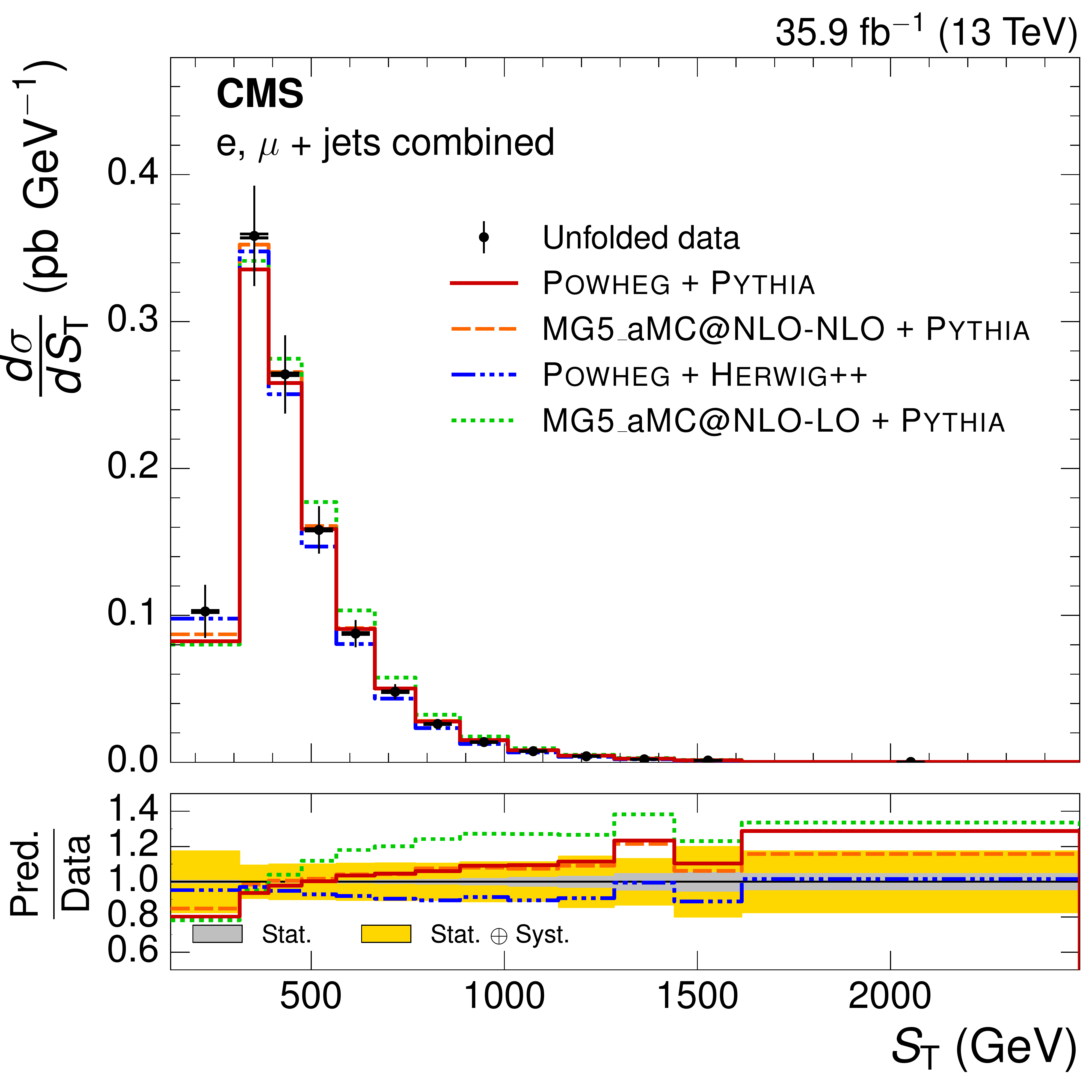}
	\includegraphics[width=0.49\linewidth]{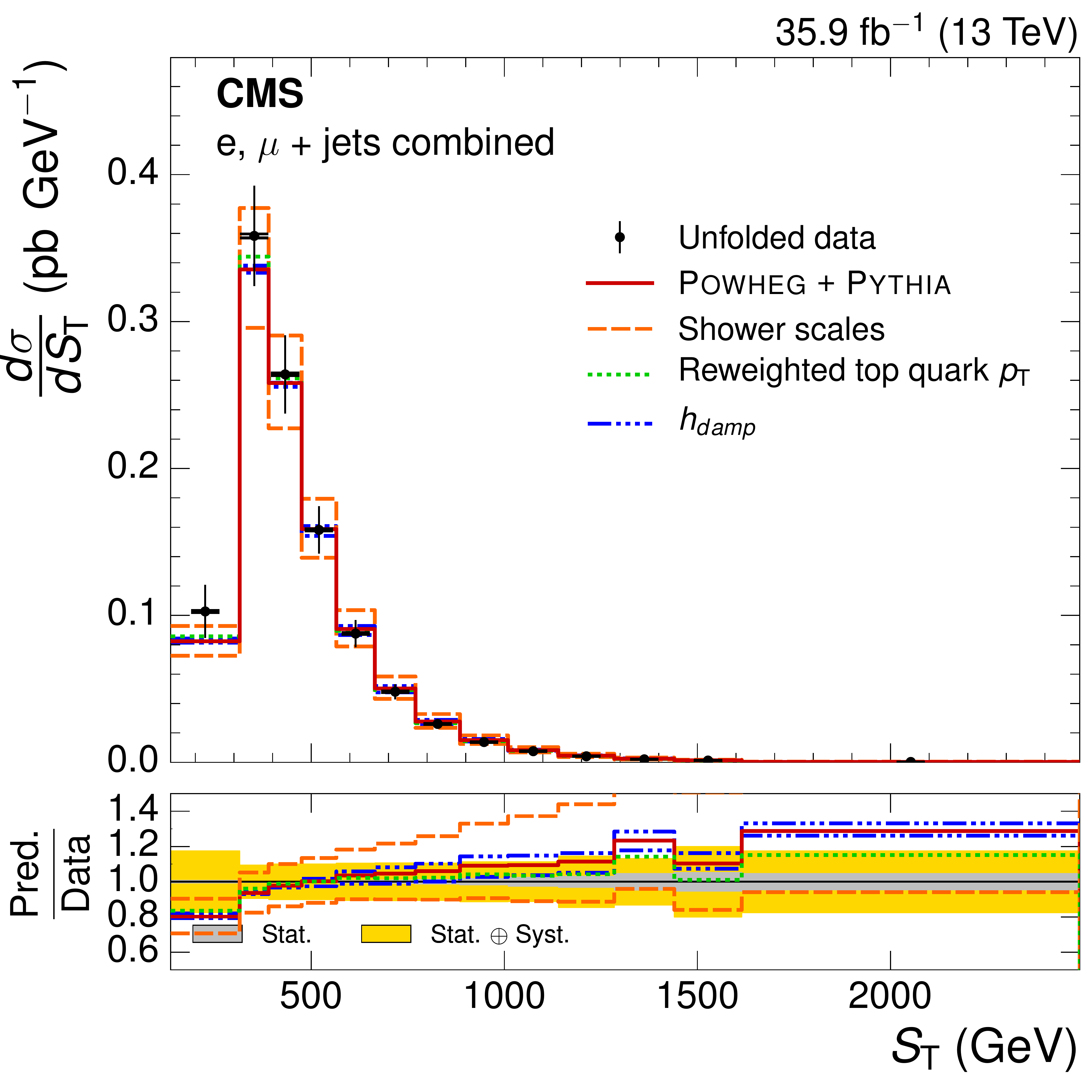}
	\caption{Absolute \HT (upper) and \ST (lower) differential \ttbar cross sections, compared to different \ttbar simulations in the left plots, and compared to the \powhegpythia simulation after varying the shower scales, and \hdamp parameter, within their uncertainties, in the right plots.  The vertical bars on the data represent the statistical and systematic uncertainties added in quadrature.  The bottom panels show the ratio of the predictions to the data.}
	\label{plt:ABSXSEC1}
\end{figure}

\begin{figure}
	\centering
	\includegraphics[width=0.49\linewidth]{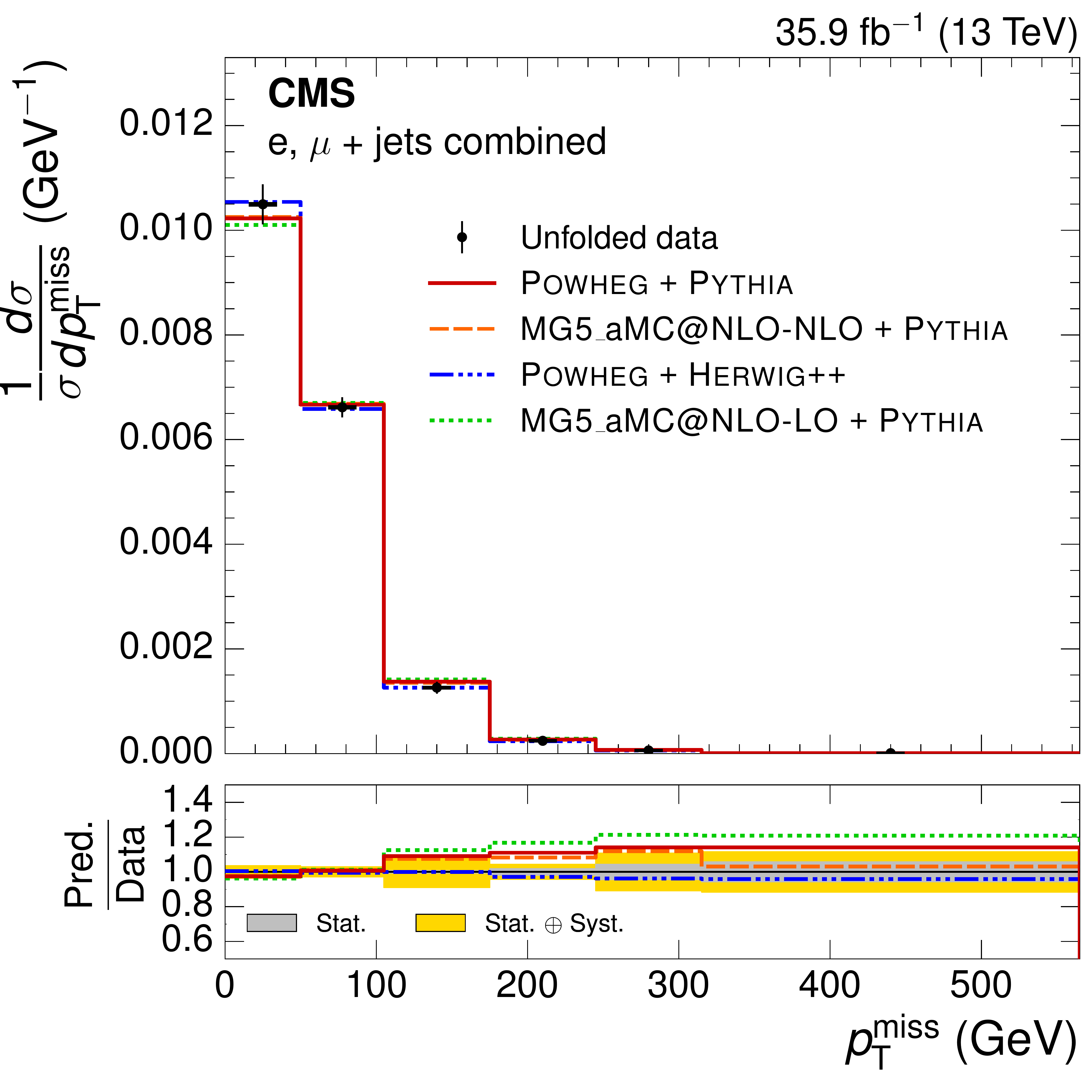}
	\includegraphics[width=0.49\linewidth]{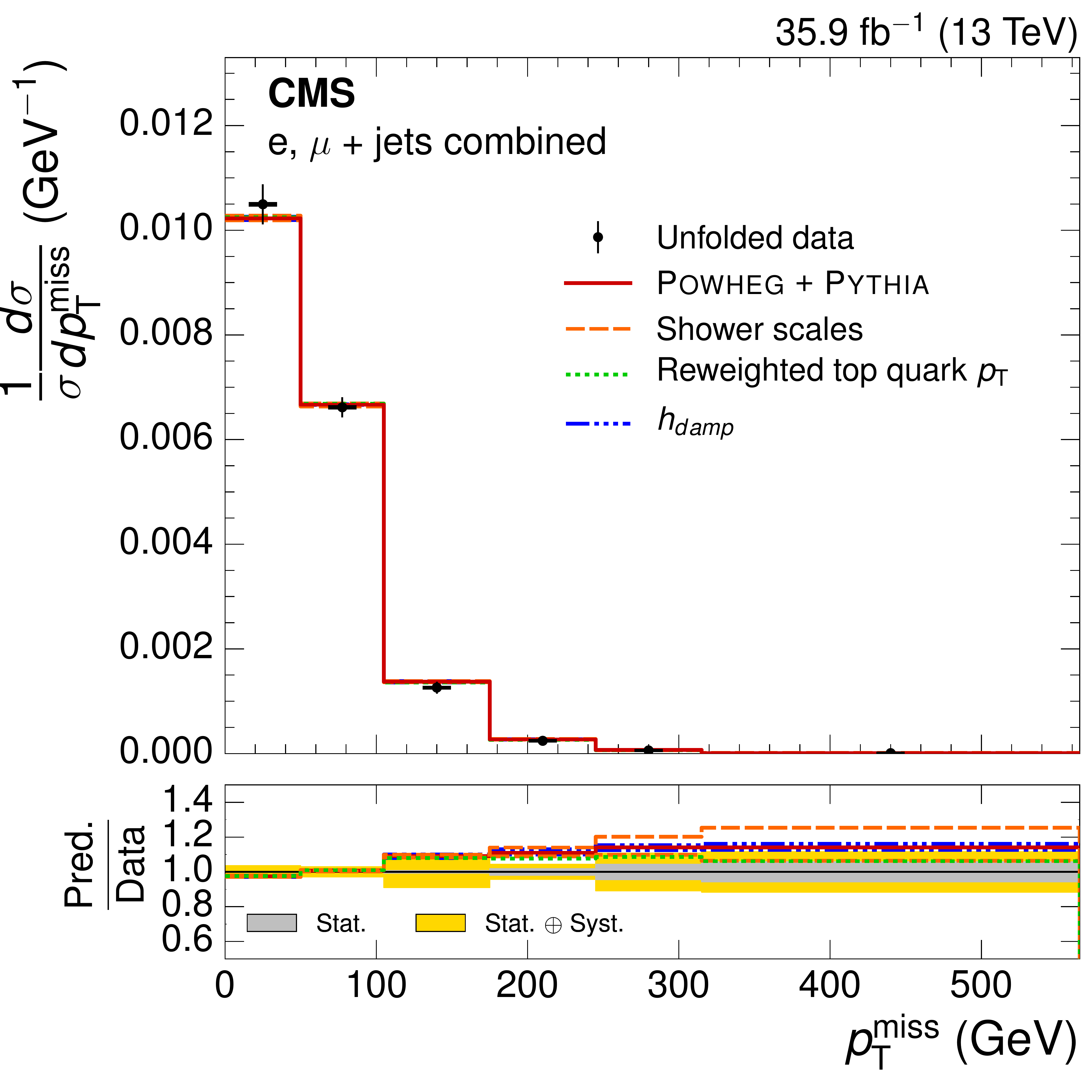} \\
	\includegraphics[width=0.49\linewidth]{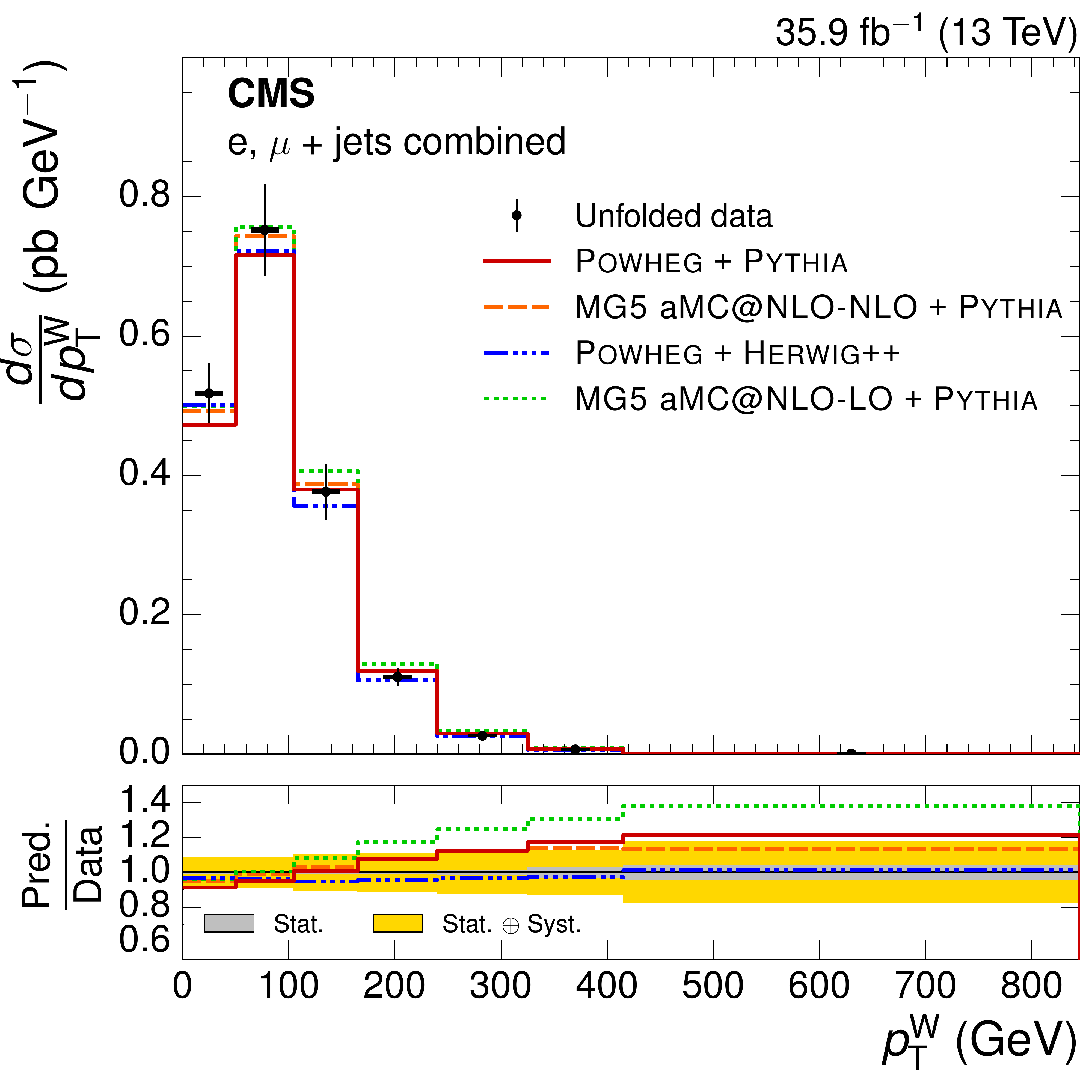}
	\includegraphics[width=0.49\linewidth]{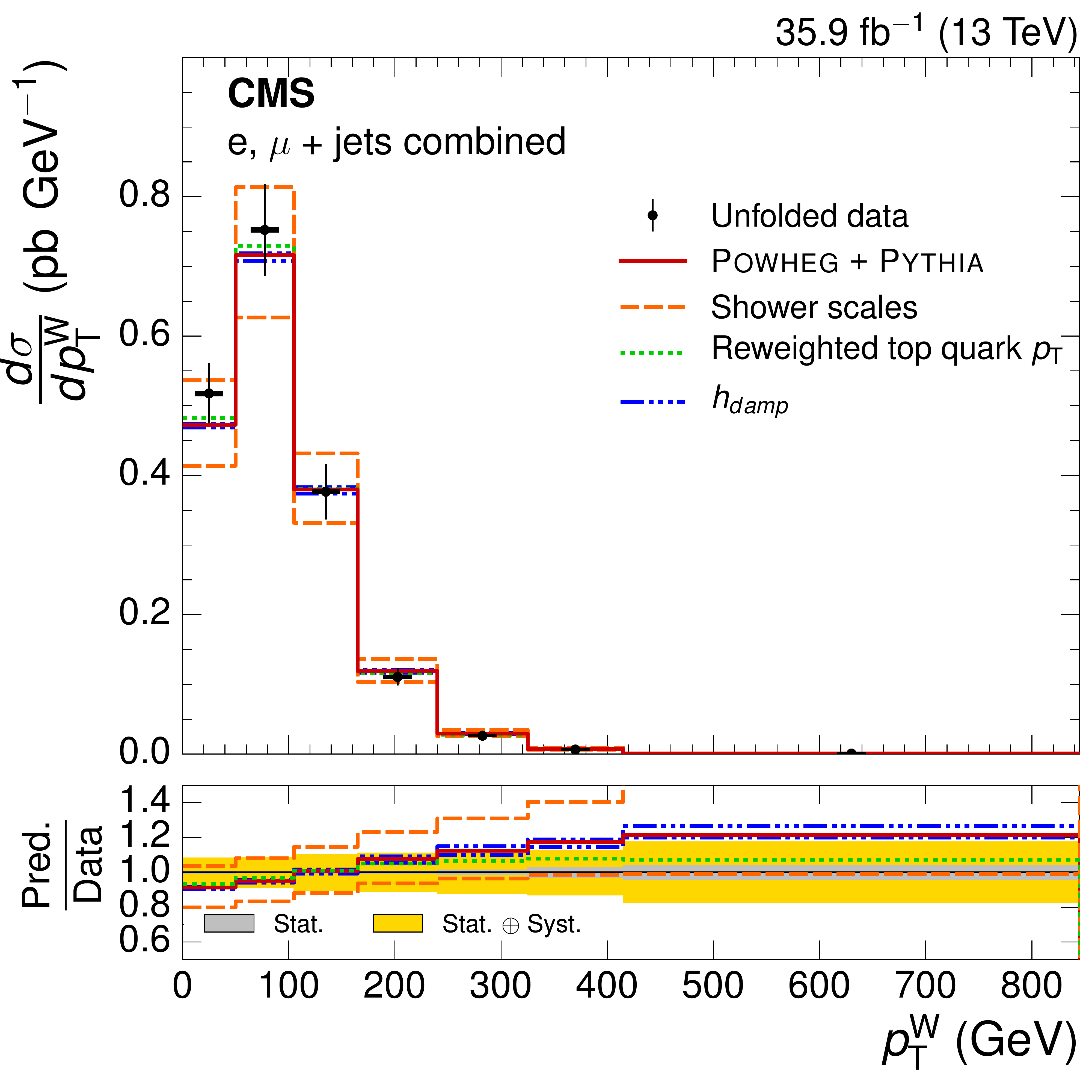}
	\caption{Absolute \ptmiss (upper) and \WPT (lower) differential \ttbar cross sections, compared to different \ttbar simulations in the left plots, and compared to the \powhegpythia simulation after varying the shower scales, and \hdamp parameter, within their uncertainties, in the right plots.  The vertical bars on the data represent the statistical and systematic uncertainties added in quadrature.  The bottom panels show the ratio of the predictions to the data.}
	\label{plt:ABSXSEC2}
\end{figure}

\begin{figure}
	\centering
	\includegraphics[width=0.49\linewidth]{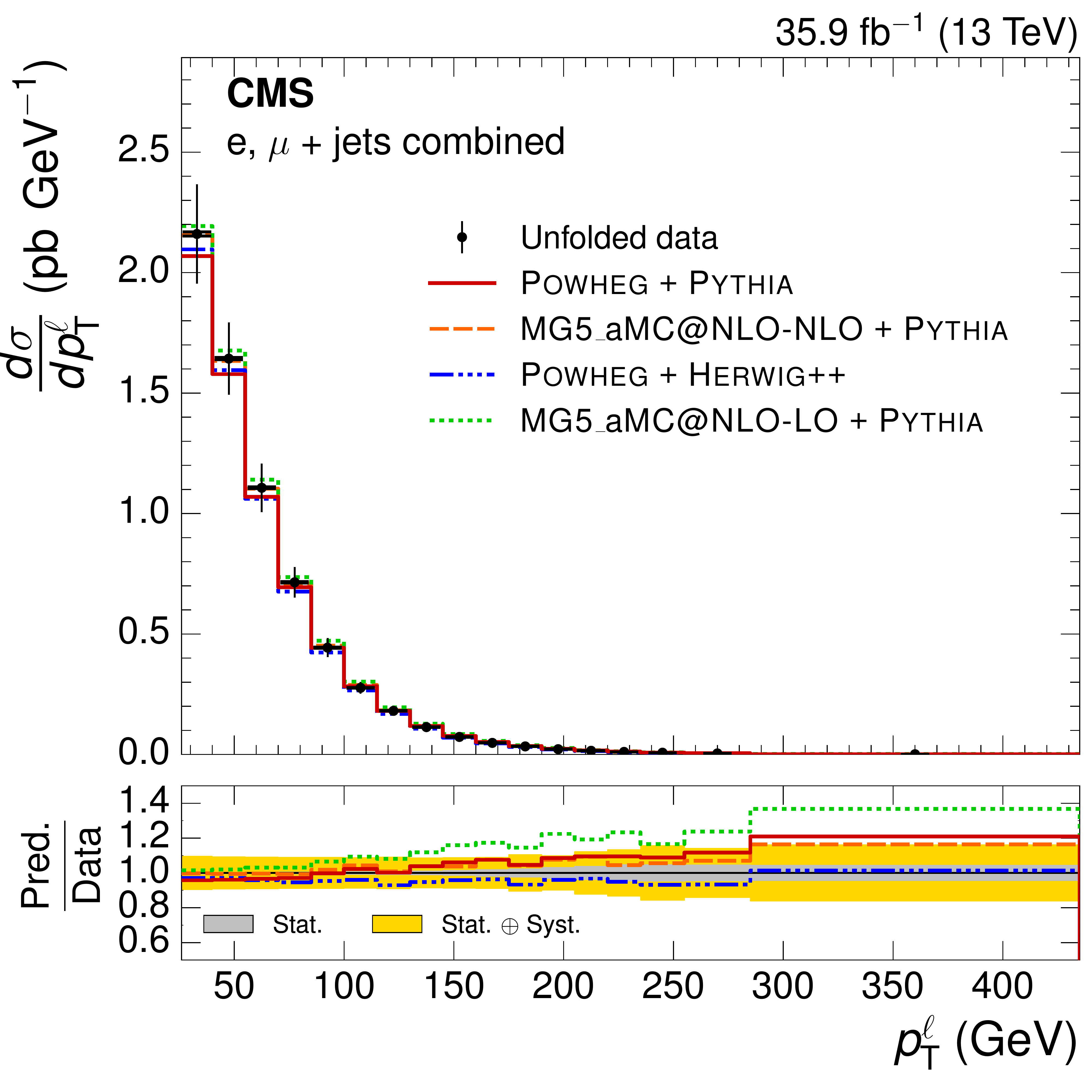}
	\includegraphics[width=0.49\linewidth]{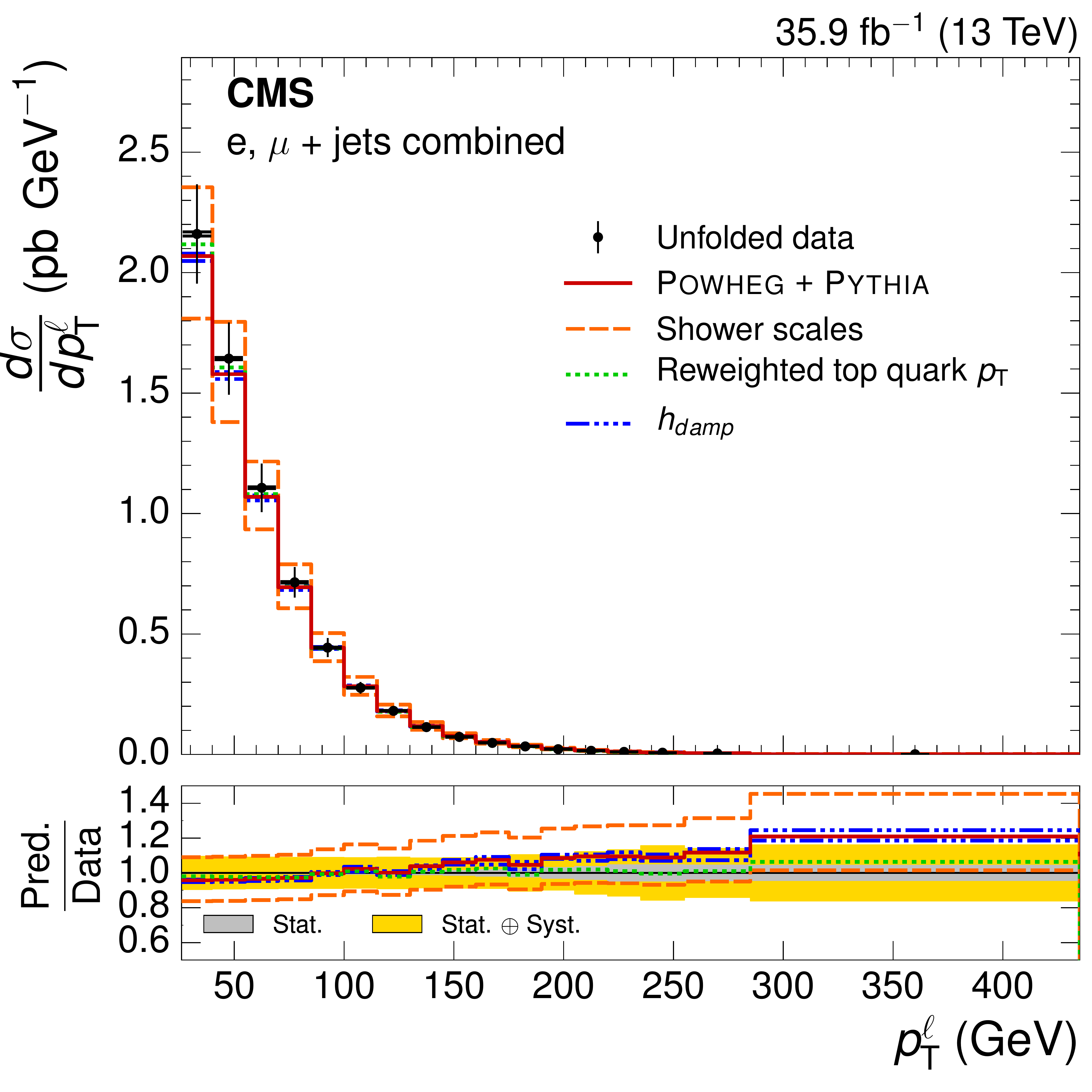} \\
	\includegraphics[width=0.49\linewidth]{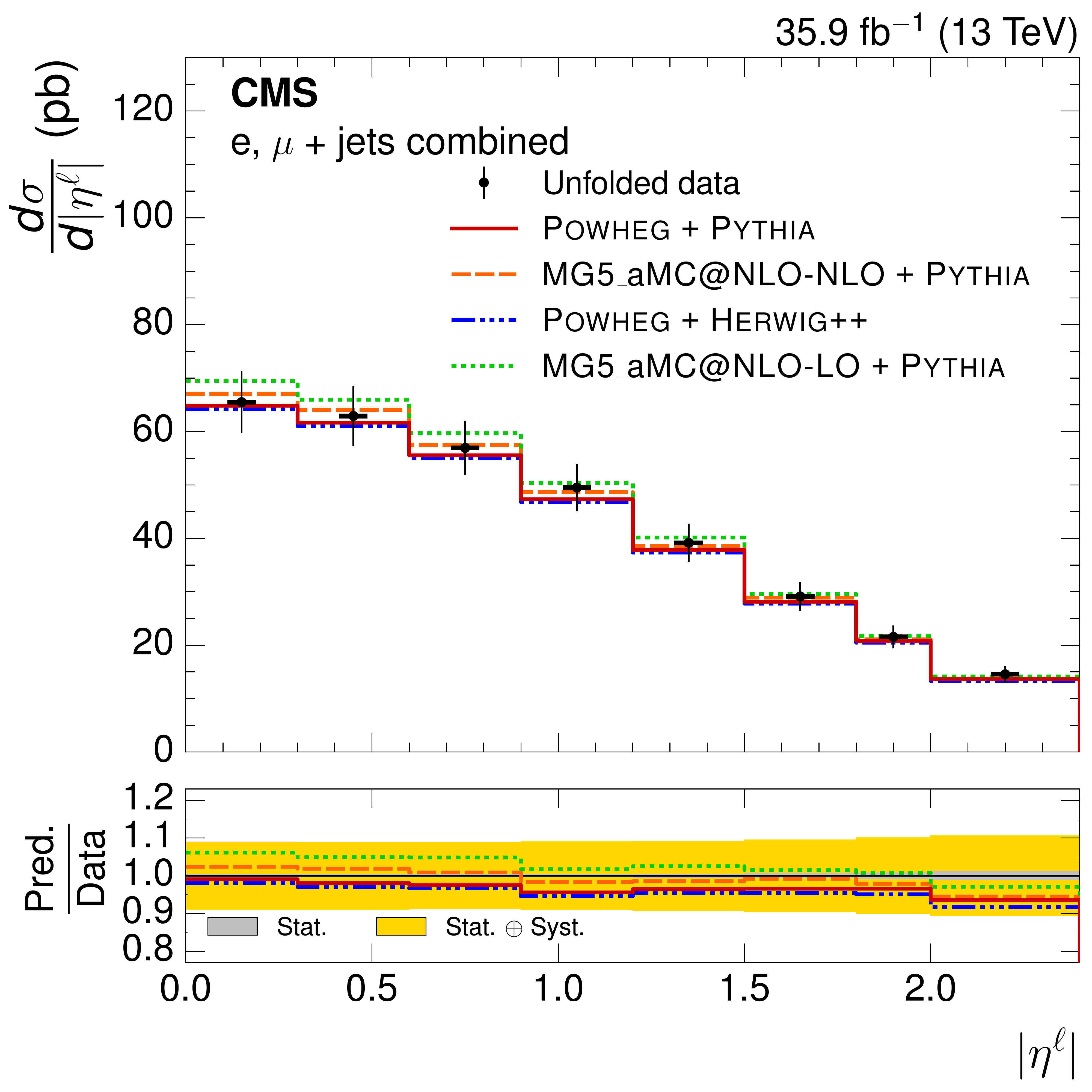}
	\includegraphics[width=0.49\linewidth]{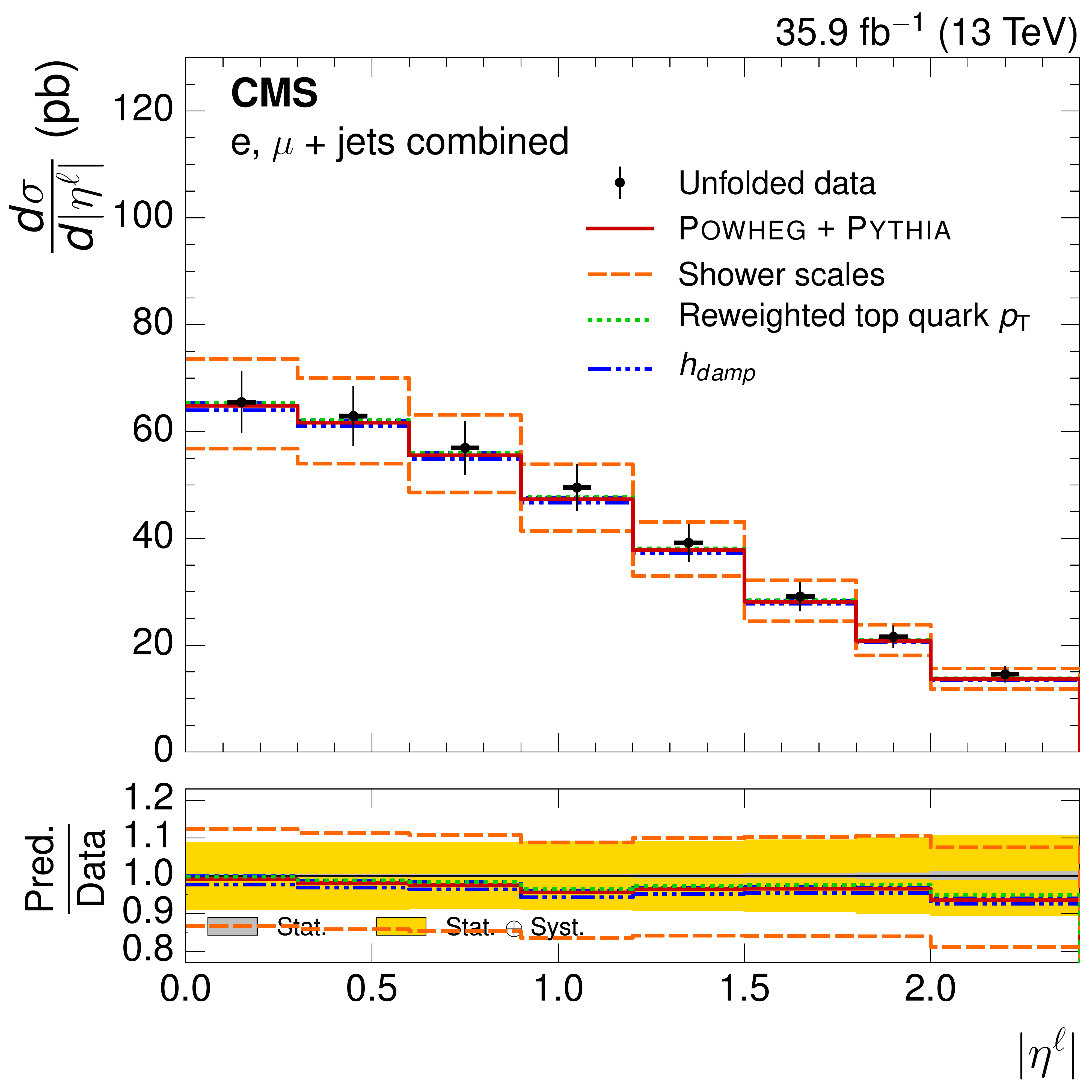}
	\caption{Absolute \LPT (upper) and \LETA (lower) differential \ttbar cross sections, compared to different \ttbar simulations in the left plots, and compared to the \powhegpythia simulation after varying the shower scales, and \hdamp parameter, within their uncertainties, in the right plots.  The vertical bars on the data represent the statistical and systematic uncertainties added in quadrature.  The bottom panels show the ratio of the predictions to the data. }
	\label{plt:ABSXSEC3}
\end{figure}
\clearpage

\section{Summary}
{\tolerance=1200
Normalized and absolute differential \ttbar production cross sections with respect to several kinematic event variables are measured at the particle level in a visible phase space region.
The results are based on proton-proton collision data at \com, collected by the CMS experiment with an integrated luminosity of \Lum.
The total cross section is observed to be consistent with previous results and next-to-next-to-leading-order calculations, and the differential measurements are compared to several \ttbar production models: \powhegpythia, \powhegherwig, \mgamcMLM, and \mgamcFxFx.
\par}

The \powhegpythia simulation is found to be generally consistent with the data, with residual differences covered by theoretical uncertainties.
The jet multiplicity distribution is particularly well-modeled, having been tuned on LHC 8\TeV data.
The \powhegherwig and \mgamcFxFx models are shown to be consistent with data for most kinematic event variables, while
the \mgamcMLM model does not provide an accurate description of any variable measured in the data.

It is expected that the results presented here will be useful for tuning \ttbar generators and models in the future.
To facilitate this, the measurements presented here have been implemented in the RIVET framework and will be available to the wider community.

\begin{acknowledgments}
\hyphenation{Rachada-pisek} We congratulate our colleagues in the CERN accelerator departments for the excellent performance of the LHC and thank the technical and administrative staffs at CERN and at other CMS institutes for their contributions to the success of the CMS effort. In addition, we gratefully acknowledge the computing centers and personnel of the Worldwide LHC Computing Grid for delivering so effectively the computing infrastructure essential to our analyses. Finally, we acknowledge the enduring support for the construction and operation of the LHC and the CMS detector provided by the following funding agencies: BMWFW and FWF (Austria); FNRS and FWO (Belgium); CNPq, CAPES, FAPERJ, and FAPESP (Brazil); MES (Bulgaria); CERN; CAS, MoST, and NSFC (China); COLCIENCIAS (Colombia); MSES and CSF (Croatia); RPF (Cyprus); SENESCYT (Ecuador); MoER, ERC IUT, and ERDF (Estonia); Academy of Finland, MEC, and HIP (Finland); CEA and CNRS/IN2P3 (France); BMBF, DFG, and HGF (Germany); GSRT (Greece); OTKA and NIH (Hungary); DAE and DST (India); IPM (Iran); SFI (Ireland); INFN (Italy); MSIP and NRF (Republic of Korea); LAS (Lithuania); MOE and UM (Malaysia); BUAP, CINVESTAV, CONACYT, LNS, SEP, and UASLP-FAI (Mexico); MBIE (New Zealand); PAEC (Pakistan); MSHE and NSC (Poland); FCT (Portugal); JINR (Dubna); MON, RosAtom, RAS, RFBR and RAEP (Russia); MESTD (Serbia); SEIDI, CPAN, PCTI and FEDER (Spain); Swiss Funding Agencies (Switzerland); MST (Taipei); ThEPCenter, IPST, STAR, and NSTDA (Thailand); TUBITAK and TAEK (Turkey); NASU and SFFR (Ukraine); STFC (United Kingdom); DOE and NSF (USA).

Individuals have received support from the Marie-Curie program and the European Research Council and Horizon 2020 Grant, contract No. 675440 (European Union); the Leventis Foundation; the A. P. Sloan Foundation; the Alexander von Humboldt Foundation; the Belgian Federal Science Policy Office; the Fonds pour la Formation \`a la Recherche dans l'Industrie et dans l'Agriculture (FRIA-Belgium); the Agentschap voor Innovatie door Wetenschap en Technologie (IWT-Belgium); the F.R.S.-FNRS and FWO (Belgium) under the "Excellence of Science - EOS" - be.h project n. 30820817; the Ministry of Education, Youth and Sports (MEYS) of the Czech Republic; the Council of Science and Industrial Research, India; the HOMING PLUS program of the Foundation for Polish Science, cofinanced from European Union, Regional Development Fund, the Mobility Plus program of the Ministry of Science and Higher Education, the National Science Center (Poland), contracts Harmonia 2014/14/M/ST2/00428, Opus 2014/13/B/ST2/02543, 2014/15/B/ST2/03998, and 2015/19/B/ST2/02861, Sonata-bis 2012/07/E/ST2/01406; the National Priorities Research Program by Qatar National Research Fund; the Programa Severo Ochoa del Principado de Asturias; the Thalis and Aristeia programs cofinanced by EU-ESF and the Greek NSRF; the Rachadapisek Sompot Fund for Postdoctoral Fellowship, Chulalongkorn University and the Chulalongkorn Academic into Its 2nd Century Project Advancement Project (Thailand); the Welch Foundation, contract C-1845; and the Weston Havens Foundation (USA).
\end{acknowledgments}

\clearpage
\appendix
\section{Tabulated normalized differential \texorpdfstring{\ttbar}{tt} production cross sections}
\label{ap:tablesOfResults}

\begin{table}[!htbp]
	\topcaption{Results of the normalised differential cross sections with relative uncertainties in the combined channel with respect to \NJET.}
	\label{tb:xsection_normalised_NJets_combined}
	\centering
	\begin{tabular}{cccc}
		\NJET &  \ensuremath{ \frac{ 1 }{ \rd\sigma } \frac{ \rd\sigma }{ \rd\NJET } }  & Stat. unc. & Syst. unc. \\
		 &  & (\%) & (\%) \\
		\hline
		3.5--4.5 	& 0.405 	& 0.27 	& 2.6 \\
		4.5--5.5 	& 0.318 	& 0.27 	& 0.65 \\
		5.5--6.5 	& 0.168 	& 0.47 	& 3.0 \\
		6.5--7.5 	& 7.19$\times 10^{ -2 }$ 	& 1.0 	& 5.3 \\
		7.5--8.5 	& 2.60$\times 10^{ -2 }$ 	& 3.0 	& 9.6 \\
		8.5--10.5 	& 5.89$\times 10^{ -3 }$ 	& 3.9 	& 7.8 \\
	\end{tabular}
\end{table}

\begin{table}[!htbp]
	\topcaption{Results of the normalised differential cross sections with relative uncertainties in the combined channel with respect to \HT.}
	\label{tb:xsection_normalised_HT_combined}
	\centering
	\begin{tabular}{cccc}
		\HT &  \ensuremath{ \frac{ 1 }{ \rd\sigma } \frac{ \rd\sigma }{ \rd\HT } }  & Stat. unc. & Syst. unc. \\
		(\GeVns) & (\ensuremath{\GeVns^{-1}}) & (\%) & (\%) \\
		\hline
		110--220 	& 1.80$\times 10^{ -3 }$ 	& 0.54 	& 14 \\
		220--275 	& 4.13$\times 10^{ -3 }$ 	& 0.37 	& 2.7 \\
		275--340 	& 3.16$\times 10^{ -3 }$ 	& 0.34 	& 3.5 \\
		340--410 	& 2.00$\times 10^{ -3 }$ 	& 0.49 	& 4.1 \\
		410--485 	& 1.18$\times 10^{ -3 }$ 	& 0.69 	& 4.2 \\
		485--570 	& 6.73$\times 10^{ -4 }$ 	& 0.89 	& 4.9 \\
		570--660 	& 3.75$\times 10^{ -4 }$ 	& 1.2 	& 5.5 \\
		660--760 	& 2.05$\times 10^{ -4 }$ 	& 1.5 	& 5.2 \\
		760--870 	& 1.15$\times 10^{ -4 }$ 	& 1.9 	& 7.2 \\
		870--990 	& 6.23$\times 10^{ -5 }$ 	& 2.5 	& 5.6 \\
		990--1115 	& 3.28$\times 10^{ -5 }$ 	& 3.6 	& 13 \\
		1115--1250 	& 1.79$\times 10^{ -5 }$ 	& 4.5 	& 12 \\
		1250--1925 	& 4.78$\times 10^{ -6 }$ 	& 3.3 	& 12 \\
	\end{tabular}
\end{table}
\clearpage

\begin{table}[!htbp]
	\topcaption{Results of the normalised differential cross sections with relative uncertainties in the combined channel with respect to \ST.}
	\label{tb:xsection_normalised_ST_combined}
	\centering
	\begin{tabular}{cccc}
		\ST &  \ensuremath{ \frac{ 1 }{ \rd\sigma } \frac{ \rd\sigma }{ \rd\ST } }  & Stat. unc. & Syst. unc. \\
		(\GeVns) & (\ensuremath{\GeVns^{-1}}) & (\%) & (\%) \\
		\hline
		136--315 	& 9.99$\times 10^{ -4 }$ 	& 0.66 	& 17 \\
		315--390 	& 3.49$\times 10^{ -3 }$ 	& 0.37 	& 2.8 \\
		390--475 	& 2.57$\times 10^{ -3 }$ 	& 0.32 	& 3.7 \\
		475--565 	& 1.54$\times 10^{ -3 }$ 	& 0.49 	& 4.6 \\
		565--665 	& 8.52$\times 10^{ -4 }$ 	& 0.71 	& 5.1 \\
		665--770 	& 4.68$\times 10^{ -4 }$ 	& 1.0 	& 5.5 \\
		770--885 	& 2.54$\times 10^{ -4 }$ 	& 1.3 	& 5.4 \\
		885--1010 	& 1.34$\times 10^{ -4 }$ 	& 1.8 	& 6.4 \\
		1010--1140 	& 7.36$\times 10^{ -5 }$ 	& 2.5 	& 6.1 \\
		1140--1285 	& 3.98$\times 10^{ -5 }$ 	& 3.2 	& 11 \\
		1285--1440 	& 1.96$\times 10^{ -5 }$ 	& 4.6 	& 9.9 \\
		1440--1615 	& 1.16$\times 10^{ -5 }$ 	& 5.3 	& 17 \\
		1615--2490 	& 2.35$\times 10^{ -6 }$ 	& 4.6 	& 14 \\
	\end{tabular}
\end{table}

\begin{table}[!htbp]
	\topcaption{Results of the normalised differential cross sections with relative uncertainties in the combined channel with respect to \ptmiss.}
	\label{tb:xsection_normalised_MET_combined}
	\centering
	\begin{tabular}{cccc}
		\ptmiss &  \ensuremath{ \frac{ 1 }{ \rd\sigma } \frac{ \rd\sigma }{ \rd\ptmiss } }  & Stat. unc. & Syst. unc. \\
		(\GeVns) & (\ensuremath{\GeVns^{-1}}) & (\%) & (\%) \\
		\hline
		0--50 	& 1.05$\times 10^{ -2 }$ 	& 0.16 	& 3.7 \\
		50--105 	& 6.62$\times 10^{ -3 }$ 	& 0.24 	& 2.9 \\
		105--175 	& 1.26$\times 10^{ -3 }$ 	& 0.75 	& 9.1 \\
		175--245 	& 2.43$\times 10^{ -4 }$ 	& 2.0 	& 4.4 \\
		245--315 	& 5.93$\times 10^{ -5 }$ 	& 4.5 	& 11 \\
		315--565 	& 7.63$\times 10^{ -6 }$ 	& 5.8 	& 12 \\
	\end{tabular}
\end{table}
\clearpage

\begin{table}[!htbp]
	\topcaption{Results of the normalised differential cross sections with relative uncertainties in the combined channel with respect to \WPT.}
	\label{tb:xsection_normalised_WPT_combined}
	\centering
	\begin{tabular}{cccc}
		\WPT &  \ensuremath{ \frac{ 1 }{ \rd\sigma } \frac{ \rd\sigma }{ \rd\WPT } }  & Stat. unc. & Syst. unc. \\
		(\GeVns) & (\ensuremath{\GeVns^{-1}}) & (\%) & (\%) \\
		\hline
		0--50 	& 5.12$\times 10^{ -3 }$ 	& 0.36 	& 3.9 \\
		50--105 	& 7.44$\times 10^{ -3 }$ 	& 0.22 	& 0.81 \\
		105--165 	& 3.72$\times 10^{ -3 }$ 	& 0.37 	& 3.2 \\
		165--240 	& 1.09$\times 10^{ -3 }$ 	& 0.69 	& 4.8 \\
		240--325 	& 2.59$\times 10^{ -4 }$ 	& 1.4 	& 5.4 \\
		325--415 	& 6.32$\times 10^{ -5 }$ 	& 2.8 	& 7.7 \\
		415--845 	& 5.47$\times 10^{ -6 }$ 	& 4.1 	& 13 \\
	\end{tabular}
\end{table}

\begin{table}[!htbp]
	\topcaption{Results of the normalised differential cross sections with relative uncertainties in the combined channel with respect to \LPT.}
	\label{tb:xsection_normalised_lepton_pt_combined}
	\centering
	\begin{tabular}{cccc}
		\LPT &  \ensuremath{ \frac{ 1 }{ \rd\sigma } \frac{ \rd\sigma }{ \rd\LPT } }  & Stat. unc. & Syst. unc. \\
		(\GeVns) & (\ensuremath{\GeVns^{-1}}) & (\%) & (\%) \\
		\hline
		26--40 	& 2.14$\times 10^{ -2 }$ 	& 0.30 	& 1.2 \\
		40--55 	& 1.63$\times 10^{ -2 }$ 	& 0.26 	& 0.89 \\
		55--70 	& 1.10$\times 10^{ -2 }$ 	& 0.32 	& 0.69 \\
		70--85 	& 7.08$\times 10^{ -3 }$ 	& 0.41 	& 1.0 \\
		85--100 	& 4.40$\times 10^{ -3 }$ 	& 0.52 	& 1.4 \\
		100--115 	& 2.74$\times 10^{ -3 }$ 	& 0.68 	& 1.8 \\
		115--130 	& 1.79$\times 10^{ -3 }$ 	& 0.86 	& 3.0 \\
		130--145 	& 1.13$\times 10^{ -3 }$ 	& 1.1 	& 2.6 \\
		145--160 	& 7.21$\times 10^{ -4 }$ 	& 1.4 	& 2.6 \\
		160--175 	& 4.76$\times 10^{ -4 }$ 	& 1.9 	& 4.2 \\
		175--190 	& 3.31$\times 10^{ -4 }$ 	& 2.3 	& 7.4 \\
		190--205 	& 2.17$\times 10^{ -4 }$ 	& 3.0 	& 6.3 \\
		205--220 	& 1.51$\times 10^{ -4 }$ 	& 3.8 	& 8.8 \\
		220--235 	& 1.06$\times 10^{ -4 }$ 	& 4.7 	& 9.6 \\
		235--255 	& 7.28$\times 10^{ -5 }$ 	& 4.6 	& 13 \\
		255--285 	& 4.25$\times 10^{ -5 }$ 	& 4.5 	& 11 \\
		285--435 	& 9.45$\times 10^{ -6 }$ 	& 4.4 	& 13 \\
	\end{tabular}
\end{table}
\clearpage

\begin{table}[!htbp]
	\topcaption{Results of the normalised differential cross sections with relative uncertainties in the combined channel with respect to \LETA.}
	\label{tb:xsection_normalised_abs_lepton_eta_coarse_combined}
	\centering
	\begin{tabular}{cccc}
		\LETA &  \ensuremath{ \frac{ 1 }{ \rd\sigma } \frac{ \rd\sigma }{ \rd\LETA } }  & Stat. unc. & Syst. unc. \\
		 &  & (\%) & (\%) \\
		\hline
		0.00--0.30 	& 0.648 	& 0.28 	& 1.0 \\
		0.30--0.60 	& 0.622 	& 0.28 	& 0.89 \\
		0.60--0.90 	& 0.563 	& 0.30 	& 0.86 \\
		0.90--1.20 	& 0.490 	& 0.33 	& 0.69 \\
		1.20--1.50 	& 0.388 	& 0.40 	& 1.1 \\
		1.50--1.80 	& 0.288 	& 0.54 	& 1.9 \\
		1.80--2.00 	& 0.213 	& 0.74 	& 3.0 \\
		2.00--2.40 	& 0.144 	& 1.0 	& 4.4 \\
	\end{tabular}
\end{table}
\clearpage

\newpage
\section{Tabulated absolute differential \texorpdfstring{\ttbar}{tt} production cross sections}
\label{ap:tablesOfAbsResults}

\begin{table}[!htbp]
	\topcaption{Results of the absolute differential cross sections with relative uncertainties in the combined channel with respect to \NJET.}
	\label{tb:xsection_absolute_NJets_combined}
	\centering
	\begin{tabular}{cccc}
		\NJET &  \ensuremath{ \frac{ \rd\sigma }{ \rd\NJET } }  & Stat. unc. & Syst. unc. \\
		 & (pb) & (\%) & (\%) \\
		\hline
		3.5--4.5 	& 40.6 	& 0.33 	& 9.3 \\
		4.5--5.5 	& 31.9 	& 0.29 	& 8.7 \\
		5.5--6.5 	& 16.8 	& 0.47 	& 9.9 \\
		6.5--7.5 	& 7.22 	& 1.0 	& 11 \\
		7.5--8.5 	& 2.60 	& 3.0 	& 13 \\
		8.5--10.5 	& 0.591 	& 3.9 	& 12 \\
	\end{tabular}
\end{table}

\begin{table}[!htbp]
	\topcaption{Results of the absolute differential cross sections with relative uncertainties in the combined channel with respect to \HT.}
	\label{tb:xsection_absolute_HT_combined}
	\centering
	\begin{tabular}{cccc}
		\HT &  \ensuremath{ \frac{ \rd\sigma }{ \rd\HT } }  & Stat. unc. & Syst. unc. \\
		(\GeVns) & (pb \ensuremath{\GeVns^{-1}}) & (\%) & (\%) \\
		\hline
		110--220 	& 0.184 	& 0.62 	& 15 \\
		220--275 	& 0.424 	& 0.38 	& 9.7 \\
		275--340 	& 0.324 	& 0.35 	& 10 \\
		340--410 	& 0.205 	& 0.50 	& 10 \\
		410--485 	& 0.121 	& 0.70 	& 10 \\
		485--570 	& 6.91$\times 10^{ -2 }$ 	& 0.89 	& 11 \\
		570--660 	& 3.85$\times 10^{ -2 }$ 	& 1.2 	& 11 \\
		660--760 	& 2.10$\times 10^{ -2 }$ 	& 1.5 	& 10 \\
		760--870 	& 1.18$\times 10^{ -2 }$ 	& 2.0 	& 12 \\
		870--990 	& 6.39$\times 10^{ -3 }$ 	& 2.5 	& 11 \\
		990--1115 	& 3.37$\times 10^{ -3 }$ 	& 3.6 	& 16 \\
		1115--1250 	& 1.84$\times 10^{ -3 }$ 	& 4.5 	& 14 \\
		1250--1925 	& 4.91$\times 10^{ -4 }$ 	& 3.3 	& 16 \\
	\end{tabular}
\end{table}
\clearpage

\begin{table}[!htbp]
	\topcaption{Results of the absolute differential cross sections with relative uncertainties in the combined channel with respect to \ST.}
	\label{tb:xsection_absolute_ST_combined}
	\centering
	\begin{tabular}{cccc}
		\ST &  \ensuremath{ \frac{ \rd\sigma }{ \rd\ST } }  & Stat. unc. & Syst. unc. \\
		(\GeVns) & (pb \ensuremath{\GeVns^{-1}}) & (\%) & (\%) \\
		\hline
		136--315 	& 0.103 	& 0.73 	& 18 \\
		315--390 	& 0.358 	& 0.38 	& 9.5 \\
		390--475 	& 0.264 	& 0.34 	& 10 \\
		475--565 	& 0.158 	& 0.50 	& 10 \\
		565--665 	& 8.76$\times 10^{ -2 }$ 	& 0.71 	& 11 \\
		665--770 	& 4.81$\times 10^{ -2 }$ 	& 1.0 	& 11 \\
		770--885 	& 2.61$\times 10^{ -2 }$ 	& 1.3 	& 11 \\
		885--1010 	& 1.38$\times 10^{ -2 }$ 	& 1.8 	& 11 \\
		1010--1140 	& 7.56$\times 10^{ -3 }$ 	& 2.5 	& 11 \\
		1140--1285 	& 4.09$\times 10^{ -3 }$ 	& 3.2 	& 15 \\
		1285--1440 	& 2.02$\times 10^{ -3 }$ 	& 4.6 	& 13 \\
		1440--1615 	& 1.20$\times 10^{ -3 }$ 	& 5.3 	& 20 \\
		1615--2490 	& 2.42$\times 10^{ -4 }$ 	& 4.6 	& 18 \\
	\end{tabular}
\end{table}

\begin{table}[!htbp]
	\topcaption{Results of the absolute differential cross sections with relative uncertainties in the combined channel with respect to \ptmiss.}
	\label{tb:xsection_absolute_MET_combined}
	\centering
	\begin{tabular}{cccc}
		\ptmiss &  \ensuremath{ \frac{ \rd\sigma }{ \rd\ptmiss } }  & Stat. unc. & Syst. unc. \\
		(\GeVns) & (pb \ensuremath{\GeVns^{-1}}) & (\%) & (\%) \\
		\hline
		0--50 	& 1.05 	& 0.21 	& 8.8 \\
		50--105 	& 0.664 	& 0.28 	& 10 \\
		105--175 	& 0.126 	& 0.76 	& 15 \\
		175--245 	& 2.44$\times 10^{ -2 }$ 	& 2.0 	& 11 \\
		245--315 	& 5.96$\times 10^{ -3 }$ 	& 4.5 	& 15 \\
		315--565 	& 7.66$\times 10^{ -4 }$ 	& 5.8 	& 17 \\
	\end{tabular}
\end{table}
\clearpage

\begin{table}[!htbp]
	\topcaption{Results of the absolute differential cross sections with relative uncertainties in the combined channel with respect to \WPT.}
	\label{tb:xsection_absolute_WPT_combined}
	\centering
	\begin{tabular}{cccc}
		\WPT &  \ensuremath{ \frac{ \rd\sigma }{ \rd\WPT } }  & Stat. unc. & Syst. unc. \\
		(\GeVns) & (pb \ensuremath{\GeVns^{-1}}) & (\%) & (\%) \\
		\hline
		0--50 	& 0.518 	& 0.41 	& 8.3 \\
		50--105 	& 0.752 	& 0.25 	& 8.7 \\
		105--165 	& 0.377 	& 0.39 	& 11 \\
		165--240 	& 0.111 	& 0.70 	& 11 \\
		240--325 	& 2.62$\times 10^{ -2 }$ 	& 1.4 	& 12 \\
		325--415 	& 6.40$\times 10^{ -3 }$ 	& 2.8 	& 13 \\
		415--845 	& 5.54$\times 10^{ -4 }$ 	& 4.1 	& 17 \\
	\end{tabular}
\end{table}

\begin{table}[!htbp]
	\topcaption{Results of the absolute differential cross sections with relative uncertainties in the combined channel with respect to \LPT.}
	\label{tb:xsection_absolute_lepton_pt_combined}
	\centering
	\begin{tabular}{cccc}
		\LPT &  \ensuremath{ \frac{ \rd\sigma }{ \rd\LPT } }  & Stat. unc. & Syst. unc. \\
		(\GeVns) & (pb \ensuremath{\GeVns^{-1}}) & (\%) & (\%) \\
		\hline
		26--40 	& 2.16 	& 0.39 	& 9.5 \\
		40--55 	& 1.64 	& 0.28 	& 9.2 \\
		55--70 	& 1.11 	& 0.33 	& 9.1 \\
		70--85 	& 0.715 	& 0.41 	& 8.8 \\
		85--100 	& 0.443 	& 0.53 	& 8.9 \\
		100--115 	& 0.277 	& 0.68 	& 8.7 \\
		115--130 	& 0.181 	& 0.86 	& 9.0 \\
		130--145 	& 0.114 	& 1.1 	& 8.6 \\
		145--160 	& 7.28$\times 10^{ -2 }$ 	& 1.4 	& 8.8 \\
		160--175 	& 4.80$\times 10^{ -2 }$ 	& 1.9 	& 9.0 \\
		175--190 	& 3.34$\times 10^{ -2 }$ 	& 2.3 	& 11 \\
		190--205 	& 2.19$\times 10^{ -2 }$ 	& 3.0 	& 9.8 \\
		205--220 	& 1.52$\times 10^{ -2 }$ 	& 3.8 	& 12 \\
		220--235 	& 1.07$\times 10^{ -2 }$ 	& 4.7 	& 13 \\
		235--255 	& 7.34$\times 10^{ -3 }$ 	& 4.6 	& 16 \\
		255--285 	& 4.29$\times 10^{ -3 }$ 	& 4.5 	& 14 \\
		285--435 	& 9.53$\times 10^{ -4 }$ 	& 4.4 	& 16 \\
	\end{tabular}
\end{table}
\clearpage

\begin{table}[!htbp]
	\topcaption{Results of the absolute differential cross sections with relative uncertainties in the combined channel with respect to \LETA.}
	\label{tb:xsection_absolute_abs_lepton_eta_coarse_combined}
	\centering
	\begin{tabular}{cccc}
		\LETA &  \ensuremath{ \frac{ \rd\sigma }{ \rd\LETA } }  & Stat. unc. & Syst. unc. \\
		 & (pb) & (\%) & (\%) \\
		\hline
		0.00--0.30 	& 65.5 	& 0.31 	& 8.9 \\
		0.30--0.60 	& 62.9 	& 0.30 	& 8.8 \\
		0.60--0.90 	& 56.9 	& 0.32 	& 8.8 \\
		0.90--1.20 	& 49.5 	& 0.35 	& 9.0 \\
		1.20--1.50 	& 39.2 	& 0.43 	& 9.1 \\
		1.50--1.80 	& 29.1 	& 0.57 	& 9.5 \\
		1.80--2.00 	& 21.6 	& 0.76 	& 10 \\
		2.00--2.40 	& 14.6 	& 1.1 	& 11 \\
	\end{tabular}
\end{table}
\clearpage

\newpage
\section{Tabulated minimum and maximum relative uncertainties for absolute cross sections}
\label{ap:absCondensed}

\begin{landscape}
\begin{table}
	\topcaption{ The upper and lower bounds, in \%, from each source of systematic uncertainty in the absolute differential cross section, over all bins of the measurement for each variable.  The bounds of the total relative uncertainty are also shown.}
	\label{tb:syst_condensed_combined_absolute}
	\centering
	\begin{tabular}{lccccccc}
		Relative uncertainty source $(\%)$	&	\NJET	&	\HT	&	\ST	&	\ptmiss	&	\WPT	&	\LPT	&	\LETA \vspace*{0.1cm}  \\
		\hline
	    \bquark tagging efficiency	&	3.1 -- 4.0	&	3.5 -- 4.5	&	3.5 -- 5.0	&	3.6 -- 4.8	&	3.6 -- 5.2	&	3.6 -- 5.4	&	3.6 -- 4.2\\
		Electron efficiency	&	1.6 -- 1.9	&	1.6 -- 2.3	&	1.4 -- 2.5	&	1.6 -- 2.4	&	1.2 -- 3.0	&	1.0 -- 3.5	&	1.0 -- 2.1\\
		Muon efficiency	&	2.0 -- 2.5	&	2.2 -- 2.5	&	2.2 -- 2.6	&	2.3 -- 2.6	&	2.3 -- 2.9	&	2.2 -- 3.3	&	2.1 -- 2.3\\
		JER	&	0.1 -- 0.8	&	0.1 -- 1.1	&	0.2 -- 2.3	&	0.4 -- 5.6	&	0.4 -- 1.6	&	0.1 -- 0.3	&	0.3 -- 0.4\\
		JES	&	1.4 -- 9.2	&	4.3 -- 12.8	&	4.3 -- 15.1	&	2.2 -- 10.9	&	2.2 -- 7.8	&	1.8 -- 4.1	&	3.8 -- 4.1\\
		Electron transverse momentum in \ptmiss	&	\NA	&	\NA	&	0.1 -- 0.3	&	0.1 -- 0.9	&	0.1 -- 0.6	&	\NA	&	\NA\\
		Muon transverse momentum in \ptmiss	&	\NA	&	\NA	&	0.1 -- 0.9	&	0.1 -- 3.5	&	0.1 -- 0.8	&	\NA	&	\NA\\
		Tau transverse momentum in \ptmiss	&	\NA	&	\NA	&	0.1 -- 1.4	&	0.1 -- 1.2	&	0.1 -- 1.4	&	\NA	&	\NA\\
		Unclustered transverse momentum in \ptmiss	&	\NA	&	\NA	&	0.1 -- 1.7	&	0.2 -- 1.9	&	0.1 -- 1.0	&	\NA	&	\NA\\
		QCD bkg cross section		&	0.1 -- 0.9	&	0.1 -- 1.6	&	0.1 -- 2.3	&	0.1 -- 0.8	&	0.1 -- 1.3	&	0.1 -- 5.0	&	0.1 -- 3.6\\
		QCD bkg shape			&	0.1 -- 0.1	&	0.1 -- 0.8	&	0.1 -- 1.0	&	0.1 -- 0.1	&	0.1 -- 1.6	&	0.1 -- 4.8	&	0.1 -- 1.5\\
		Single top quark cross section	&	1.1 -- 1.7	&	1.1 -- 3.5	&	1.1 -- 5.8	&	1.3 -- 6.3	&	1.1 -- 8.4	&	1.3 -- 7.4	&	1.4 -- 1.5\\
		V+jets cross section	&	0.7 -- 1.1	&	0.6 -- 3.4	&	0.5 -- 4.6	&	0.7 -- 2.9	&	0.7 -- 4.5	&	0.6 -- 6.3	&	0.6 -- 2.5\\
		PDF 	&	0.1 -- 0.3	&	0.1 -- 0.4	&	0.1 -- 0.6	&	0.1 -- 0.3	&	0.1 -- 0.4	&	0.1 -- 0.5	&	$<$0.1\\
		Color reconnection (Gluon move)	&	0.2 -- 2.8	&	0.2 -- 4.1	&	0.1 -- 11.8	&	0.2 -- 1.0	&	0.1 -- 1.1	&	0.2 -- 4.7	&	0.1 -- 0.5\\
		Color reconnection (QCD-based)	&	0.2 -- 2.1	&	0.1 -- 4.3	&	0.1 -- 6.7	&	0.3 -- 4.6	&	0.3 -- 3.8	&	0.1 -- 7.9	&	0.1 -- 1.7\\
		Color reconnection (Early resonance decays)	&	0.1 -- 3.9	&	0.1 -- 7.1	&	0.1 -- 4.1	&	0.1 -- 1.4	&	0.1 -- 3.8	&	0.1 -- 5.0	&	0.1 -- 1.1\\
		Fragmentation	&	0.1 -- 0.7	&	0.6 -- 1.6	&	0.5 -- 1.4	&	0.1 -- 0.7	&	0.2 -- 0.8	&	0.1 -- 0.8	&	0.3 -- 0.4\\
		\hdamp	&	0.1 -- 3.5	&	0.2 -- 3.1	&	0.1 -- 3.0	&	0.6 -- 2.1	&	0.2 -- 3.2	&	0.4 -- 3.4	&	0.5 -- 2.0\\
		Top quark mass	&	0.7 -- 2.0	&	0.3 -- 3.4	&	0.3 -- 3.7	&	0.3 -- 5.0	&	0.4 -- 2.0	&	0.3 -- 3.5	&	0.9 -- 1.7\\
		Peterson fragmentation model	&	0.3 -- 2.0	&	1.6 -- 2.7	&	1.9 -- 3.2	&	1.1 -- 2.6	&	1.3 -- 2.6	&	1.2 -- 2.9	&	1.5 -- 1.5\\
		Shower scales	&	2.7 -- 6.5	&	2.6 -- 6.4	&	3.0 -- 7.0	&	4.0 -- 6.6	&	4.6 -- 6.2	&	3.9 -- 6.2	&	4.9 -- 5.6\\
		\PB\ hadron decay semileptonic branching fraction	&	0.2 -- 0.3	&	0.1 -- 0.3	&	0.1 -- 0.3	&	0.2 -- 0.3	&	0.2 -- 0.3	&	0.2 -- 0.3	&	0.2 -- 0.3\\
		Top quark \ensuremath{\pt}	&	0.4 -- 1.2	&	0.1 -- 0.8	&	0.1 -- 0.9	&	0.1 -- 1.5	&	0.1 -- 1.0	&	0.1 -- 1.0	&	0.6 -- 0.7\\
		Underlying event tune	&	0.1 -- 2.9	&	0.2 -- 5.3	&	0.2 -- 4.4	&	0.2 -- 5.4	&	0.1 -- 2.6	&	0.1 -- 6.0	&	0.2 -- 0.9\\
		Simulated sample size	&	0.1 -- 1.6	&	0.1 -- 1.6	&	0.1 -- 1.9	&	0.1 -- 2.2	&	0.1 -- 1.4	&	0.1 -- 1.7	&	0.1 -- 0.4\\
		Additional interactions	&	0.1 -- 0.3	&	0.1 -- 0.7	&	0.1 -- 1.3	&	0.3 -- 1.3	&	0.1 -- 0.6	&	0.1 -- 0.8	&	0.1 -- 0.3\\
		Integrated luminosity	&	2.5 -- 2.5	&	2.5 -- 2.5	&	2.5 -- 2.5	&	2.5 -- 2.5	&	2.5 -- 2.5	&	2.5 -- 2.5	&	2.5 -- 2.5\\
		\hline
		Total	&	8.7 -- 13.4	&	9.7 -- 15.9	&	9.5 -- 20.0	&	8.8 -- 17.1	&	8.3 -- 17.5	&	8.6 -- 16.1	&	8.8 -- 10.6\\
	\end{tabular}
\end{table}
\end{landscape}
\clearpage
\bibliography{auto_generated}

\cleardoublepage \section{The CMS Collaboration \label{app:collab}}\begin{sloppypar}\hyphenpenalty=5000\widowpenalty=500\clubpenalty=5000\input{TOP-16-014-authorlist.tex}\end{sloppypar}
\end{document}

%% file: TOP-16-014-authorlist.tex
\textbf{Yerevan Physics Institute,  Yerevan,  Armenia}\\*[0pt]
A.M.~Sirunyan, A.~Tumasyan
\vskip\cmsinstskip
\textbf{Institut f\"{u}r Hochenergiephysik,  Wien,  Austria}\\*[0pt]
W.~Adam, F.~Ambrogi, E.~Asilar, T.~Bergauer, J.~Brandstetter, E.~Brondolin, M.~Dragicevic, J.~Er\"{o}, A.~Escalante Del Valle, M.~Flechl, M.~Friedl, R.~Fr\"{u}hwirth\cmsAuthorMark{1}, V.M.~Ghete, J.~Grossmann, J.~Hrubec, M.~Jeitler\cmsAuthorMark{1}, A.~K\"{o}nig, N.~Krammer, I.~Kr\"{a}tschmer, D.~Liko, T.~Madlener, I.~Mikulec, E.~Pree, N.~Rad, H.~Rohringer, J.~Schieck\cmsAuthorMark{1}, R.~Sch\"{o}fbeck, M.~Spanring, D.~Spitzbart, A.~Taurok, W.~Waltenberger, J.~Wittmann, C.-E.~Wulz\cmsAuthorMark{1}, M.~Zarucki
% ...
\vskip\cmsinstskip
% ...
1:~~Also at Vienna University of Technology, Vienna, Austria\\
% ...


%% file: TOP-16-014_temp.bbl
\providecommand{\href}[2]{#2}\begingroup\raggedright\begin{thebibliography}{10}%
\makeatletter
\providecommand{\hrefCMSnoop }[0]{\@secondoftwo}%
\makeatother
\providecommand{\doi}{\texttt{doi:}\begingroup \urlstyle{tt}\Url}

\bibitem{Chatrchyan:2008zzk}
\hrefCMSnoop {}{{CMS Collaboration}, ``The {CMS} experiment at the {CERN}
  {LHC}'',} \textit{ JINST} \textbf{ 3} (2008) S08004,
\href{http://dx.doi.org/10.1088/1748-0221/3/08/S08004}{\doi{10.1088/1748-0221/3/08/S08004}}.
%%CITATION = JINST,3,S08004;%%.

\bibitem{Diff7TeV}
\hrefCMSnoop {}{{CMS Collaboration}, ``{Measurement of differential top-quark
  pair production cross sections in pp collisions at $\sqrt{s}=7$ TeV}'',}
  \textit{ Eur. Phys. J. C} \textbf{ 73} (2013) 2339,
  \href{http://dx.doi.org/10.1140/epjc/s10052-013-2339-4}{\doi{10.1140/epjc/s10052-013-2339-4}},
\href{http://www.arXiv.org/abs/1211.2220}{\texttt{arXiv:1211.2220}}.
%%CITATION = ARXIV:1211.2220;%%.

\bibitem{Diff8TeV}
\hrefCMSnoop {}{{CMS Collaboration}, ``{Measurement of the differential cross
  section for top quark pair production in pp collisions at $\sqrt{s} =
  8\,\text {TeV} $}'',} \textit{ Eur. Phys. J. C} \textbf{ 75} (2015) 542,
  \href{http://dx.doi.org/10.1140/epjc/s10052-015-3709-x}{\doi{10.1140/epjc/s10052-015-3709-x}},
\href{http://www.arXiv.org/abs/1505.04480}{\texttt{arXiv:1505.04480}}.
%%CITATION = ARXIV:1505.04480;%%.

\bibitem{DileptonDoubleDiff8TeV}
\hrefCMSnoop {}{{CMS Collaboration}, ``{Measurement of double-differential
  cross sections for top quark pair production in pp collisions at $\sqrt{s} =
  8$ $\,\text {TeV}$ and impact on parton distribution functions}'',} \textit{
  Eur. Phys. J. C} \textbf{ 77} (2017) 459,
  \href{http://dx.doi.org/10.1140/epjc/s10052-017-4984-5}{\doi{10.1140/epjc/s10052-017-4984-5}},
\href{http://www.arXiv.org/abs/1703.01630}{\texttt{arXiv:1703.01630}}.
%%CITATION = ARXIV:1703.01630;%%.

\bibitem{LjetHighPtDiff8TeV}
\hrefCMSnoop {}{{CMS Collaboration}, ``{Measurement of the integrated and
  differential $t \bar t$ production cross sections for high-$p_T$ top quarks
  in pp collisions at $\sqrt s =$ 8 TeV}'',} \textit{ Phys. Rev. D} \textbf{
  94} (2016) 072002,
  \href{http://dx.doi.org/10.1103/PhysRevD.94.072002}{\doi{10.1103/PhysRevD.94.072002}},
\href{http://www.arXiv.org/abs/1605.00116}{\texttt{arXiv:1605.00116}}.
%%CITATION = ARXIV:1605.00116;%%.

\bibitem{DiffAllJet8TeV}
\hrefCMSnoop {}{{CMS Collaboration}, ``{Measurement of the $ \mathrm{ t \bar{t}
  } $ production cross section in the all-jets final state in pp collisions at
  $\sqrt{s} =$ 8 TeV}'',} \textit{ Eur. Phys. J. C} \textbf{ 76} (2015) 128,
  \href{http://dx.doi.org/10.1140/epjc/s10052-016-3956-5}{\doi{10.1140/epjc/s10052-016-3956-5}},
  \href{http://www.arXiv.org/abs/1509.06076}{\texttt{arXiv:1509.06076}}.

\bibitem{LJetsDiff13TeV}
\hrefCMSnoop {}{{CMS Collaboration}, ``{Measurement of differential cross
  sections for top quark pair production using the lepton+jets final state in
  proton-proton collisions at 13 TeV}'',} \textit{ Phys. Rev. D} \textbf{ 95}
  (2017) 092001,
  \href{http://dx.doi.org/10.1103/PhysRevD.95.092001}{\doi{10.1103/PhysRevD.95.092001}},
\href{http://www.arXiv.org/abs/1610.04191}{\texttt{arXiv:1610.04191}}.
%%CITATION = ARXIV:1610.04191;%%.

\bibitem{DileptonDiff13TeV}
\hrefCMSnoop {}{{CMS Collaboration}, ``{Measurement of normalized differential
  $\mathrm{t \bar t}$ cross sections in the dilepton channel from pp collisions
  at $\sqrt{s}=13~\mathrm{TeV}$}'',} (2017).
  \href{http://www.arXiv.org/abs/1708.07638}{\texttt{arXiv:1708.07638}}.
Submitted to {\it JHEP}.
%%CITATION = ARXIV:1708.07638;%%.

\bibitem{ATLASLJetsDiff13TeV}
\hrefCMSnoop {}{{ATLAS Collaboration}, ``{Measurements of top-quark pair
  differential cross-sections in the lepton+jets channel in $pp$ collisions at
  $\sqrt{s}=13$ TeV using the ATLAS detector}'',} \textit{ JHEP} \textbf{ 11}
  (2017) 191,
  \href{http://dx.doi.org/10.1007/JHEP11(2017)191}{\doi{10.1007/JHEP11(2017)191}},
\href{http://www.arXiv.org/abs/1708.00727}{\texttt{arXiv:1708.00727}}.
%%CITATION = ARXIV:1708.00727;%%.

\bibitem{ATLASDileptonDiff13TeV}
\hrefCMSnoop {}{{ATLAS Collaboration}, ``{Measurements of top-quark pair
  differential cross-sections in the $e\mu$ channel in pp collisions at
  $\sqrt{s} = 13$ TeV using the ATLAS detector}'',} \textit{ Eur. Phys. J. C}
  \textbf{ 77} (2017) 292,
  \href{http://dx.doi.org/10.1140/epjc/s10052-017-4821-x}{\doi{10.1140/epjc/s10052-017-4821-x}},
\href{http://www.arXiv.org/abs/1612.05220}{\texttt{arXiv:1612.05220}}.
%%CITATION = ARXIV:1612.05220;%%.

\bibitem{TOP12042}
\hrefCMSnoop {}{{CMS Collaboration}, ``{Measurement of the differential cross
  sections for top quark pair production as a function of kinematic event
  variables in pp collisions at $\sqrt s=7$ and 8 TeV}'',} \textit{ Phys. Rev.
  D} \textbf{ 94} (2016) 052006,
  \href{http://dx.doi.org/10.1103/PhysRevD.94.052006}{\doi{10.1103/PhysRevD.94.052006}},
\href{http://www.arXiv.org/abs/1607.00837}{\texttt{arXiv:1607.00837}}.
%%CITATION = ARXIV:1607.00837;%%.

\bibitem{CMSTrigger}
\hrefCMSnoop {}{{CMS Collaboration}, ``{The CMS trigger system}'',} \textit{
  JINST} \textbf{ 12} (2017) P01020,
  \href{http://dx.doi.org/10.1088/1748-0221/12/01/P01020}{\doi{10.1088/1748-0221/12/01/P01020}},
\href{http://www.arXiv.org/abs/1609.02366}{\texttt{arXiv:1609.02366}}.
%%CITATION = ARXIV:1609.02366;%%.

\bibitem{Powheg_ref2}
\hrefCMSnoop {}{S.~Frixione, P.~Nason, and C.~Oleari, ``{Matching NLO QCD
  computations with Parton Shower simulations: the POWHEG method}'',} \textit{
  JHEP} \textbf{ 11} (2007) 070,
  \href{http://dx.doi.org/10.1088/1126-6708/2007/11/070}{\doi{10.1088/1126-6708/2007/11/070}},
\href{http://www.arXiv.org/abs/0709.2092}{\texttt{arXiv:0709.2092}}.
%%CITATION = ARXIV:0709.2092;%%.

\bibitem{Powheg_ref1}
\hrefCMSnoop {}{P.~Nason, ``{A new method for combining NLO QCD with shower
  Monte Carlo algorithms}'',} \textit{ JHEP} \textbf{ 11} (2004) 040,
  \href{http://dx.doi.org/10.1088/1126-6708/2004/11/040}{\doi{10.1088/1126-6708/2004/11/040}},
\href{http://www.arXiv.org/abs/hep-ph/0409146}{\texttt{arXiv:hep-ph/0409146}}.
%%CITATION = HEP-PH/0409146;%%.

\bibitem{Powheg_ref3}
\hrefCMSnoop {}{S.~Alioli, P.~Nason, C.~Oleari, and E.~Re, ``{A general
  framework for implementing NLO calculations in shower Monte Carlo programs:
  the POWHEG BOX}'',} \textit{ JHEP} \textbf{ 06} (2010) 043,
  \href{http://dx.doi.org/10.1007/JHEP06(2010)043}{\doi{10.1007/JHEP06(2010)043}},
\href{http://www.arXiv.org/abs/1002.2581}{\texttt{arXiv:1002.2581}}.
%%CITATION = ARXIV:1002.2581;%%.

\bibitem{Powheg_hvq}
\hrefCMSnoop {}{S.~Frixione, P.~Nason, and G.~Ridolfi, ``{A positive-weight
  next-to-leading-order Monte Carlo for heavy flavour hadroproduction}'',}
  \textit{ JHEP} \textbf{ 09} (2007) 126,
  \href{http://dx.doi.org/10.1088/1126-6708/2007/09/126}{\doi{10.1088/1126-6708/2007/09/126}},
\href{http://www.arXiv.org/abs/0707.3088}{\texttt{arXiv:0707.3088}}.
%%CITATION = ARXIV:0707.3088;%%.

\bibitem{Pythia6}
\hrefCMSnoop {}{T.~Sj{\"o}strand, S.~Mrenna, and P.~Skands, ``{PYTHIA} 6.4
  physics and manual'',} \textit{ JHEP} \textbf{ 05} (2006) 026,
  \href{http://dx.doi.org/10.1088/1126-6708/2006/05/026}{\doi{10.1088/1126-6708/2006/05/026}},
\href{http://www.arXiv.org/abs/hep-ph/0603175}{\texttt{arXiv:hep-ph/0603175}}.
%%CITATION = HEP-PH/0603175;%%.

\bibitem{Pythia8}
\hrefCMSnoop {}{T.~Sj{\"o}strand, S.~Mrenna, and P.~Skands, ``A brief
  introduction to {PYTHIA 8.1}'',} \textit{ Comput. Phys. Commun.} \textbf{
  178} (2008) 852,
  \href{http://dx.doi.org/10.1016/j.cpc.2008.01.036}{\doi{10.1016/j.cpc.2008.01.036}},
\href{http://www.arXiv.org/abs/0710.3820}{\texttt{arXiv:0710.3820}}.
%%CITATION = ARXIV:0710.3820;%%.

\bibitem{CUETP8M2T4_Tune}
\href {http://cds.cern.ch/record/2235192}{{CMS Collaboration},
  ``{Investigations of the impact of the parton shower tuning in Pythia 8 in
  the modelling of $\mathrm{t\overline{t}}$ at $\sqrt{s}=8$ and 13 TeV}'',} CMS
  Physics Analysis Summary CMS-PAS-TOP-16-021, 2016.

\bibitem{Herwigpp}
M.~Bahr\hrefCMSnoop {}{ {et~al.}, ``{Herwig++} physics and manual'',} \textit{
  Eur. Phys. J. C} \textbf{ 58} (2008) 639,
  \href{http://dx.doi.org/10.1140/epjc/s10052-008-0798-9}{\doi{10.1140/epjc/s10052-008-0798-9}},
\href{http://www.arXiv.org/abs/0803.0883}{\texttt{arXiv:0803.0883}}.
%%CITATION = ARXIV:0803.0883;%%.

\bibitem{EE5C}
\hrefCMSnoop {}{S.~Gieseke, C.~Rohr, and A.~Siodmok, ``{Colour reconnections in
  Herwig++}'',} \textit{ Eur. Phys. J. C} \textbf{ 72} (2012) 2225,
  \href{http://dx.doi.org/10.1140/epjc/s10052-012-2225-5}{\doi{10.1140/epjc/s10052-012-2225-5}},
\href{http://www.arXiv.org/abs/1206.0041}{\texttt{arXiv:1206.0041}}.
%%CITATION = ARXIV:1206.0041;%%.

\bibitem{MGamc}
J.~Alwall\hrefCMSnoop {}{ {et~al.}, ``{The automated computation of tree-level
  and next-to-leading order differential cross sections, and their matching to
  parton shower simulations}'',} \textit{ JHEP} \textbf{ 07} (2014) 079,
  \href{http://dx.doi.org/10.1007/JHEP07(2014)079}{\doi{10.1007/JHEP07(2014)079}},
\href{http://www.arXiv.org/abs/1405.0301}{\texttt{arXiv:1405.0301}}.
%%CITATION = ARXIV:1405.0301;%%.

\bibitem{CUETP8M1}
\hrefCMSnoop {}{{CMS Collaboration}, ``{Event generator tunes obtained from
  underlying event and multiparton scattering measurements}'',} \textit{ Eur.
  Phys. J. C} \textbf{ 76} (2016) 155,
  \href{http://dx.doi.org/10.1140/epjc/s10052-016-3988-x}{\doi{10.1140/epjc/s10052-016-3988-x}},
\href{http://www.arXiv.org/abs/1512.00815}{\texttt{arXiv:1512.00815}}.
%%CITATION = ARXIV:1512.00815;%%.

\bibitem{MLM}
J.~Alwall\hrefCMSnoop {}{ {et~al.}, ``{Comparative study of various algorithms
  for the merging of parton showers and matrix elements in hadronic
  collisions}'',} \textit{ Eur. Phys. J. C} \textbf{ 53} (2008) 473,
  \href{http://dx.doi.org/10.1140/epjc/s10052-007-0490-5}{\doi{10.1140/epjc/s10052-007-0490-5}},
\href{http://www.arXiv.org/abs/0706.2569}{\texttt{arXiv:0706.2569}}.
%%CITATION = ARXIV:0706.2569;%%.

\bibitem{FxFx}
\hrefCMSnoop {}{R.~Frederix and S.~Frixione, ``{Merging meets matching in
  MC@NLO}'',} \textit{ JHEP} \textbf{ 12} (2012) 061,
  \href{http://dx.doi.org/10.1007/JHEP12(2012)061}{\doi{10.1007/JHEP12(2012)061}},
\href{http://www.arXiv.org/abs/1209.6215}{\texttt{arXiv:1209.6215}}.
%%CITATION = ARXIV:1209.6215;%%.

\bibitem{nnpdf}
\hrefCMSnoop {}{{NNPDF} Collaboration, ``{Parton distributions for the LHC Run
  II}'',} \textit{ JHEP} \textbf{ 04} (2015) 040,
  \href{http://dx.doi.org/10.1007/JHEP04(2015)040}{\doi{10.1007/JHEP04(2015)040}},
\href{http://www.arXiv.org/abs/1410.8849}{\texttt{arXiv:1410.8849}}.
%%CITATION = ARXIV:1410.8849;%%.

\bibitem{ttxsec_1}
\hrefCMSnoop {}{M.~Beneke, P.~Falgari, S.~Klein, and C.~Schwinn, ``{Hadronic
  top-quark pair production with NNLL threshold resummation}'',} \textit{ Nucl.
  Phys. B} \textbf{ 855} (2012) 695,
  \href{http://dx.doi.org/10.1016/j.nuclphysb.2011.10.021}{\doi{10.1016/j.nuclphysb.2011.10.021}},
\href{http://www.arXiv.org/abs/1109.1536}{\texttt{arXiv:1109.1536}}.
%%CITATION = ARXIV:1109.1536;%%.

\bibitem{ttxsec_2}
M.~Cacciari\hrefCMSnoop {}{ {et~al.}, ``{Top-pair production at hadron
  colliders with next-to-next-to-leading logarithmic soft-gluon
  resummation}'',} \textit{ Phys. Lett. B} \textbf{ 710} (2012) 612,
  \href{http://dx.doi.org/10.1016/j.physletb.2012.03.013}{\doi{10.1016/j.physletb.2012.03.013}},
\href{http://www.arXiv.org/abs/1111.5869}{\texttt{arXiv:1111.5869}}.
%%CITATION = ARXIV:1111.5869;%%.

\bibitem{ttxsec_3}
\hrefCMSnoop {}{P.~B{\"a}rnreuther, M.~Czakon, and A.~Mitov,
  ``Percent-level-precision physics at the {Tevatron}: Next-to-next-to-leading
  order {QCD} corrections to $\mathrm{q \bar{q}} \to \mathrm{t \bar{t}} +
  \mathrm{X}$'',} \textit{ Phys. Rev. Lett.} \textbf{ 109} (2012) 132001,
  \href{http://dx.doi.org/10.1103/PhysRevLett.109.132001}{\doi{10.1103/PhysRevLett.109.132001}},
\href{http://www.arXiv.org/abs/1204.5201}{\texttt{arXiv:1204.5201}}.
%%CITATION = ARXIV:1204.5201;%%.

\bibitem{ttxsec_4}
\hrefCMSnoop {}{M.~Czakon and A.~Mitov, ``{NNLO corrections to top-pair
  production at hadron colliders: the all-fermionic scattering channels}'',}
  \textit{ JHEP} \textbf{ 12} (2012) 054,
  \href{http://dx.doi.org/10.1007/JHEP12(2012)054}{\doi{10.1007/JHEP12(2012)054}},
\href{http://www.arXiv.org/abs/1207.0236}{\texttt{arXiv:1207.0236}}.
%%CITATION = ARXIV:1207.0236;%%.

\bibitem{ttxsec_5}
\hrefCMSnoop {}{M.~Czakon and A.~Mitov, ``{NNLO corrections to top pair
  production at hadron colliders: the quark-gluon reaction}'',} \textit{ JHEP}
  \textbf{ 01} (2013) 080,
  \href{http://dx.doi.org/10.1007/JHEP01(2013)080}{\doi{10.1007/JHEP01(2013)080}},
\href{http://www.arXiv.org/abs/1210.6832}{\texttt{arXiv:1210.6832}}.
%%CITATION = ARXIV:1210.6832;%%.

\bibitem{ttxsec_6}
\hrefCMSnoop {}{M.~Czakon, P.~Fiedler, and A.~Mitov, ``{Total top-quark
  pair-production cross section at hadron colliders through
  O($\alpha^4_S$)}'',} \textit{ Phys. Rev. Lett.} \textbf{ 110} (2013) 252004,
  \href{http://dx.doi.org/10.1103/PhysRevLett.110.252004}{\doi{10.1103/PhysRevLett.110.252004}},
\href{http://www.arXiv.org/abs/1303.6254}{\texttt{arXiv:1303.6254}}.
%%CITATION = ARXIV:1303.6254;%%.

\bibitem{ttxsec_7}
\hrefCMSnoop {}{M.~Czakon and A.~Mitov, ``{Top++: A program for the calculation
  of the top-pair cross-section at hadron colliders}'',} \textit{ Comput. Phys.
  Commun.} \textbf{ 185} (2014) 2930,
  \href{http://dx.doi.org/10.1016/j.cpc.2014.06.021}{\doi{10.1016/j.cpc.2014.06.021}},
\href{http://www.arXiv.org/abs/1112.5675}{\texttt{arXiv:1112.5675}}.
%%CITATION = ARXIV:1112.5675;%%.

\bibitem{stxsec_1}
M.~Aliev\hrefCMSnoop {}{ {et~al.}, ``{HATHOR: HAdronic Top and Heavy quarks
  crOss section calculatoR}'',} \textit{ Comput. Phys. Commun.} \textbf{ 182}
  (2011) 1034,
  \href{http://dx.doi.org/10.1016/j.cpc.2010.12.040}{\doi{10.1016/j.cpc.2010.12.040}},
\href{http://www.arXiv.org/abs/1007.1327}{\texttt{arXiv:1007.1327}}.
%%CITATION = ARXIV:1007.1327;%%.

\bibitem{stxsec_2}
P.~Kant\hrefCMSnoop {}{ {et~al.}, ``{HATHOR} for single top-quark production:
  Updated predictions and uncertainty estimates for single top-quark production
  in hadronic collisions'',} \textit{ Comput. Phys. Commun.} \textbf{ 191}
  (2015) 74,
  \href{http://dx.doi.org/10.1016/j.cpc.2015.02.001}{\doi{10.1016/j.cpc.2015.02.001}},
\href{http://www.arXiv.org/abs/1406.4403}{\texttt{arXiv:1406.4403}}.
%%CITATION = ARXIV:1406.4403;%%.

\bibitem{Powheg_ST_tch_4f}
\hrefCMSnoop {}{R.~Frederix, E.~Re, and P.~Torrielli, ``{Single-top $t$-channel
  hadroproduction in the four-flavour scheme with POWHEG and aMC@NLO}'',}
  \textit{ JHEP} \textbf{ 09} (2012) 130,
  \href{http://dx.doi.org/10.1007/JHEP09(2012)130}{\doi{10.1007/JHEP09(2012)130}},
\href{http://www.arXiv.org/abs/1207.5391}{\texttt{arXiv:1207.5391}}.
%%CITATION = ARXIV:1207.5391;%%.

\bibitem{Powheg_ST_sch}
\hrefCMSnoop {}{S.~Alioli, P.~Nason, C.~Oleari, and E.~Re, ``{NLO single-top
  production matched with shower in POWHEG: $s$- and $t$-channel
  contributions}'',} \textit{ JHEP} \textbf{ 09} (2009) 111,
  \href{http://dx.doi.org/10.1088/1126-6708/2009/09/111}{\doi{10.1088/1126-6708/2009/09/111}},
  \href{http://www.arXiv.org/abs/0907.4076}{\texttt{arXiv:0907.4076}}.
[Erratum: \DOI{10.1007/JHEP02(2010)011}].
%%CITATION = ARXIV:0907.4076;%%.

\bibitem{Powheg_ST_tW}
\hrefCMSnoop {}{E.~Re, ``{Single-top Wt-channel production matched with parton
  showers using the POWHEG method}'',} \textit{ Eur. Phys. J. C} \textbf{ 71}
  (2011) 1547,
  \href{http://dx.doi.org/10.1140/epjc/s10052-011-1547-z}{\doi{10.1140/epjc/s10052-011-1547-z}},
\href{http://www.arXiv.org/abs/1009.2450}{\texttt{arXiv:1009.2450}}.
%%CITATION = ARXIV:1009.2450;%%.

\bibitem{DiagramRemoval}
S.~Frixione\hrefCMSnoop {}{ {et~al.}, ``{Single-top hadroproduction in
  association with a W boson}'',} \textit{ JHEP} \textbf{ 07} (2008) 029,
  \href{http://dx.doi.org/10.1088/1126-6708/2008/07/029}{\doi{10.1088/1126-6708/2008/07/029}},
\href{http://www.arXiv.org/abs/0805.3067}{\texttt{arXiv:0805.3067}}.
%%CITATION = ARXIV:0805.3067;%%.

\bibitem{FEWZ}
\hrefCMSnoop {}{Y.~Li and F.~Petriello, ``{Combining QCD and electroweak
  corrections to dilepton production in FEWZ}'',} \textit{ Phys. Rev. D}
  \textbf{ 86} (2012) 094034,
  \href{http://dx.doi.org/10.1103/PhysRevD.86.094034}{\doi{10.1103/PhysRevD.86.094034}},
\href{http://www.arXiv.org/abs/1208.5967}{\texttt{arXiv:1208.5967}}.
%%CITATION = ARXIV:1208.5967;%%.

\bibitem{GEANT}
\hrefCMSnoop {}{{GEANT4} Collaboration, ``{\GEANTfour}---a simulation
  toolkit'',} \textit{ Nucl. Instrum. Meth. A} \textbf{ 506} (2003) 250,
\href{http://dx.doi.org/10.1016/S0168-9002(03)01368-8}{\doi{10.1016/S0168-9002(03)01368-8}}.
%%CITATION = NUIMA,A506,250;%%.

\bibitem{CMS-PRF-14-001}
\hrefCMSnoop {}{{CMS Collaboration}, ``{{Particle-flow reconstruction and
  global event description with the CMS detector}}'',} \textit{ JINST} \textbf{
  12} (2017) P10003,
  \href{http://dx.doi.org/10.1088/1748-0221/12/10/P10003}{\doi{10.1088/1748-0221/12/10/P10003}},
\href{http://www.arXiv.org/abs/1706.04965}{\texttt{arXiv:1706.04965}}.
%%CITATION = ARXIV:1706.04965;%%.

\bibitem{eIDDescription}
\hrefCMSnoop {}{{CMS Collaboration}, ``{Performance of electron reconstruction
  and selection with the CMS detector in proton-proton collisions at
  ${\sqrt{s}} = 8\,\text{TeV} $}'',} \textit{ JINST} \textbf{ 10} (2015)
  P06005,
  \href{http://dx.doi.org/10.1088/1748-0221/10/06/P06005}{\doi{10.1088/1748-0221/10/06/P06005}},
\href{http://www.arXiv.org/abs/1502.02701}{\texttt{arXiv:1502.02701}}.
%%CITATION = ARXIV:1502.02701;%%.

\bibitem{muIDDescription}
\hrefCMSnoop {}{{CMS Collaboration}, ``{Performance of CMS muon reconstruction
  in pp collision events at $\sqrt{s}=7$ TeV}'',} \textit{ JINST} \textbf{ 7}
  (2012) P10002,
  \href{http://dx.doi.org/10.1088/1748-0221/7/10/P10002}{\doi{10.1088/1748-0221/7/10/P10002}},
\href{http://www.arXiv.org/abs/1206.4071}{\texttt{arXiv:1206.4071}}.
%%CITATION = ARXIV:1206.4071;%%.

\bibitem{TnP}
\hrefCMSnoop {}{{CMS Collaboration}, ``{Measurement of the inclusive $W$ and
  $Z$ production cross sections in pp collisions at $\sqrt{s}=7$ TeV with the
  CMS experiment}'',} \textit{ JHEP} \textbf{ 10} (2011) 132,
  \href{http://dx.doi.org/10.1007/JHEP10(2011)132}{\doi{10.1007/JHEP10(2011)132}},
\href{http://www.arXiv.org/abs/1107.4789}{\texttt{arXiv:1107.4789}}.
%%CITATION = ARXIV:1107.4789;%%.

\bibitem{Cacciari:2008gp}
\hrefCMSnoop {}{M.~Cacciari, G.~P. Salam, and G.~Soyez, ``The anti-\kt jet
  clustering algorithm'',} \textit{ JHEP} \textbf{ 04} (2008) 063,
  \href{http://dx.doi.org/10.1088/1126-6708/2008/04/063}{\doi{10.1088/1126-6708/2008/04/063}},
  \href{http://www.arXiv.org/abs/0802.1189}{\texttt{arXiv:0802.1189}}.

\bibitem{FASTJET}
\hrefCMSnoop {}{M.~Cacciari, G.~P. Salam, and G.~Soyez, ``{FastJet user
  manual}'',} \textit{ Eur. Phys. J. C} \textbf{ 72} (2012) 1896,
  \href{http://dx.doi.org/10.1140/epjc/s10052-012-1896-2}{\doi{10.1140/epjc/s10052-012-1896-2}},
\href{http://www.arXiv.org/abs/1111.6097}{\texttt{arXiv:1111.6097}}.
%%CITATION = ARXIV:1111.6097;%%.

\bibitem{JECJERUnc}
\hrefCMSnoop {}{{CMS Collaboration}, ``{Jet energy scale and resolution in the
  CMS experiment in pp collisions at 8 TeV}'',} \textit{ JINST} \textbf{ 12}
  (2017) P02014,
  \href{http://dx.doi.org/10.1088/1748-0221/12/02/P02014}{\doi{10.1088/1748-0221/12/02/P02014}},
\href{http://www.arXiv.org/abs/1607.03663}{\texttt{arXiv:1607.03663}}.
%%CITATION = ARXIV:1607.03663;%%.

\bibitem{BTag}
\hrefCMSnoop {}{{CMS Collaboration}, ``{Identification of b-quark jets with the
  CMS experiment}'',} \textit{ JINST} \textbf{ 8} (2013) P04013,
  \href{http://dx.doi.org/10.1088/1748-0221/8/04/P04013}{\doi{10.1088/1748-0221/8/04/P04013}},
\href{http://www.arXiv.org/abs/1211.4462}{\texttt{arXiv:1211.4462}}.
%%CITATION = ARXIV:1211.4462;%%.

\bibitem{BTagRun2}
\hrefCMSnoop {}{{CMS Collaboration}, ``{Identification of heavy-flavour jets
  with the CMS detector in pp collisions at 13 TeV}'',} (2017).
  \href{http://www.arXiv.org/abs/1712.07158}{\texttt{arXiv:1712.07158}}.
Submitted to {\it JINST}.
%%CITATION = ARXIV:1712.07158;%%.

\bibitem{Buckley:2010ar}
A.~Buckley\hrefCMSnoop {}{ {et~al.}, ``{RIVET user manual}'',} \textit{ Comput.
  Phys. Commun.} \textbf{ 184} (2013) 2803,
  \href{http://dx.doi.org/10.1016/j.cpc.2013.05.021}{\doi{10.1016/j.cpc.2013.05.021}},
\href{http://www.arXiv.org/abs/1003.0694}{\texttt{arXiv:1003.0694}}.
%%CITATION = ARXIV:1003.0694;%%.

\bibitem{Collaboration:2267573}
\href {https://cds.cern.ch/record/2267573}{{CMS Collaboration}, ``{Object
  definitions for top quark analyses at the particle level}'',} Technical
  Report CMS-NOTE-2017-004, 2017.

\bibitem{TUnfold}
\hrefCMSnoop {}{{S Schmitt}, ``{{TUnfold, an algorithm for correcting migration
  effects in high energy physics}}'',} \textit{ {JINST}} \textbf{ {7}} ({2012})
  \href{http://dx.doi.org/{10.1088/1748-0221/7/10/T10003}}{\doi{{10.1088/1748-0221/7/10/T10003}}}.

\bibitem{CMS-PAS-LUM-17-001}
\href {http://cds.cern.ch/record/2257069}{{CMS Collaboration}, ``{CMS}
  luminosity measurements for the 2016 data taking period'',} CMS Physics
  Analysis Summary CMS-PAS-LUM-17-001, 2017.

\bibitem{PileupUnc}
\hrefCMSnoop {}{{CMS Collaboration}, ``{Measurement of the inelastic
  proton-proton cross section at $\sqrt{s}=13~\mathrm{TeV}$}'',} (2018).
  \href{http://www.arXiv.org/abs/1802.02613}{\texttt{arXiv:1802.02613}}.
Submitted to {\it JHEP}.
%%CITATION = ARXIV:1802.02613;%%.

\bibitem{tchannelxsec}
\hrefCMSnoop {}{{CMS Collaboration}, ``{Cross section measurement of t-channel
  single top quark production in pp collisions at $\sqrt s =$ 13 TeV}'',}
  \textit{ Phys. Lett. B} \textbf{ 772} (2017) 752,
  \href{http://dx.doi.org/10.1016/j.physletb.2017.07.047}{\doi{10.1016/j.physletb.2017.07.047}},
\href{http://www.arXiv.org/abs/1610.00678}{\texttt{arXiv:1610.00678}}.
%%CITATION = ARXIV:1610.00678;%%.

\bibitem{Wbb8TeV}
\hrefCMSnoop {}{{CMS Collaboration}, ``{Measurement of the production cross
  section of a W boson in association with two b jets in pp collisions at
  $\sqrt{s} = 8{\,\mathrm{{TeV}}} $}'',} \textit{ Eur. Phys. J. C} \textbf{ 77}
  (2017) 92,
  \href{http://dx.doi.org/10.1140/epjc/s10052-016-4573-z}{\doi{10.1140/epjc/s10052-016-4573-z}},
\href{http://www.arXiv.org/abs/1608.07561}{\texttt{arXiv:1608.07561}}.
%%CITATION = ARXIV:1608.07561;%%.

\bibitem{Zbb8TeV}
\hrefCMSnoop {}{{CMS Collaboration}, ``{Measurements of the associated
  production of a Z boson and b jets in pp collisions at ${\sqrt{s}} = 8\,\text
  {TeV} $}'',} \textit{ Eur. Phys. J. C} \textbf{ 77} (2017) 751,
  \href{http://dx.doi.org/10.1140/epjc/s10052-017-5140-y}{\doi{10.1140/epjc/s10052-017-5140-y}},
\href{http://www.arXiv.org/abs/1611.06507}{\texttt{arXiv:1611.06507}}.
%%CITATION = ARXIV:1611.06507;%%.

\bibitem{pdg2016}
\hrefCMSnoop {}{{Particle Data Group}, C.~Patrignani {et~al.}, ``Review of
  particle physics'',} \textit{ Chin. Phys. C} \textbf{ 40} (2016) 100001,
  \href{http://dx.doi.org/10.1088/1674-1137/40/10/100001}{\doi{10.1088/1674-1137/40/10/100001}}.

\bibitem{fsrTuning}
\hrefCMSnoop {}{P.~Skands, S.~Carrazza, and J.~Rojo, ``{Tuning PYTHIA 8.1: the
  Monash 2013 Tune}'',} \textit{ Eur. Phys. J. C} \textbf{ 74} (2014) 3024,
  \href{http://dx.doi.org/10.1140/epjc/s10052-014-3024-y}{\doi{10.1140/epjc/s10052-014-3024-y}},
\href{http://www.arXiv.org/abs/1404.5630}{\texttt{arXiv:1404.5630}}.
%%CITATION = ARXIV:1404.5630;%%.

\bibitem{PetersonFragmentation}
\hrefCMSnoop {}{C.~Peterson, D.~Schlatter, I.~Schmitt, and P.~M. Zerwas,
  ``Scaling violations in inclusive $\mathrm{{e}^{+}{e}}^{\ensuremath{-}}$
  annihilation spectra'',} \textit{ Phys. Rev. D} \textbf{ 27} (1983) 105,
  \href{http://dx.doi.org/10.1103/PhysRevD.27.105}{\doi{10.1103/PhysRevD.27.105}}.

\bibitem{QCDCR}
\hrefCMSnoop {}{J.~R. Christiansen and P.~Z. Skands, ``{String formation beyond
  leading colour}'',} \textit{ JHEP} \textbf{ 08} (2015) 003,
  \href{http://dx.doi.org/10.1007/JHEP08(2015)003}{\doi{10.1007/JHEP08(2015)003}},
\href{http://www.arXiv.org/abs/1505.01681}{\texttt{arXiv:1505.01681}}.
%%CITATION = ARXIV:1505.01681;%%.

\bibitem{GluonMoveCR}
\hrefCMSnoop {}{S.~Argyropoulos and T.~Sj{\"o}strand, ``{Effects of color
  reconnection on $\mathrm{t\bar{t}}$ final states at the LHC}'',} \textit{
  JHEP} \textbf{ 11} (2014) 043,
  \href{http://dx.doi.org/10.1007/JHEP11(2014)043}{\doi{10.1007/JHEP11(2014)043}},
\href{http://www.arXiv.org/abs/1407.6653}{\texttt{arXiv:1407.6653}}.
%%CITATION = ARXIV:1407.6653;%%.

\end{thebibliography}\endgroup
